\documentclass[12pt]{spieman}
\usepackage[T1]{fontenc}
\usepackage[utf8]{inputenc}
\usepackage{amsmath,amsfonts,amssymb}
\usepackage{graphicx}
\usepackage{setspace}
\usepackage{tocloft}
\usepackage{tabularx}
\usepackage{float}
\usepackage{pdfpages}
\usepackage{twemojis}
\usepackage{newunicodechar}
\newunicodechar{‐}{-}
\newunicodechar{×}{\ensuremath{\times}}
\makeatletter
\def\oc@movep{\@tempa}
\let\@citey\relax

\makeatother

\title{A Comprehensive Review of Large Language Models for Nanophotonics: From Surrogate Modeling to Autonomous Design}
\author[a]{Huanshu Zhang}
\author[b]{Kegeng Tang}
\author[a]{Lei Kang}
\author[a]{Sawyer D. Campbell}
\author[b]{Zihao Wang}
\author[a,*]{Douglas H. Werner}
\affil[a]{The Pennsylvania State University, Department of Electrical Engineering, University Park, PA 16802, USA}
\affil[b]{University of Tennessee at Chattanooga, Department of Computer Science and Engineering, Chattanooga, TN 37403, USA}

\cftpagenumbersoff{figure}
\cftpagenumbersoff{table}
\begin{document}
\maketitle
\begin{abstract}
Metasurfaces have revolutionized the development of photonic devices by enabling unprecedented precision in light manipulation. However, their design processes are often constrained by computationally expensive simulations and complex high-dimensional design spaces. Although deep learning has accelerated the design process by serving as a surrogate model, it remains constrained by task-specific architectures and lacks universal reasoning capabilities. This review surveys how Large Language Models (LLMs) are adding semantic interfaces, code generation, and tool orchestration to established numerical nanophotonic workflows. We first outline the development from classical neural networks to transformer-based models and their applications in nanophotonic design. We then review the emergence of LLM-related methods in nanophotonics and organize them into two operational modes: surrogate models that treat structure-spectrum mapping as a language task, and agentic systems that have been demonstrated to generate code, orchestrate selected simulation steps, and support closed-loop optimization. Furthermore, to identify future cross-disciplinary opportunities, we briefly explore applications of LLMs in research fields such as materials science and wireless communications. This review concludes by looking ahead to the next generation of multimodal foundation models with physical perception capabilities. In this vision, artificial intelligence is evolving from passive tools into active collaborators, participating in autonomous scientific discovery.
\end{abstract}
{\noindent\footnotesize\textbf{*}Corresponding Author, E-mail: \linkable{dhw@psu.edu}}
\begin{spacing}{2}

\section{Introduction}

Over the past two decades, human-engineered nanomaterials, known as metamaterials, have achieved new levels of control over electromagnetic radiation, realizing properties not found in natural materials\cite{ref001}. Early microwave metamaterials achieved negative refractive indices and many other properties through subwavelength structural design\cite{ref002,ref003,ref004}. Metamaterials have also inspired revolutionary concepts such as cloaking, which were first experimentally verified in the microwave band in 2006\cite{ref005}. By scaling these structures to optical wavelengths, researchers have developed a wide variety of optical metamaterials, including photonic crystals\cite{ref006} and plasmonic nanostructures\cite{ref007,ref008}, enabling applications such as ultrathin lenses (metalenses)\cite{ref009} and holography\cite{ref010}. Optical metamaterials form a platform enabling high-precision control over the phase, amplitude, and polarization of light, thereby opening new possibilities for next-generation photonic devices.

Metasurfaces have gained increasing attention since the concept was first introduced in 2011 along with the generalized Snell’s law\cite{ref011}. Despite their origins in metamaterials, metasurfaces have shown great success in controlling light-matter interaction via single-layer or few-layer planar artificial structures\cite{ref012,ref013}. In contrast to metamaterials in which properties of light vary due to their bulk effective material properties, metasurfaces can manipulate light by engineering local interfacial properties. These characteristics make metasurfaces an attractive ultracompact platform for light control, enabling a number of exotic phenomena including near-arbitrary beam reflection/refraction\cite{ref011,ref014}, vortex beam generation\cite{ref015,ref016}, and holography\cite{ref017,ref018}. Beyond the linear regime, the resonant nature of metasurfaces has also been exploited to demonstrate enhanced nonlinearities with tailored local nonlinear phase profiles\cite{ref019,ref020,ref021}.

Although metasurfaces have been demonstrated as a powerful system for light manipulation, the corresponding designs have generally been considered a challenge, especially for those that are expected to realize complex functionalities, such as aberration-corrected lensing\cite{ref022,ref023} and chromatic holography\cite{ref024,ref025}. Conventionally, the design of a metasurface starts with unit cell modeling in which periodic boundary conditions are applied to mimic an infinitely large array. While sufficient for simple functionalities, this approach becomes computationally prohibitive for high-performance devices that must satisfy multiple simultaneous constraints. Rigorous demands such as wide-field uniformity\cite{ref026,ref027}, broadband operation\cite{ref028,ref029}, and the mitigation of parasitic near-field coupling\cite{ref030,ref031}, render brute-force simulation and iterative tuning both time-consuming and suboptimal. To accelerate this process, researchers typically employ optimization algorithms\cite{ref032}. Adjoint and topology optimization retain an explicit Maxwell solver in every iteration and obtain gradients at a cost that is largely independent of the number of design variables, enabling free-form, high-degree-of-freedom structures and direct incorporation of minimum-feature-size, discreteness, and robustness constraints\cite{ref033,ref034,ref035}. Nevertheless, these iterative, target-specific methods can remain computationally demanding and sensitive to initialization in nonconvex design spaces\cite{ref036,ref037}, particularly for multiobjective, broadband, or large-area problems\cite{ref028,ref038,ref039,ref040}. Therefore, there is an urgent need for specialized design strategies that can significantly enhance computational efficiency.

To address this gap, deep learning (DL) has emerged as a promising approach, offering substantially greater efficiency than traditional numerical simulations or optimization methods. Existing reviews have established broad foundations for AI-assisted photonics, metasurface design, and nanophotonic inverse design\cite{ref041,ref042,ref043,ref044,ref045,ref046}. Feng et al. survey the bidirectional relationship between photonics and AI across design, imaging, communications, and optical computing\cite{ref041}; Saifullah et al. focus on DL-enabled metasurface design and adaptive metadevices\cite{ref042}; Kim et al. organize optimization- and DL-based nanophotonic inverse design and implementation tools\cite{ref047}; and Fu et al. cover AI methods across advanced metasurface research\cite{ref048}. Our earlier review adopts a model-centric view of AI-enabled metadevices, including high-degree-of-freedom and fabrication-aware design, with only a brief treatment of emerging LLM assistance\cite{ref045} (Table 1). There is no dedicated review specifically addressing the emerging capabilities of large language models (LLMs), which play roles such as surrogate models, semantic reasoners, code-generating agents, and workflow coordinators in the nanophotonics design process. This review has a narrower and different objective: we define the boundary among task-specific transformers, domain language-model-style generators, adapted pre-trained LLMs, RAG systems, and agents. We also compare representations, model scales, data, metrics, autonomy, and validation, and propose physical-validity and reporting protocols. We first trace the evolutionary trajectory from classical machine learning (ML) to transformer architecture, which is the foundation of LLMs (Section 2). Next, we review the specific applications of transformer models in metasurface design (Section 3), before turning to the core theme of this paper: the application of LLMs in nanophotonics (Section 4). We also broaden our scope to include recent advances in other relevant fields (Section 5) to explore interdisciplinary opportunities that may inspire LLM-enabled nanophotonics in the future. Finally, we offer a forward-looking perspective on this rapidly evolving field. The organization and scope of this transition are summarized in Figure 1.

\begin{table}[H]
\caption{Recent Related Reviews}
\label{tab:related-reviews}
\begin{tabularx}{\textwidth}{@{}p{0.14\textwidth}X@{}}
\hline
\textbf{Review} & \textbf{Primary scope} \\
41 & Broad two-way interaction: classical DNNs for photonic design, imaging, communication, and materials \\
42 & Classical DNNs for metasurface prediction/design and adaptive metadevices \\
47 & Optimization, adjoint methods, discriminative/generative DL, devices, tools, and foundries \\
48 & Classical AI methods across metasurface elements and optical systems \\
45 & Model-centric AI-enabled metadevice design, high-DoF geometry, robustness, fabrication limitations, and transformers \\
49 & AI across accelerated electromagnetic modeling and inverse design, optical-data characterization, end-to-end imaging, and autonomous metasurface systems \\
50 & Bidirectional AI-metasurface relationship: DL-assisted design and metasurface-based optical and microwave physical neural networks \\
Current review & Task-specific transformers; domain token generators; adapted LLMs; VLMs; foundation models; RAG; agents; adjacent-field transfer \\
\hline
\end{tabularx}
\end{table}
\nocite{ref049,ref050}

\begin{figure}[H]
\centering
\includegraphics[width=\textwidth,height=0.62\textheight,keepaspectratio]{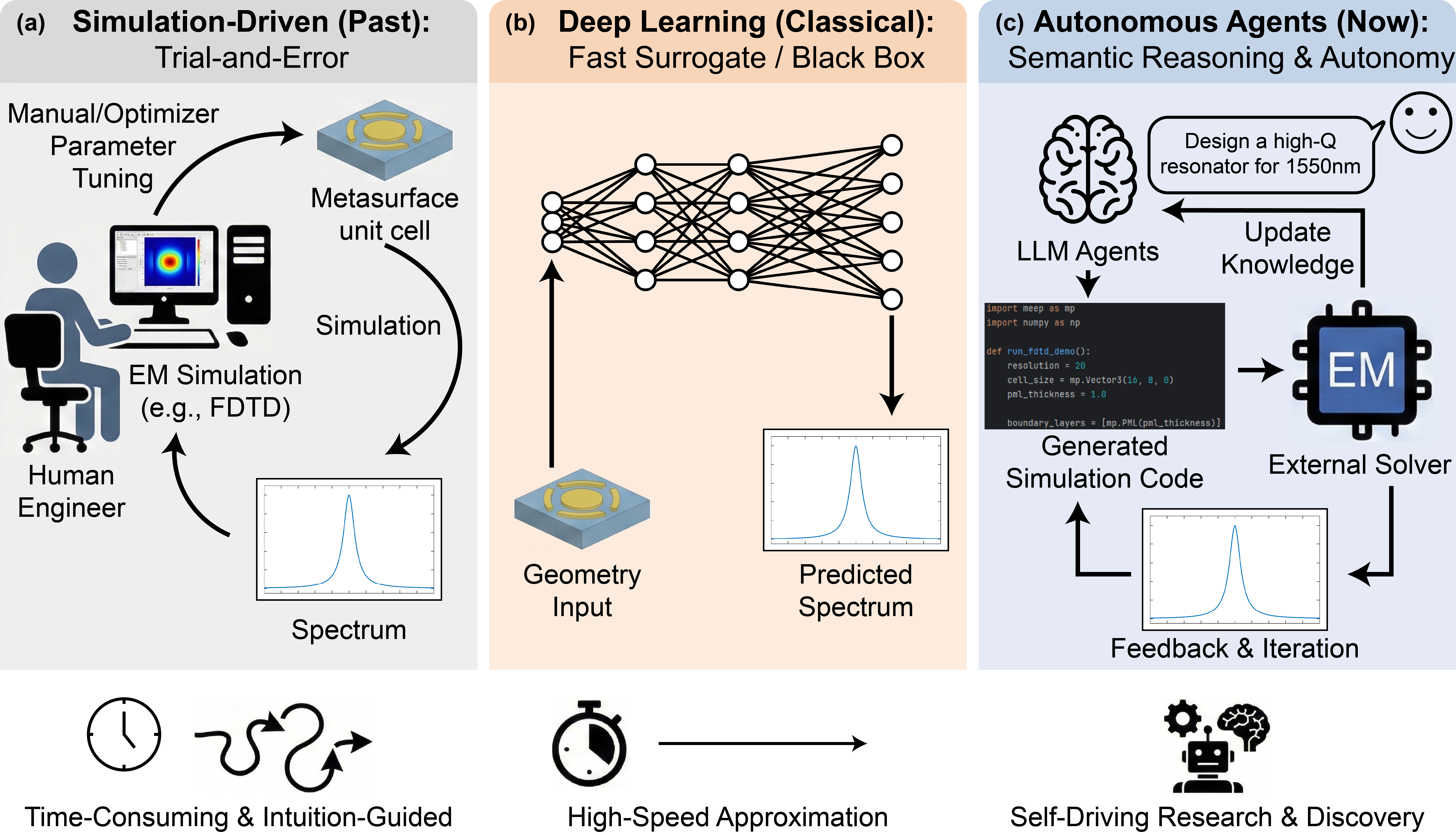}
\caption{The paradigm shift in the design of nanophotonic devices.}
\label{fig:1}
\end{figure}

\section{Deep Learning and Large Language Models}

Addressing the fundamental design bottlenecks necessitates a transition from intuition-guided trial-and-error to high-throughput, data-centric methodologies. There is a growing demand for approaches that can uncover non-intuitive correlations between geometric parameters and optical responses within vast, multi-dimensional parameter spaces\cite{ref045,ref051,ref052}. Ideally, such a framework would not only accelerate the search for optimal structures but also predict performance without relying exclusively on resource-intensive full-wave simulations for every iteration. The demand for versatile design engines has led to the use of advanced computational tools. These tools learn complex physics, effectively accelerating the transition from design targets to final devices. 

To avoid conflating architecture, training scale, modality, and system function, we use the following terminology throughout this review. A transformer is an attention-based neural network architecture first introduced by Vaswani et al.\mbox{ }\cite{ref053}. A language model learns probability distributions over token sequences. In this review, an LLM denotes a broadly pre-trained language model used across tasks through prompting or fine-tuning. A foundation model is defined by broad pre-training and adaptability across downstream tasks and may be linguistic, visual, or multimodal. A vision-language model (VLM) specifically connects visual and language representations. These labels describe different dimensions, including architecture, modality, training scale, and workflow role, and are therefore not mutually exclusive. For example, a VLM may also be a foundation model, and an agent may use an LLM for planning while relying on a non-LLM surrogate or Maxwell solver for physical prediction. Conversely, a decoder-only transformer trained from scratch on domain tokens is language-model-like in architecture but is not treated here as a broadly pre-trained LLM. A surrogate model approximates a physical input-output mapping and is classified by its role rather than its parameter count. Retrieval-augmented generation (RAG) grounds generation in externally retrieved evidence. An autonomous agent places a model within an iterative loop involving planning, memory, tools, observations, and stopping criteria. Accordingly, the studies reviewed below are separated into: (i) task-specific/domain transformer surrogates trained from scratch, (ii) pre-trained LLMs adapted by prompting or fine-tuning, and (iii) agentic systems. For each work, we identify the architecture, pre-training and adaptation strategy, input and output modalities, role in the design loop, and validation method.

\subsection{Machine learning and deep learning}

Machine learning (ML) enables computational models to infer relationships from data rather than relying exclusively on manually encoded rules\cite{ref054,ref055,ref056}. Classical methods such as support vector machines (SVMs) can be effective for classification and regression\cite{ref057,ref058} and remain useful for low-dimensional, highly parameterized nanophotonic components\cite{ref059}. Their dependence on hand-selected features limits their scalability to high-dimensional geometries\cite{ref060}. Deep learning (DL) addresses this limitation by learning hierarchical representations directly from vectors, grids, or sequences\cite{ref061}. The appropriate architecture depends on the data representation and task\cite{ref062}: multilayer perceptrons (MLPs) are an early and important class of fully connected network for fixed-length parameter vectors\cite{ref063}; convolutional neural networks (CNNs) efficiently capture local spatial structure in images and pixelated device topologies\cite{ref064,ref065,ref066}; and recurrent neural networks (RNNs), including long short-term memory (LSTM)\cite{ref067,ref068} and Gated Recurrent Unit (GRU)\cite{ref069}, model ordered sequences. These network architectures are already well-established for nanophotonics applications\cite{ref036,ref070,ref071,ref072,ref073} (Figure 2). For example, Peurifoy et al. trained a forward surrogate for multilayer-particle scattering and performed inverse design by backpropagating through the learned model\cite{ref074} (Figure 2b), while Liu et al. introduced a tandem forward-inverse architecture to address the nonuniqueness and data inconsistency of inverse design (Figure 2c)\cite{ref075}. Residual networks use identity or shortcut paths to improve gradient propagation and make residual mappings easier to optimize\cite{ref076}, while U-Nets preserve multiscale spatial information through encoder–decoder and lateral connections\cite{ref077}. Generative adversarial networks (GANs)\cite{ref078} and diffusion models\cite{ref079} can represent one-to-many design distributions when several structures satisfy the same target. An early metasurface demonstration by Liu et al. coupled a generative network to a learned spectral simulator to produce candidate patterns conditioned on target optical responses\cite{ref080} (Figure 2d). A parallel branch of scientific machine learning focuses on embedding or learning the governing field equations. Physics-informed neural networks (PINNs) incorporate partial-differential-equation residuals into training\cite{ref081}, and MaxwellNet specialized this principle to Maxwell’s equations to predict electromagnetic fields without requiring supervised field labels (Figure 2e)\cite{ref082}. Neural-operator architectures such as the Fourier neural operator\cite{ref083} and DeepONet\cite{ref084} learn mappings between function spaces, while electromagnetic neural operators have subsequently been used as surrogate solvers for free-form 3D inverse design\cite{ref085} (Figure 2f). These specialist methods and their nanophotonic applications have already been reviewed extensively\cite{ref041,ref047,ref048,ref086,ref087}.

\begin{figure}[H]
\centering
\includegraphics[width=\textwidth,height=0.62\textheight,keepaspectratio]{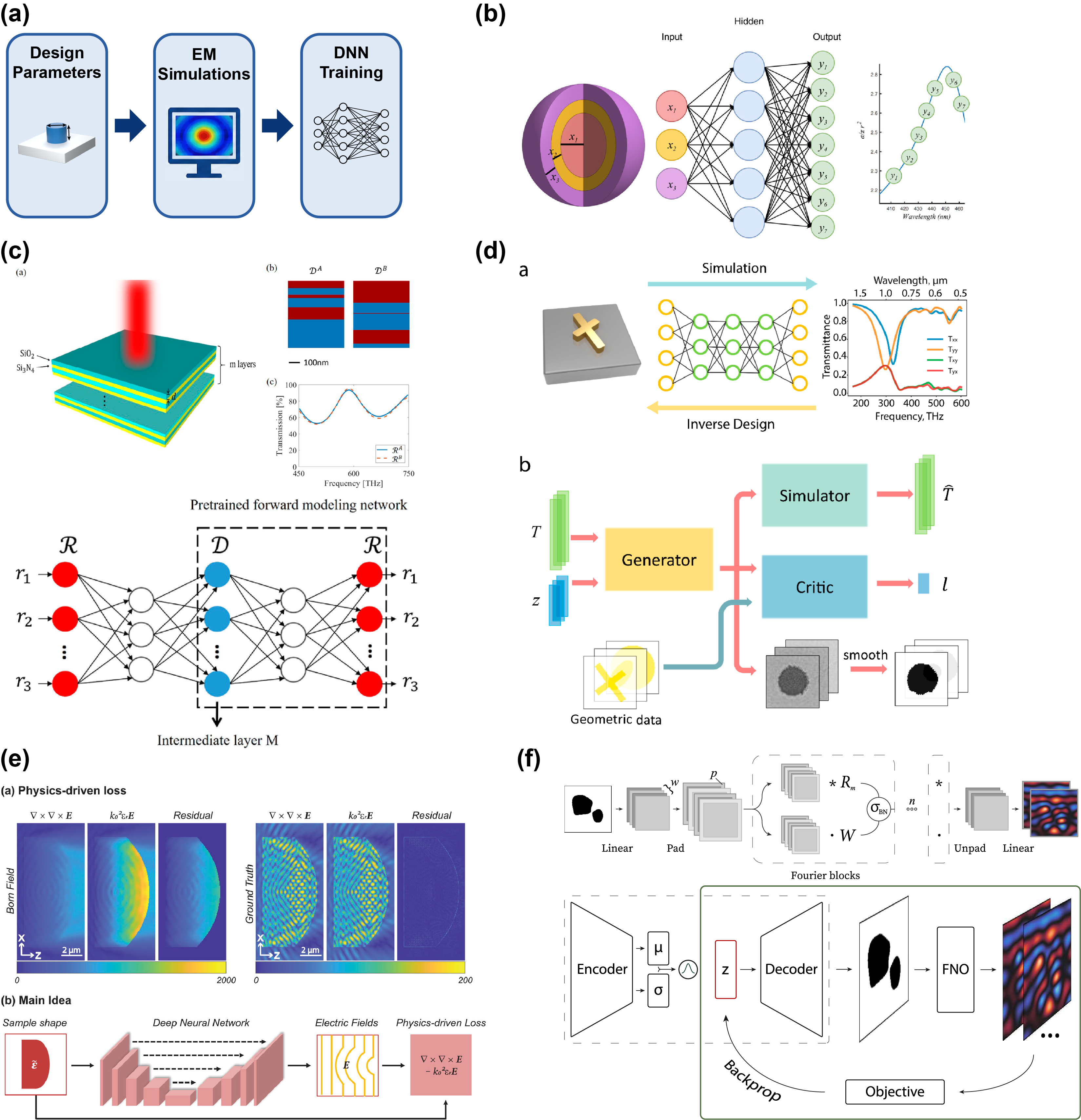}
\caption{Several preliminary and foundational works of DL-enabled nanophotonics. (a) The general workflow of DL-enabled nanophotonics. (b) Forward prediction and inverse design of multilayer-particle scattering. Reprinted with permission from\cite{ref074}, Copyright 2018, The American Association for the Advancement of Science. (c) Tandem network for addressing nonuniqueness and data inconsistency of inverse design. Reproduced with permission from\cite{ref075}. Copyright 2018 American Chemical Society. (d) A generative network for producing candidate patterns conditioned on target optical responses. Reprinted with permission from\cite{ref080}. Copyright 2018, American Chemical Society. (e) MaxwellNet. Reprinted with permission from\cite{ref082}. This article is distributed under a CC BY license. (f) Neural operator-based surrogate solver for free-form metasurface inverse design. Reprinted with permission from\cite{ref085}, Copyright 2023 American Chemical Society.}
\label{fig:2}
\end{figure}

\subsection{From neural language processing to transformers}

Natural language processing (NLP) seeks to enable computers to understand, interpret, and generate human language\cite{ref088}. While RNNs and LSTMs provided an early way to process text as a sequence of tokens, they were fundamentally constrained by their sequential nature, which made them hard to parallelize during training and limited their ability to capture very long-range dependencies\cite{ref089}. The transformer addresses these limitations by replacing recurrence with self-attention, which lets the model dynamically weigh the importance of different parts of the input data regardless of their separation in the sequence\cite{ref053,ref090} (Figure 3a). In nanophotonics, a token may represent a wavelength interval, spatial patch, material layer, geometric parameter, or polarization channel. Attention is therefore useful when distant spectral or spatial elements jointly determine the electromagnetic response.

The original transformer contains an encoder that represents the input and a decoder that generates the output, both constructed from multi-head attention and position-wise feed-forward blocks\cite{ref053} (Figure 3a). Encoder-only models such as BERT use bidirectional context and are suited to representation learning and discriminative prediction\cite{ref091}, which can align with forward mappings from a device description to an optical response. Decoder-only models such as the GPT family generate tokens autoregressively\cite{ref092} (Figure 3b), making them suitable for variable-length outputs such as material-layer sequences, symbolic geometries, or code. Architecture alone, however, does not establish that a model is an LLM: an encoder or decoder trained from scratch on nanophotonic data remains a task-specific or domain transformer under the taxonomy adopted in this review.

Large-scale transformer language models are commonly trained through self-supervised objectives such as masked-token or next-token prediction on broad text and code corpora\cite{ref092}. Empirical scaling laws relate performance to model size, data, and compute within the regimes studied\cite{ref093,ref094}, while broad pre-training provides a reusable starting point for downstream tasks\cite{ref095} (Figure 3c). A pre-trained model can then be specialized through fine-tuning\cite{ref089} using domain examples such as paired metasurface geometries and spectra\cite{ref089,ref096,ref097}. This process may reduce task-specific architecture development, but pre-training does not guarantee accurate material knowledge, Maxwell consistency, or generalization outside the adaptation data. These distinctions separate task-specific transformers discussed in Section 3 from the broadly pre-trained and subsequently adapted LLMs discussed in Section 4.

\subsection{Large language models}

The scaling of decoder-only transformer architectures to hundreds of billions of parameters has contributed to the emergence of LLMs, a class of artificial intelligence distinct not only in size but in capability. While structurally similar to their smaller predecessors, LLMs exhibit a qualitative leap in performance often described as "emergent behavior," where the model acquires complex problem-solving abilities that were not explicitly encoded in the training objective\cite{ref098}. Unlike earlier NLP tools restricted to specific tasks like translation or classification, modern LLMs function as general-purpose computational models. They demonstrate a capacity for high-level reasoning and knowledge representation, enabling them to navigate interdisciplinary domains ranging from creative writing to complex scientific analysis, without necessitating task-specific architectural modifications (Figure 3c). For nanophotonics, this capability shifts AI from a fixed regression tool toward a semantic interface that can interpret design intent (such as specs, constraints, and fabrication limits), propose candidate implementations, and draft the computational steps needed to validate them. 

Training at this scale depends on distributed systems that combine data parallelism with forms of model parallelism and memory optimization to fit and efficiently update models with very large parameter counts. Practical toolchains such as intra-layer model parallelism for transformers and optimizer-state partitioning have been central to pushing model sizes into the multi-billion and trillion-parameter regimes\cite{ref099,ref100}.

Pure scaling, however, has clear limitations. Training and inference costs grow rapidly, and the environmental and social externalities of ever-larger models have been highlighted as key concerns\cite{ref101}. Moreover, larger models are not automatically aligned with user intent or reliability goals, motivating alignment methods such as instruction tuning with human feedback\cite{ref102}. Finally, hallucination remains a persistent failure mode in open-ended generation, underscoring the need for better evaluation, calibration, and mitigation beyond simply increasing scale\cite{ref103}.

\subsection{Fine-tuning strategies}

While the pre-training and fine-tuning paradigm offers a clear path to domain specialization of LLMs, the practical execution of this phase presents significant computational hurdles. Conventionally, adaptation involves full parameter fine-tuning (FFT), where every weight in the model is updated via backpropagation. For modern LLMs containing billions of parameters, FFT is highly resource-intensive, requiring GPU memory comparable to the initial pre-training phase to store optimizer states and gradients\cite{ref104}. Furthermore, modifying all parameters risks "catastrophic forgetting," where the model loses the general reasoning capabilities acquired during pre-training\cite{ref105}. To mitigate these costs, the field has shifted toward Parameter-Efficient Fine-Tuning (PEFT)\cite{ref105}. Rather than retraining the full model, PEFT typically freezes the entire backbone of pre-trained weights to preserve their general knowledge and significantly reduce memory requirements (Figure 3d). Adaptation is then achieved by injecting small, trainable neural network layers (called adapters) between the frozen transformer blocks, appending learnable continuous vectors (soft prompts) to the input sequence, or exclusively unfreezing and updating a minute fraction of specific components, such as the bias terms, while keeping the rest of the model static\cite{ref104,ref105,ref106}.  

Among the various PEFT strategies, Low-Rank Adaptation (LoRA) has emerged as a particularly widely used technique. LoRA operates on the hypothesis that the change in weights during model adaptation has a low "intrinsic rank" (often from single digits to a few tens, tuned per task), meaning the necessary updates can be represented effectively by much smaller matrices. Instead of updating the full dense weight matrix directly, LoRA decomposes the weight update into two low-rank matrices, optimizing only these small matrices while keeping the original pre-trained weights frozen\cite{ref107} (Figure 3d). This method can reduce the number of trainable parameters by several orders of magnitude compared to FFT\cite{ref108}. Practically, PEFT (especially LoRA) makes it feasible for photonics groups to adapt an off-the-shelf LLM into a domain-specific surrogate or code-assistant method using lab-scale datasets and modest computational resources.

\subsection{Operational paradigms for engineering and design}

While fine-tuning strategies like LoRA significantly lower the barrier to specializing models, modifying weights is not always necessary or feasible for dynamic engineering workflows. An alternative approach leverages the pre-trained model’s inherent adaptability through prompt engineering and In-Context Learning (ICL)\cite{ref109,ref110}. In this paradigm, the burden of adaptation shifts from the model parameters to the input context window (Figure 3e). By strategically designing the textual prompt, engineers can guide the LLM to reason through complex problems using "zero-shot" instructions or "few-shot" examples without a single gradient update\cite{ref111}. For instance, an engineer might provide the geometric parameters and transmission spectra of three known metasurface unit cells within the prompt itself. The model, utilizing its attention mechanism, identifies the underlying correlations in these examples to predict the behavior of the fourth geometry. This method allows for rapid prototyping and logic verification, enabling researchers to exploit the model's generalized physics knowledge for specific tasks purely through natural language interaction.  

However, a reliance on internal parametric knowledge alone exposes a critical vulnerability: the potential for hallucination, particularly regarding specific physical constants or recent experimental data not present in the training corpus. To address this, Retrieval-Augmented Generation (RAG) provides a framework for grounding the model’s reasoning in verifiable external data\cite{ref112,ref113}. Rather than forcing the LLM to memorize the dispersive properties of every optical material, a RAG system intercepts the user’s query, such as a request for a silicon nitride waveguide design, and first searches a connected knowledge base for relevant documents or data tables\cite{ref113} (Figure 3e). This retrieved information, which could be drawn from verified databases, is then injected into the prompt context. Retrieval grounding can reduce unsupported factual claims, but retrieved documentation does not establish that generated geometry, code, or fields satisfy Maxwell’s equations. Physical validity therefore requires an independent executable or numerical test rather than LLM self-consistency alone.

A further operational paradigm is the AI agent, which transitions the LLM from a passive informational tool to an active participant in the design loop. Unlike simple question-answering systems, agents can be configured to use external tools autonomously to achieve a high-level goal\cite{ref114}. For instance, an agent can be instructed to optimize a beam-steering metasurface for a specific figure of merit (FOM). The agent can decompose this objective into executable steps: writing a Python script to define the geometry, triggering an electromagnetic solver, and parsing the resulting output data. To facilitate the complex interoperability required between the LLM and these disparate local or remote software environments, the Model Context Protocol (MCP) has emerged as a standard interface, allowing the model to discover and expose tools and resources through a standardized interface without custom integrations for every tool\cite{ref115}. Through this standardized connection, the agent can coordinate an iterative loop in which an external solver and acceptance checks determine whether the target has been met, effectively automating the trial-and-error components of the engineering workflow. The broader LLM-agent literature provides several reusable patterns for such workflows. ReAct interleaves reasoning traces with actions and observations\cite{ref116}, while Toolformer studies self-supervised selection and use of external tools\cite{ref117}. Reflexion\cite{ref118} and Self-Refine\cite{ref119} use feedback to revise previous outputs, although this feedback does not by itself provide independent physical verification. Tree of Thoughts introduces explicit branching, evaluation, and backtracking for tasks that require search\cite{ref120}, and AutoGen provides a framework for configurable multi-agent conversations\cite{ref121}. For code-generating agents, Codex and the HumanEval benchmark established execution-based pass@k evaluation rather than relying only on textual similarity\cite{ref122}. These methods motivate planning, tool-use, retry, and executable-test metrics for nanophotonic agents.

\begin{figure}[H]
\centering
\includegraphics[width=\textwidth,height=0.62\textheight,keepaspectratio]{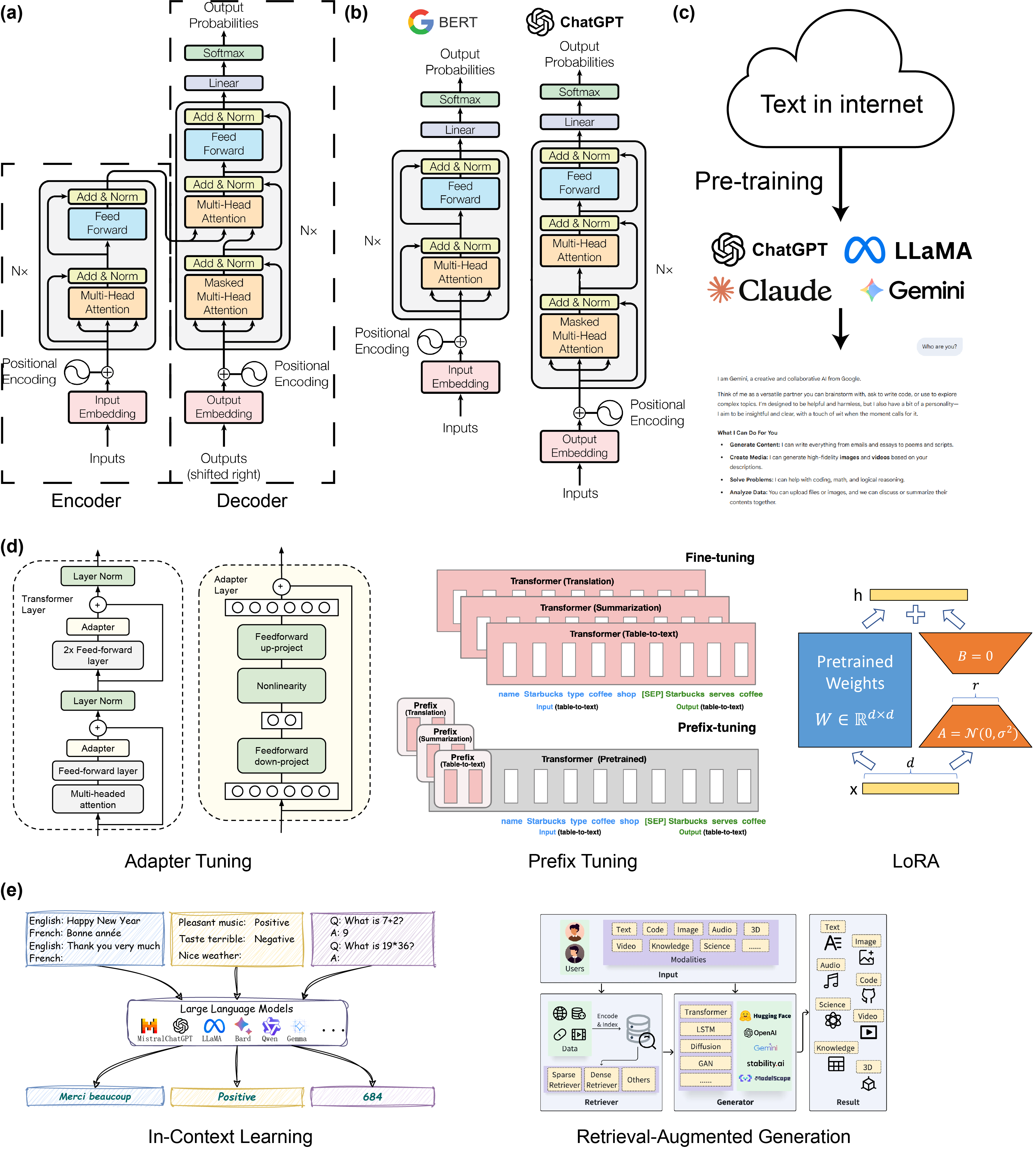}
\caption{Transformers and LLMs. (a) Transformer architecture. Reproduced from\cite{ref053} with permission from Google, which grants reproduction of tables and figures for scholarly works provided that proper attribution is given. (b) Encoder-only model BERT and decoder-only model GPT. (c) Pre-training of LLMs. (d) PEFT methods. Reproduced from\cite{ref107,ref123,ref124}, all licensed under CC BY 4.0. (e) ICL and RAG. Reprinted with permission from\cite{ref125,ref126}. Ref.\mbox{ }\cite{ref125} is licensed under CC BY 4.0,\mbox{ }\cite{ref126} is licensed under CC BY-NC-ND 4.0. }
\label{fig:3}
\end{figure}

\section{Transformers for Nanophotonics}

The attention mechanism allows the transformer to effectively capture long-range dependencies in data, and it inherently exhibits scaling laws. These properties allow performance to be predictably and continuously enhanced as model size and data volume increase, which has become the foundation for the success of large-scale transformer models and further advancements in LLMs\cite{ref093,ref094}. Transformer architectures have been widely adopted in NLP tasks, as understanding long-range semantic dependencies is crucial in language comprehension and generation\cite{ref127}. 

Building upon these capabilities, current nanophotonics research employs transformer architectures in several mainstream paradigms. Spectral patches or physics-aware channel tokens are most useful when distant wavelengths, resonant peaks, or polarization components must be related. CNN-transformers and windowed-vision encodings (Swin Transformer\cite{ref128}) better preserve local structure in pixelated layouts. Autoregressive structure tokens are advantageous for variable-length multilayers or symbolic layouts. These distinctions motivate the comparison of encoder-only, decoder-only, hybrid, and generative architectures included below.

\subsection{Encoder-only transformer models}

Many transformer-based studies have adopted encoder-only architectures as this approach provides a simpler, more data-efficient, and computationally stable entry point for learning global electromagnetic correlations. In a pioneering study, Chen et al. (2023) proposed an encoder-only transformer for inverse design of thin-film broadband solar metamaterial absorbers (SMAs)\cite{ref129}. The framework processes spectral data by dividing the optical spectrum into discrete wavelength segments and treating each segment as a token sequence (Figure 4b). This enables the model to capture global spectral correlations through self-attention mechanisms, eliminating reliance on point-by-point regression methods. This spectrum-patching strategy has since been widely adopted in later transformer-based photonic and metamaterial design studies. More broadly, this work signals a paradigm shift toward transformer-driven AI tools for scalable, data-efficient metamaterial and energy-device engineering. Later, Niu et al. proposed an encoder-only transformer macroscopic design framework that directly maps user-defined far-field performance specifications to physically realizable metasurfaces\cite{ref130}. The workflow consists of a discrete variational autoencoder (dVAE) that first encodes equivalent source distributions into a latent space, while the encoder-only transformer operates on this latent token sequence to learn the mapping from far-field specification to the source. Cai et al. proposed an improved transformer-based neural network for the inverse design of tunable broadband THz reflective metamaterials incorporating monolayer graphene\cite{ref131}. Their model combines a transformer encoder with a stack of four fully connected layers to address the limited generalization and slow convergence of traditional MLP- and CNN-based approaches. In late 2023, Chen et al. extended their transformer framework to design all-dielectric surface-enhanced Raman scattering (SERS) metasurfaces\cite{ref132}. Also in 2023, Lin et al. employed a convex-deep (CODE) neural network to enable stable learning from extremely limited training data\cite{ref133}. The model adopts a U-Net architecture composed of multiple transformer encoder blocks.

Stepping into 2024, Gao et al. proposed the Metaformer, an explainable encoder-only transformer-based DL framework for the intelligent design of high-Q metasurface sensors, addressing a long-standing limitation of conventional “black-box” neural networks in capturing sharp spectral features critical for sensing applications\cite{ref134} (Figure 4c). Adopting a spectrum-splitting strategy, the Metaformer achieves \textasciitilde{}99\% reduction in model parameters while simultaneously reducing prediction MSE by \textasciitilde{}99\%. Importantly, the attention maps offer model-level interpretability, revealing how different attention heads focus on resonant peaks versus off-resonant regions, thereby offering new insights into the physics-AI connection in meta-sensing. Zhang et al. showed that an encoder-only transformer can achieve comparable inverse-design accuracy to a bidirectional GRU when applied to the design of microwave resonant metasurfaces\cite{ref135}. Toward the end of 2024, Sun et al. proposed a MetaE-former, an inverse-design framework based on an encoder-only transformer that acts as an ultrafast, high-accuracy surrogate electromagnetic solver for large-scale metasurfaces, which can directly predict full complex electric-field responses of high-degree-of-freedom metasurfaces\cite{ref136}. At about the same time, Ma et al. presented a lightweight encoder-only transformer-based inverse-design framework for quad-band metasurface absorbers\cite{ref137}.

In 2025, an increasing number of studies began adopting encoder-only transformers in metasurface design. Yin et al. proposed an encoder-only transformer model for predicting the electromagnetic response of high-Q resonators, alongside a CNN incorporating self-attention layers for inverse design\cite{ref138}. Belonovskii et al. introduced an encoder-only transformer architecture with MLPs for photonics and proposed physics-informed "hybrid embeddings" to encode specific physical quantities and structural contexts into the model\cite{ref139}. Yan et al. presented MetasurfaceViT, a universal foundation model based on vision transformers for metasurface inverse design\cite{ref140} (Figure 4d). Their method no longer trains separate networks for fixed wavelengths or polarizations. Instead, it uses the full Jones matrix, which varies with wavelength, to represent optical responses. The model also employs a masked pre-training strategy, learning correlations between different wavelength and polarization channels from a large-scale dataset containing 60 million samples. This study overcomes the longstanding lack of universality in DL-driven nanophotonics and demonstrates a Vision-Transformer-based foundation-model approach in metasurface design. Chen et al. introduced a "reality-infused" framework\cite{ref141} which is centered on an encoder-only transformer-based model that bridges the longstanding gap between theoretical simulations and experimental realities in quasi-optical Fourier surface design. The authors generated a large-scale training dataset comprising over 25,000 measured spectra by utilizing high-throughput angle-resolved imaging spectroscopy (ARS). This enables the model to capture sources of real-world variation and optical dispersion characteristics that simulations typically overlook. This work represents an important step toward real-world data-centric optical design. Building on their earlier approach\cite{ref137}, Ma et al. integrated a pyramid attention mechanism that progressively extracts features from global to local scales. Concurrently, they integrated a Learnable Feature Fusion (LFF) module that dynamically adjusts weights to prioritize attention toward these high-value peak regions\cite{ref142}. Li and Bogdanov proposed MetaDiT, a generative framework that combines a pre-trained spectral encoder via contrastive learning with a diffusion transformer for inverse design\cite{ref143}. This method enables joint optimization of all structural parameters while satisfying high-resolution spectral constraints. Liao et al. developed an encoder-only transformer that utilizes a dual-branch attention mechanism to capture both long-range dependencies and fine local details in spectral data\cite{ref144}. Building upon this foundation, Bian et al. added a gating mechanism that allows the model to simultaneously learn multiple distinct structural configurations\cite{ref145}. Xin et al. developed a model based on the Swin transformer\cite{ref128} architecture designed to predict the optical properties of surface-emitting lasers in photonic crystals\cite{ref146}. Chu et al. (2026) developed CSSformer, a classification-assisted transformer-encoder framework that uses spectrum splitting to jointly identify multilayer core-shell configurations and geometric parameters, outperforming an MLP in forward and inverse design\cite{ref147}.

\subsection{Other transformer architectures and attention mechanisms}

Despite these efforts, encoder-only architectures often face challenges when handling generative tasks, such as synthesizing variable-length structures or complex image topologies, while also struggling to maintain robustness when confronted with shifts in data distribution. To address these issues, other studies have explored different network architectures to achieve more flexible and accurate inverse design. In 2023, Zeng et al. addressed data shifts in AI-based electromagnetic solvers\cite{ref148}. A data shift occurs when a network trained on random structures performs poorly on optimized, high-efficiency structures drawn from a different distribution. Taking a one-dimensional grating coupler as a case study, the authors proposed a "mixed training" strategy that injects shifted data into the training set. This approach was combined with an enhanced ResNet architecture incorporating multi-head attention layers to strengthen global feature extraction capabilities. In 2024, Huang et al. proposed an inverse-design framework for THz multi-resonant graphene metasurfaces\cite{ref149}. This framework combines an enhanced encoder-decoder transformer network with a conditional GAN (CGAN) to directly generate metasurface structure images from target absorption spectra. In the same year, Yuan et al. proposed a multitask DL framework that combines encoder-decoder transformers and CNNs to decouple the classification of phase-change-material crystallinity from geometric regression, thereby enabling the precise, on-demand inverse design of active, tunable mid-infrared bandpass filters embedded in Fabry-Perot cavities\cite{ref150}. Mao et al. (2024) introduced a transformer-based inverse-design framework for ultra-broadband THz polarization converter design\cite{ref151}. The model employs a transformer architecture featuring an input head with 1D convolutions, a core module utilizing encoder-decoder stacks with self-attention mechanisms, and an output head to accurately map target spectra to meta-atom geometric parameters. In 2024, Ma et al. introduced OptoGPT, a decoder-only transformer trained for multilayer thin-film optics, as one of the first decoder-only transformer-based inverse-design tools and foundation models\cite{ref152} (Figure 4e). Conventional AI workflows for designing multilayer thin-film metasurfaces often fix the material set/order, the number of layers, the illumination conditions (e.g., incident angle), and DNN’s input size, meaning a network trained for a 5-layer structure cannot process a 10-layer structure. OptoGPT addressed this by employing “structure tokens” and “structure serialization” to represent each layer as a combination of material and thickness (e.g., TiO2\_10), then encoding these representations as text tokens and training the model to generate corresponding designs based on the target spectrum. Because the transformer generates sequences autoregressively, OptoGPT is capable of generating variable-length layer sequences, which enables a single model to design thin-film stacks with different numbers of layers. This work represents a significant step toward decoder-only transformer-driven nanophotonic design, demonstrating that a language-model-style transformer can directly generate thin-film designs from target optical spectra. Dang et al. (2025) developed Meta-GPT and a METASTRINGS encoding method to represent metasurface layouts as text sequences\cite{ref153}. Similar to OptoGPT’s “structure tokens” and “structure serialization”, the authors formalized metasurface designs as ordered symbolic strings. These strings explicitly encode information such as materials, layer thicknesses, lattice structures, and geometric parameters. The success of Meta-GPT underscores the importance of structured prompt engineering in bridging the gap between human understanding and AI-driven photonic design. Wu et al. (2025) combined CNN-based local feature extraction with Transformer encoders in a surrogate-assisted particle swarm optimization framework for reconfigurable liquid-crystal THz metasurface design\cite{ref154}. In 2025, Shan and Jiang proposed a ViT inverse hybrid genetic particle swarm optimization framework for the dynamic design of absorption and phase-cancellation metasurfaces aimed at broadband radar cross-section reduction\cite{ref155} (Figure 4f). The core network architecture combined a 2D convolutional vision module with a transformer encoder-decoder, where pixelated metasurface unit cells were first encoded into rich 2D features and then processed by multi-head attention to predict broadband complex electromagnetic responses. Marzban et al. (2025) introduced HiLAB, a hybrid inverse-design framework that combines early-terminated adjoint topology optimization with a pre-trained ViT-based VAE and Bayesian optimization, compressing high-dimensional freeform metasurface geometries into an eight-dimensional latent space while jointly optimizing physical parameters\cite{ref156}. For an achromatic RGB beam deflector, HiLAB achieved a 24.7\% worst-case diffraction efficiency using about 1,400 electromagnetic simulations, compared with approximately 14,000 forward solves for conventional multi-start topology optimization, while allowing the learned VAE representation to be reused for related design objectives. Supplementary Table S1 compares the architectures, representations, datasets, reported performance, computational resources, availability, validation levels, and scope boundaries of these transformer-based studies.

\begin{figure}[H]
\centering
\includegraphics[width=\textwidth,height=0.62\textheight,keepaspectratio]{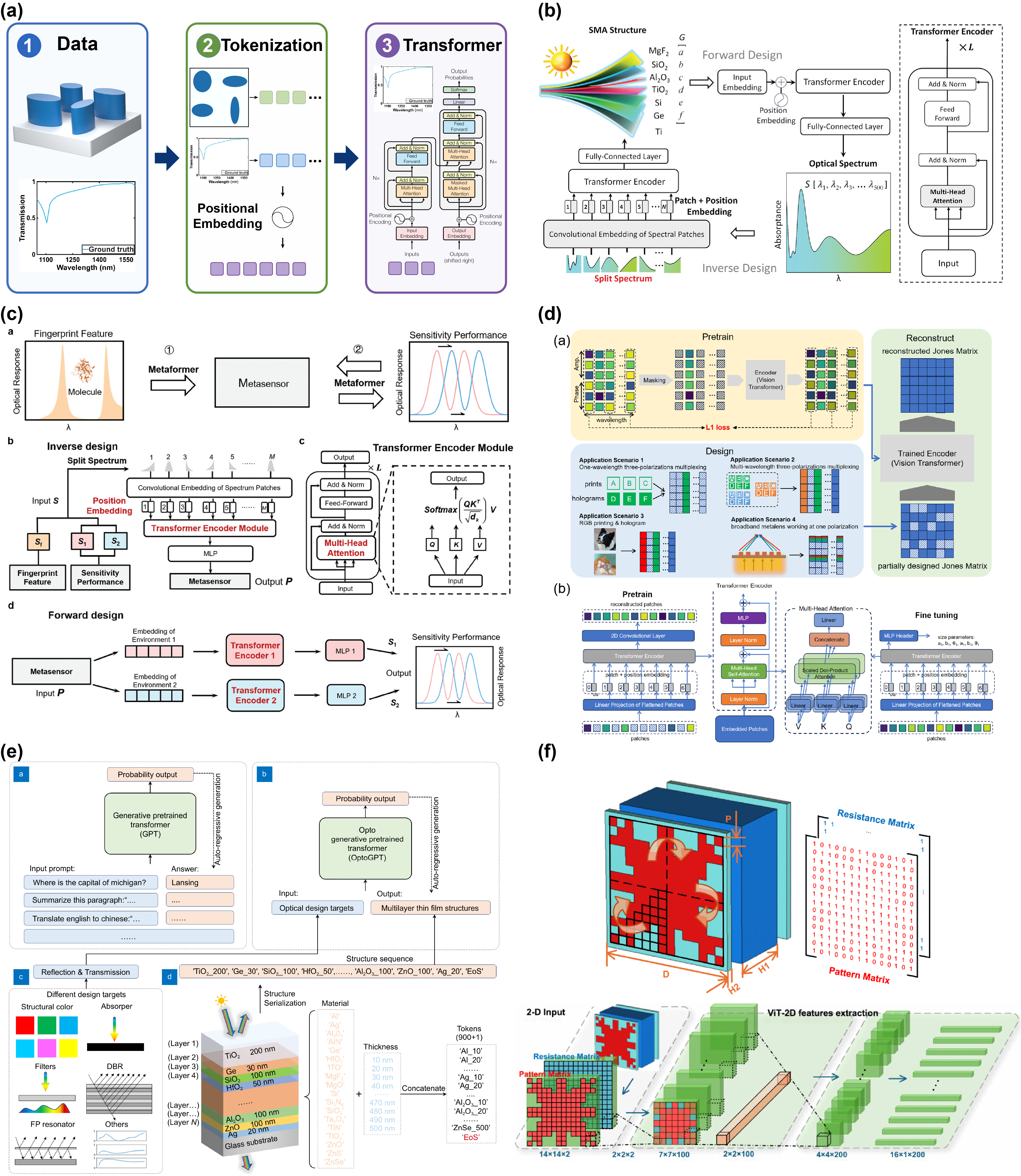}
\caption{Transformers for nanophotonics. (a) General workflow of transformers for nanophotonics. (b) Encoder-only transformers for the design of broadband solar metamaterial absorbers. Reproduced from\cite{ref129}, licensed under CC BY 4.0. (c) Metaformer for development of metasurface sensors. Reproduced from\cite{ref134}, licensed under CC BY 4.0. (d) MetasurfaceViT. Reproduced from\cite{ref140}, licensed under CC BY 4.0. (e) OptoGPT. Reproduced from\cite{ref152}, licensed under CC BY 4.0. (f) Vision transformer for the design of absorption and phase-cancellation metasurfaces. Reproduced from\cite{ref155}. Copyright 2026 IEEE.}
\label{fig:4}
\end{figure}

The rapid advancement of transformer-based architectures in metasurface design indicates that electromagnetic modeling is undergoing a fundamental shift from local feature extraction to global context awareness. The evidence supports a task- and representation-dependent choice rather than a universal transformer implementation. Transformers are most compelling for long-range spectral or cross-channel dependencies\cite{ref133,ref139,ref143}, CNN or Swin hybrids for local 2D free-form geometries\cite{ref146,ref155}, decoder-only tokens for variable-length discrete stacks or layouts\cite{ref152,ref153}, and diffusion models for one-to-many topology generation\cite{ref143}. Explicit modeling of non-local interactions is particularly beneficial for resonant metasurfaces, whose optical response is dominated by collective interference between meta-atoms. This has led to significant efficiency gains, as demonstrated by several previous studies that utilized far fewer parameters than classical neural networks while achieving similar accuracy. Furthermore, advances in tokenization have transformed metasurface design into a flexible sequential modeling task, making it possible to unify different data representations spanning thin-film layer sequences, dielectric shapes, and plasmonic metasurface layouts. 

However, the application of transformers also comes with several significant challenges. A major limitation is their lack of inductive bias. CNNs encode locality and translation equivariance, while RNNs encode sequential order. Transformers must learn positional semantics from scratch. Consequently, they typically require significantly larger datasets to converge, as evidenced by the use of millions of training samples in numerous studies. This "data hunger" has created a bottleneck in nanophotonics, as generating high-fidelity simulation data is computationally expensive. Moreover, this creates a paradox in certain scenarios where data can be generated rapidly. If physical solvers (e.g., the transfer matrix method for thin-film metasurfaces) are already sufficiently efficient to produce millions of training samples within a reasonable time, the necessity for transformer surrogate models diminishes substantially. This may reduce the value of a surrogate, as the computational cost of training a complex model to bypass an already fast simulation yields negligible acceleration benefits. Additionally, the computational complexity and memory requirements of self-attention mechanisms exhibit quadratic growth (O(N\textsuperscript{2})), posing significant obstacles when processing high-resolution spectral data or tackling large-scale topology optimization tasks. Consequently, to maintain computational feasibility, special partitioning strategies (such as the spectrum-splitting strategy proposed by Chen et al.\cite{ref129}) or more aggressive down-sampling are often needed. Beyond data and computational resources, implementation barriers are equally significant. Stabilizing transformer training often requires complex and meticulous tuning of hyperparameters, including learning rate, model dimensions, warm-up schedules, positional encoding choices, and attention masking. These factors are more sensitive and harder to master compared to classical regression neural networks. Thus, this process typically demands substantial human effort and specialized expertise, potentially limiting its accessibility within the broader physics community.

Several mitigation strategies might address different parts of the attention bottleneck. Sparse or local attention restricts each token to a neighborhood or selected global tokens, reducing complexity for long sequences. Longformer, for example, combines windowed local attention with task-specific global attention and scales linearly with sequence length\cite{ref157}. Kernelized linear-attention methods such as Performer approximate softmax attention with random features and achieve linear time and space in sequence length, with an approximation-versus-fidelity trade-off\cite{ref158}. FlashAttention computes exact attention but tiles the operation to reduce reads and writes between GPU memory levels, which improves wall-clock time and memory use\cite{ref159}. Representation-level compression is equally important. Vision Transformers replace per-pixel tokens with image patches, reducing token count at the cost of potentially losing features below the patch scale\cite{ref160}. Perceiver-style architectures cross-attend large inputs into a fixed-size learned latent array, so the expensive latent self-attention no longer scales quadratically with the raw input size\cite{ref161}. Flamingo uses a Perceiver Resampler to convert variable-size visual features into a fixed number of visual tokens before conditioning a language model, showing how pixel-based encoders and language reasoning can be combined without serializing every pixel as text\cite{ref162}. Several of these mitigations already appear in the studies reviewed above. For example, refs.\mbox{ }\cite{ref129,ref134} patch optical spectra to shorten broadband or high-Q spectral sequences and reduce model size, ref.\mbox{ }\cite{ref130} converts far-field specifications and equivalent-source distributions into learned dVAE tokens, and refs.\mbox{ }\cite{ref136,ref146} use neighborhood or shifted-window attention. 

Looking forward, pre-trained LLMs may complement task-specific transformer frameworks for applications in which machine-learning expertise, human effort, or task-specific data are limited, by shifting the paradigm from "training" to "prompting." While standard transformers require training and hyperparameter optimization for specific tasks, pre-trained LLMs possess emergent reasoning capabilities. During their extensive pre-training process, these models incorporate vast amounts of cross-domain knowledge, including physical principles and material properties. This potentially lowers the required number of training samples, thereby largely reducing the need for users to run an impractical number of simulations or customize network architectures for each new task. Beyond this, the pre-configured nature of LLMs offers a distinct “off-the-shelf” advantage. By alleviating the need to design model architectures and/or manage complex training schedules, they can reduce reliance on specialized human expertise\cite{ref163,ref164}. Moreover, LLMs lower the technical barrier by providing natural language interfaces, enabling physicists to iterate designs using semantic descriptions without writing complex code. For example, “Optimize high-Q factor in the near-infrared". Ultimately, by combining LLM-based language processing with external physics solvers in an agentic workflow, LLMs can use tools to revise outputs in response to tool feedback and coordinate the validation of candidate designs with less human intervention. This raises a broader question: does greater model complexity necessarily lead to better engineering outcomes, or can it be justified by the simpler and more accessible user experience it provides?

\section{LLMs for Nanophotonics}

The integration of LLMs into nanophotonics represents an important step forward in classical DNN workflows. Although DNNs have proven to be effective surrogate solvers, one of their inherent limitations is that researchers often need to redesign the network architecture and training strategy for each new task. In contrast, LLMs offer a promising path toward reducing task-specific architecture redesign by reusing a fixed pre-trained backbone. Their pre-trained knowledge base enables them to achieve few-shot generalization under various physical conditions. Additionally, their variable-length tokenization supports the design of diverse topological structures without requiring model architecture redesign. In agentic systems, LLMs have evolved from passive computational tools into active agents capable of writing executable code, planning experiments, and interacting with external physics solvers, thereby enabling semi-automation of selected design steps. In nanophotonics, the application of LLMs for intelligent metasurface design is still in its early stages, with studies emerging in 2024 and rapidly expanding to a variety of application scenarios.  

To date, the integration of LLMs into nanophotonics has developed along several directions, each leveraging the unique capabilities of LLMs. First, LLMs are trained like DNNs to accelerate optimization by learning the mappings between geometric structures and optical responses\cite{ref165}. Second, LLMs can work as autonomous agents that coordinate multi-step design loops, such as planning, generating simulation code, screening candidates, and even operating within multi-agent frameworks paired with Maxwell solvers for faster metasurface discovery\cite{ref166}. Third, LLMs can serve as agents that ingest literature databases and utilize RAG to propose and optimize candidate nanostructures based on existing scientific knowledge\cite{ref167}.

\subsection{LLM surrogates for optical metasurface simulation}

Similar to DNNs, early studies have demonstrated that LLMs can serve as powerful surrogate models by directly learning the mapping between a metasurface’s structure and its optical response\cite{ref045}. Unlike classical DNNs, LLMs use language-based encoding to capture design parameters, which allows for more adaptable and varied metasurface simulations. Lu et al. (2025) fine-tuned GPT-3.5, which was able to achieve comparable accuracy to customized DNNs in predicting dielectric metasurface absorption spectra\cite{ref165} (Figure 5b). This study cast both the metasurface geometries and their corresponding spectra into English-text sequences so that the LLM can process these numerical data. The corresponding results showed that LLMs became competitive surrogates once the training set surpasses \textasciitilde{}1,000 samples, achieving performance metrics comparable to traditional DNNs, and further improving as more data was added. This work marks a further step into LLM-enabled metasurface design and serves as guidance for future studies by highlighting two practical takeaways: (1) accuracy is not significantly influenced by the level of detail in the description and (2) with larger training sets, lower sampling temperatures tend to reduce errors. In early 2025, Kim et al. fine-tuned Llama-3.1-8B into a surrogate model to characterize a TiO2 cuboid on a SiO2 substrate system by training it on text-encoded geometry-spectrum pairs\cite{ref164} (Figure 5c). By simply reversing the input and output, the same method can perform inverse design and generate multiple candidate structures for the same objective without specialized approaches used in DNNs, such as tandem networks. In addition, the authors showed that providing an LLM with domain-specific notes and worked examples through ICL enables it to generate and revise accurate simulation and optimization code without model retraining or substantial user expertise. While their approaches provide a more accessible “no-code” data-driven photonics design, the demonstrated design tasks remain limited to relatively simple multilayer thin films and canonical, low-dimensional metasurface geometries.

Building on these advances, researchers have begun using LLMs to tackle more complex design tasks. Zhang et al. (2025) demonstrated a “Chat-to-Chip” process for arbitrarily shaped metasurfaces\cite{ref163} (Figure 5d). Their key contribution lies in demonstrating that LLMs based on 1D token streams could accurately predict broadband spectra of 2D free-form metasurface patterns through mappings between patterns and sequences\cite{ref168}, and in turn generate candidate geometric structures that match the target spectra. They further fine-tuned and compared multiple open-source LLMs, providing a practical benchmark for model selection in future metasurface design workflows. Their method showed that when flattening an image is not practical, LLMs can still serve as surrogates for free-form shape design by mapping images to sequences. Similarly, Li et al. (2026) applied a fine-tuned Llama-3-7B model to design a free-form nanophotonic power beam splitter\cite{ref169} (Figure 5e). The design region is pixelated to a 20 × 20 binary matrix (0 = etched/air, 1 = unetched/silicon). Instead of mapping shape to sequence as Zhang et al. did, they directly flattened the 20 × 20 matrix to a 1 × 400 vector. Compared to Zhang’s method, this representation incurs substantially higher GPU memory consumption and fine-tuning cost, since the attention matrix of LLMs scales quadratically with the input sequence length. In this case, the text used to describe the shape contained about 400 tokens, whereas Zhang’s method uses only on the order of 16 tokens. Yang et al. (2025) pushed this field further to include reconfigurable multi-state metamaterials, introducing a method called CoSP\cite{ref170} (Figure 5f). They performed contrastive pre-training on spectral data from devices with multiple ON/OFF states, built a spectral encoder, and fed it as input to GPT-2. The resulting model can analyze the target’s multi-state optical response and generate a single metamaterial design that exhibits all of these desired states, whereas previous DL-based inverse-design methods could typically only handle one fixed state at a time. More recently, Zhang et al. (2026) extended the LLM-surrogate paradigm to heterogeneous multi-family meta-optics\cite{zhang2026universalmetaopticssolverlarge}. They encoded geometry identity, wavelength range, structural parameters, and family-specific optical-response channels in a shared instruction-following format and LoRA fine-tuned Gemma-2-9B across eight structurally distinct heterogeneous metasurface families. The joint forward model achieved a global MSE of $1.1\times10^{-3}$ and reduced the MSE of every family relative to the selected single-family LLM baselines, with an average relative reduction of 56.5\%. The same sequence-based representation was also used for inverse design, yielding an overall MSE of $2.06\times10^{-3}$. This work demonstrates that a single LLM backbone can accommodate heterogeneous parameter counts, spectral ranges, and response formats without family-specific input/output architecture redesign, providing a practical path toward a universal meta-optics surrogate solver.

Supplementary Table S2 quantifies the trade-offs among these LLM-surrogate studies. The available evidence favors compact, physically meaningful encodings: low-dimensional parameter strings, low-dimensional wavelength-indexed spectra, and short shape codes retain the relevant design variables without the quadratic attention cost and weakened 2D locality of long flattened-pixel sequences. Natural-language context is most useful for code generation and conversational specification, as demonstrated by the prompt-guided TMM study. Tokenized numerical sequences are most useful when the output is sequential or one-to-many, such as variable-length material stacks, multi-state conditions, or multiple inverse candidates. When geometry cannot be compressed without losing spatial information, CNNs or image-native models are generally more efficient. In general, LLM surrogates are easier to train and use for inference than classical DNN surrogates (such as MLP, CNN, or transformers) when researchers only have limited AI expertise (Figure 5a). However, this ease of implementation may come at the cost of accuracy. For fixed-dimensional forward regression, LLMs often show lower prediction accuracy than matched MLP/DNN baselines. 

\begin{figure}[H]
\centering
\includegraphics[width=\textwidth,height=0.62\textheight,keepaspectratio]{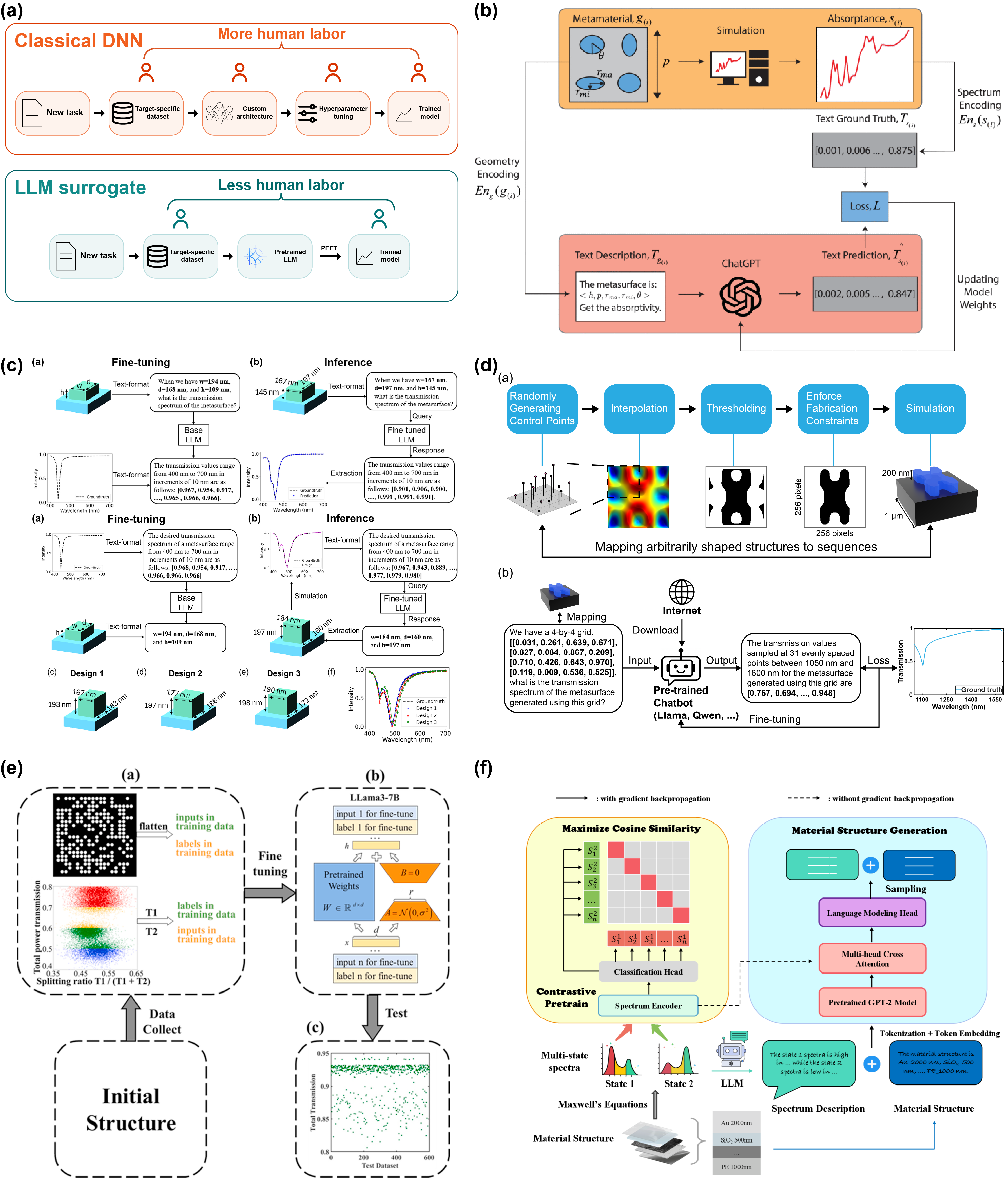}
\caption{LLM surrogates. (a) Comparison of classical DNN and LLM surrogates. (b) Fine-tuning GPT-3.5 for dielectric metasurface. Reproduced from\cite{ref165}, licensed under CC BY 4.0. (c) Fine-tuned Llama-3.1-8B as a surrogate predictor/inverse designer. Reproduced from\cite{ref164}, licensed under CC BY 4.0. (d) Fine-tuned Llama-3.1-8B to map 4 × 4 control-point grids to 31-point transmission spectra. Reproduced from\cite{ref163}, licensed under CC BY 4.0. (e) Fine-tuned Llama-3-7B to predict and inversely design silicon power beam splitter geometries. Reproduced from\cite{ref169}. Copyright 2025 Elsevier. (f) CoSP. Reproduced with permission from\cite{ref170}. }
\label{fig:5}
\end{figure}

\subsection{LLM design agents and human-AI co-design}

A subsequent direction for LLMs in photonics is to make metasurface design more accessible to users without specialized metasurface expertise and to expand what metasurface engineers can accomplish through agentic design frameworks. A recent example is MetaChat, introduced by Lupoiu et al. in late 2025\cite{ref166} (Figure 6b). MetaChat is a multi-agent metasurface design framework consisting of an Agentic Iterative Monologue (AIM) Design Agent and an AIM Materials Expert Agent. The Design Agent (GPT-4o) interprets semantically specified photonic design objectives, requests missing constraints, consults the Materials Expert Agent, and orchestrates external design APIs, numerical tools, optimization algorithms, and electromagnetic solvers. The AIM paradigm implements an iterative reasoning-action-feedback loop in which agents refine their decisions through self-monologue and interactions with tools, other agents, and human designers. They also benchmarked 7 AIM driver models (Supplementary Table S3) among which GPT-4o achieved the highest score. Rather than using LLMs as surrogate solvers, MetaChat uses a classical DNN (FiLM WaveY-Net) surrogate for full-wave simulations while LLM agents coordinate planning, tool use, and optimization. Beyond accelerating metasurface optimization, the broader significance of MetaChat lies in its demonstration of a system-level approach to scientific intelligence. The framework establishes an interface between high-level design objectives and low-level software-based operations, while assigning semantic reasoning, knowledge retrieval, numerical simulation, and gradient-based optimization to specialized components. Nevertheless, the reported results are primarily computational, therefore MetaChat is best viewed as a proof of concept and architectural blueprint for agent-mediated scientific CAD rather than a general autonomous photonic designer. Wu et al. (2026) introduced MetaDesigner, a self-correcting multi-agent system that translates natural-language optical objectives into long-horizon design workflows by coordinating planning, knowledge retrieval, electromagnetic simulation, optimization, code generation, and a dedicated Verifier that detects and repairs errors in reasoning, code, numerical results, and technical reports\cite{ref171}. Across RGB metalens design, six-plane full-color holography, and optoelectronic image style transfer, the system autonomously completed 74-136 reasoning steps and corrected substantive errors during execution, demonstrating a shift from agent-assisted optimization toward verifiable end-to-end orchestration of complex metasurface design workflows.

Meanwhile, Lu et al. (2025) developed an LLM-based agentic framework that autonomously executes several stages of metamaterial inverse-design workflow\cite{ref172} (Figure 6c). In their software-based system, a Planner (OpenAI o4-mini) coordinates three specialized agents (Input Verifier, Forward Modeler, and Inverse Designer) together with memory and API tools for file checking, adaptive data acquisition, DNN training code generation, and neural-adjoint optimization. The training loop dynamically decides whether to expand the dataset, test the current model, generate a new DNN architecture, or terminate based on performance history and resource limits. On an all-dielectric metasurface benchmark, the agent-designed forward models achieved the lowest reported MSEs of 1.3 × 10\textsuperscript{-3}, close to the 1.2 × 10\textsuperscript{-3} human-designed baselines, while producing inverse designs with the best MSE of 1.4 × 10\textsuperscript{-3}. Li et al. (2025) introduced LLM4Laser, a human-AI co-design framework in which GPT-4 supports the design and optimization of photonic-crystal surface-emitting lasers (PCSELs) through conversational interaction\cite{ref173} (Figure 6d). The workflow divides the task into conceptualization, code generation and debugging, and simulation and optimization, with GPT-4 generating FDTD simulation code and PyTorch DQN code while a researcher provides physical specifications and execution feedback. The resulting DQN agent interacts with FDTD to optimize PCSEL design parameters. This approach establishes an end-to-end software workflow that spans from initial concept generation to simulation and computational optimization. Zhao et al. (2025) proposed an LLM “translator” that converts intuitive natural language prompts into numerical inputs for a trained diffusion model\cite{ref174}. Lai et al. (2025) proposed PriM, a principle-guided multi-agent framework that uses GPT-4o for its language-based agents and applies the system to the discovery of chiral nano-helices\cite{ref175}. By adopting a Hypothesis Agent that uses physicochemical constraints via chain-of-principles prompting, and integrating an Optimizer Agent that performs Monte Carlo tree search, the system establishes a closed loop between semantic scientific reasoning and numerical validation. This methodology distinguishes itself by ensuring the exploration of chemical space doesn’t rely solely on statistical patterns but also incorporates mechanism-aware logic. As a result, it achieves improved performance optimization while also providing a transparent and human-understandable decision-making process.

However, due to issues such as hallucinations that have not yet been adequately resolved, LLMs remain unreliable when generating executable research code. For example, they may call non-existent functions or misuse APIs, which limits their feasibility for direct use as research agents. Huang et al. (2025) introduced an MCP-enabled workflow using Claude Sonnet 4 on Claude Desktop to mitigate these issues\cite{ref176} (Figure 6e). By providing Claude Sonnet 4 with structured access to TorchRDIT (a differentiable solver for rigorous diffraction theory), the authors enabled the LLM to construct valid inverse-design scripts, while TorchRDIT performed the electromagnetic modeling, automatic differentiation, and numerical optimization. Through MCP, the LLM can query dedicated servers to retrieve verified code templates and documentation. This hybrid approach significantly reduces errors and computational costs compared to standard RAG (Supplementary Table S3), demonstrating a robust method for integrating LLMs with physics-based engines. Building on this solver-grounded approach, Huang et al. (2026) developed a self-evolving agentic framework in which a coding agent programs a fixed differentiable electromagnetic solver while a meta-agent updates explicit human-readable skill files from deterministic physics-based feedback, enabling reusable optimization strategies to improve without modifying the LLM weights or the solver\cite{ref177}. Skill evolution increased same-type task success from 38\% to 74\% and, in the harder New-type-B setting, from 20\% to 90\%, while simultaneously reducing the average number of design attempts.

Finally, the concept of the "agent" has been transitioned from software to the direct control of experimental electromagnetic hardware by Hu et al. (2025)\cite{ref178} through their metaAgent prototype (Figure 6f). The FPGA-controlled programmable metasurfaces physically reconfigured microwave fields to localize people, phones, and robots, monitor respiration and heartbeat, enhance wireless links, and communicate with a mobile robot. LLMs served as a supervisory layer, interpreting open-ended commands such as “check Alice’s breathing,” decomposing them into ordered subtasks mapped to available hardware functions, and generating executable Python code. The underlying electromagnetic sensing and control operations continued to rely on conventional algorithms. Its LLM-based “cerebrum” comprised sensing, planning, grounding, and coding experts, with GPT-4 achieving the highest reported success rate. This work shifts programmable metamaterials from task-specific adaptive devices to embodied agents that can interpret semantic goals, plan over long horizons, and act autonomously in physical environments. More broadly, it establishes a blueprint for language-grounded intelligent materials that directly couple foundation-model reasoning with electromagnetic hardware, sensing, and robotics.

\begin{figure}[H]
\centering
\includegraphics[width=\textwidth,height=0.62\textheight,keepaspectratio]{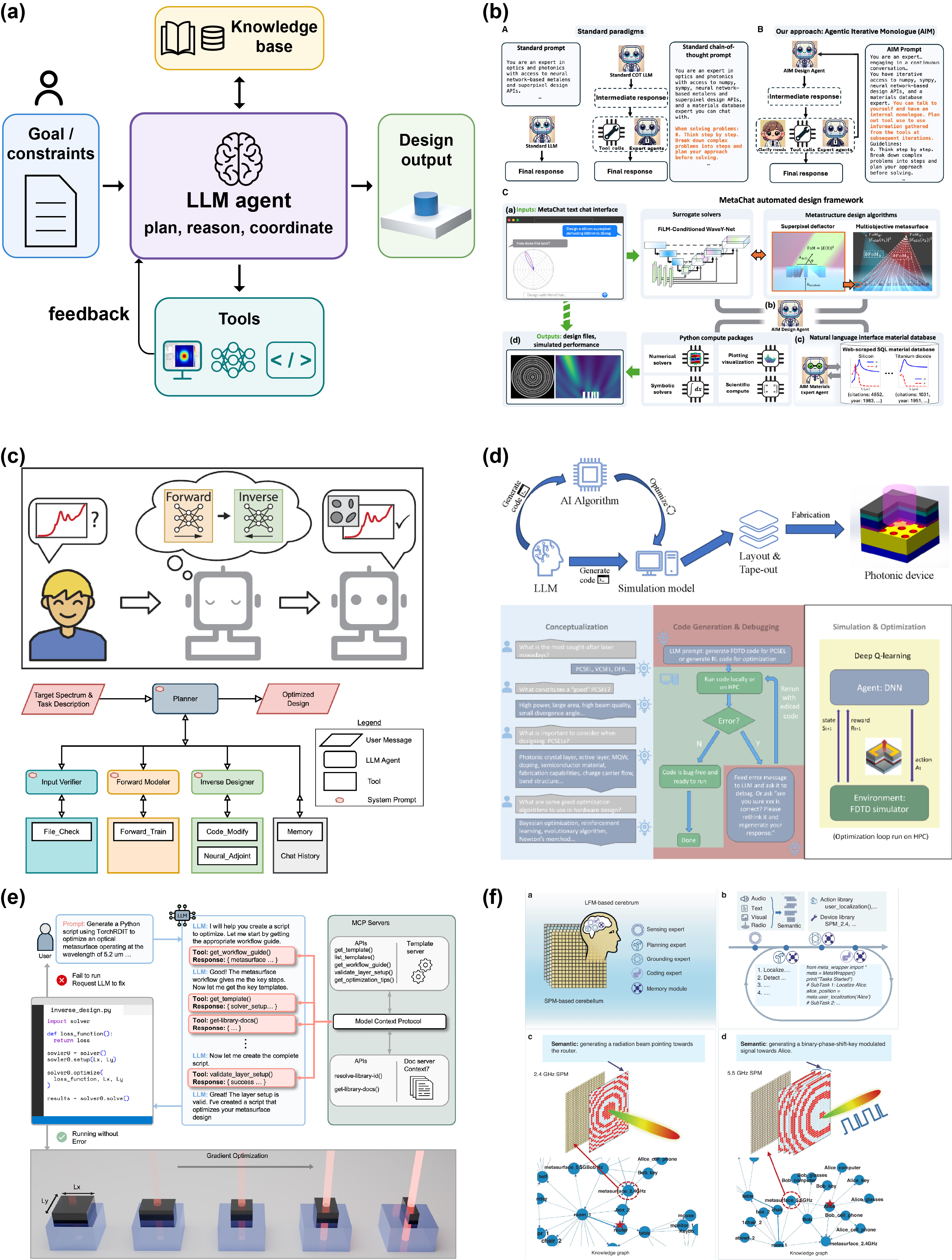}
\caption{LLM agents. (a) Agent-driven design workflow. (b) MetaChat. Reproduced from\cite{ref166}, licensed under CC BY 4.0. (c) An agentic framework for metamaterial inverse design. Reproduced from\cite{ref172}. Copyright 2025 American Chemical Society. (d) A human-AI co-design paradigm for photonic-crystal surface-emitting lasers. Reproduced from\cite{ref173}, licensed under CC BY-ND 4.0. (e) MCP-enabled LLM for meta-optics inverse design. Reproduced from\cite{ref176}, licensed under CC BY-ND 4.0. (f) Electromagnetic metamaterial agent. Reproduced from\cite{ref178}, licensed under CC BY-ND 4.0. }
\label{fig:6}
\end{figure}

In addition to serving as surrogates or planning tools, recent studies have also used LLMs to support the search for new algorithms and exploring scientific literature to augment the design process. Iwanaga et al. (2025) developed a RAG system built on a corpus of 5,000 scientific papers\cite{ref167}. The goal was to conduct "non-empirical" exploration, discovering new designs by synthesizing knowledge from literature rather than simply tweaking parameters. The system uses different generative models to extract information from text (using Titan Text G1) and images (using Claude 3.5 Sonnet). It then generates Python scripts to construct candidate geometries for single-layer circular dichroic (CD) metasurfaces. This approach led to the discovery of a near-perfect CD metasurface structure that was previously difficult to achieve. Another creative use of LLMs is in algorithm discovery for photonics. Yin et al. (2025) viewed LLMs as optimizers that can design novel optimization algorithms tailored to photonic problems\cite{ref179}. Given a task described through structured prompts (such as optimizing a Bragg mirror or solar cell coating), the LLM was asked to generate pseudocode for an evolutionary strategy. In the reported benchmarks, LLM-generated optimization strategies outperformed selected human-designed baselines after iterative evaluation and revision. Along the same line of LLM-enabled algorithmic discovery, Maman et al. (2026) introduced a “prompt-to-prescription” framework that utilizes Claude Sonnet 4.5 as a semantic controller to translate natural language requirements into physically valid, diffraction-limited refractive optical prescriptions\cite{ref180}. They adopt a reasoning-first RAG strategy that retrieves relevant exemplars from a curated library of \textasciitilde{}1,700 validated lens designs, so the LLM can synthesize a grounded initialization by analogy. The candidate prescription is then refined by a differentiable ray-tracing engine (DiffOptics) using gradient-based optimization with physics/specification penalties, producing industry-standard designs without expert “cold-start” trial-and-error. They validated the approach across telecentric finite-conjugate metrology systems, NIR/SWIR/LWIR objectives, and high-dimensional aspheric mobile lenses (via curriculum-style optimization), highlighting how knowledge-grounded LLM reasoning can bridge user intent and rigorous optical engineering.

Supplementary Table S3 compares the LLM backbone, agent role, external executor or verifier, workflow inputs and outputs, level of human intervention, evaluation and validation evidence, reported performance, resources, availability, and scope boundaries of these knowledge-augmented and agentic systems. Overall, these agentic and tool-augmented frameworks show that LLMs can support functions beyond surrogate modeling, including code generation, software orchestration, optimization, and, in some cases, interaction with physical hardware. Current demonstrations primarily automate selected stages of research workflows rather than replacing complete experimental workflows. Future work is expected to expand these capabilities through more extensive hardware implementations.

Recent demonstrations extend some nanophotonic workflows from numerical prediction to language-mediated code generation and software orchestration. By treating metasurface design as a linguistic problem and encoding complex topological structures, material properties, and spectral targets into token sequences, LLMs effectively lower the technical barrier to entry. They complement task-specific DNN architectures by providing flexible, prompt-driven interfaces. This evolution has profound implications for the research process: the focus is shifting from simply speeding up individual design steps to automating several stages of the research process. As demonstrated by the emergence of multi-agent frameworks, LLMs no longer merely serve as alternative solvers. Instead, they are becoming research orchestrators capable of planning experiments, writing code, and debugging simulation feedback loops, thereby helping to address aspects of the “human bottleneck” in design iterations.

However, this transition introduces new complexities that both reflect and amplify the challenges observed in transformers. Just as standard transformers face a "data hunger" paradox where the cost of training may outweigh the benefits of simulation, LLMs face a "tokenization efficiency" paradox. Mapping high-dimensional, pixelated nanophotonic structures to 1D text sequences (as illustrated by the difference between “shape mapping”\cite{ref163} and “matrix flattening”\cite{ref169} methods) often incurs high memory and computational costs due to the quadratic complexity of attention mechanisms\cite{ref181}. This can make general-purpose LLMs computationally more intensive than specialized CNNs or standard transformers. Specifically, when a high-dimensional 2D image cannot be compressed into a substantially shorter token sequence without losing important spatial information, CNNs are generally more memory-efficient for grid representations than LLM- or transformer-based methods, particularly under limited VRAM budgets\cite{ref182,ref183}. Thus, a critical trade-off emerges. While the overarching goal is to democratize design by lowering the technical barrier to entry, the high memory and compute requirements these foundation models place on VRAM and computational resources inadvertently raise the infrastructure barrier, shifting the burden from human capital to high-performance computing resources. Furthermore, the issue of the reliability problem shifts from "accuracy" to "hallucination." While standard DNNs may only produce quantitative errors, LLM agents may generate simulation scripts or manufacturing masks that are syntactically correct but physically invalid. In this context, the distinction between self-revision and external verification is essential. Simulator error messages can help an LLM correct syntax and API misuse, but successful execution establishes only that the program runs; functional and physical acceptance still require independent checks of the requested optical response and the governing constraints. Therefore, this reliance on feedback loops suggests that current LLMs are better suited to act as “translators” that convert between human intent and the outputs of rigorous numerical solvers, rather than as “physicists” operating independently.

Looking ahead, future developments in this field will likely involve combining the “numerical intuition” of standard transformers or classical DNNs with the “broad reasoning capabilities” of LLMs. We are entering the early stages of “Physics AI Foundation Models” and “Autonomous Physical Agents”. Through multimodal training on large-scale experimental and simulated datasets, these processes may reintroduce the inductive biases that are lacking in pure language models. These approaches could contribute to future self-driving-laboratory workflows that couple LLM-based planning with numerical solvers, experimental feedback, and human oversight.

\section{Beyond Nanophotonics: LLMs in Scientific Discovery}

LLM-enabled methods are advancing across a range of scientific and engineering fields. This section examines representative developments in adjacent fields to offer a broader, cross-disciplinary perspective on how analogous challenges are addressed. Across these domains, several strategies offer practical templates for improving the generality, reliability, and automation of future LLM-enabled nanophotonic systems.

In the field of acoustic metasurfaces, Jiang et al. (2026) established two frameworks for forward prediction and inverse design of sound absorption\cite{ref184} (Figure 7a). In the first framework, the user uploads CSV data to ChatGPT o4-mini-high and instructs the agent, entirely through dialogue, to train and apply compact machine-learning models. The second framework uses LoRA to fine-tune DeepSeek-R1-1.5B on JSON-formatted structure-response pairs so that the adapted language model directly performs forward prediction or inverse design. Because acoustic and nanophotonic metamaterial design share analogous data-driven structure-response mappings, these LLM-based interaction and fine-tuning strategies could be tried in future nanophotonic design studies, subject to domain-specific physical validation.

Parallel developments in wireless communications and antenna engineering provide several transferable strategies for LLM-enabled nanophotonics. Tong et al. (2026) proposed WirelessAgent, which organizes an LLM agent around perception, memory, planning, action, knowledge retrieval, and external tools, and stores expert-validated graph-based workflows for repeated execution\cite{ref185} (Figure 7b). In nanophotonics, future studies could build reusable workflow libraries with persistent global state and experimental memory, allowing validated design or measurement procedures to be executed consistently across recurring tasks rather than regenerated from scratch. Zheng and Dai (2026) developed a multi-task architecture in which natural-language task instructions and lightweight task-specific networks connect several heterogeneous numerical problems to a shared LLM backbone, which reduced the cost of maintaining a separate large model for every task\cite{ref186}. This suggests that a common nanophotonic backbone could support multiple different nanophotonic tasks through small task adapters, promoting knowledge sharing while limiting training and deployment costs. 

The field of mechanical metamaterials also offers transferable strategies for LLM-enabled nanophotonic design. CrossMatAgent (2026) combines multimodal agent collaboration with a domain-adapted generative model to translate natural-language or image-based specifications into geometrically coherent, fabrication-oriented patterns\cite{ref187} (Figure 7c). In nanophotonics, future agentic systems could couple language and vision agents with photonics-specific generation, electromagnetic simulation, and fabrication-rule validation so that proposed meta-atoms or metasurfaces are physically usable rather than merely visually plausible. UniMate (2025) instead aligns structural, conditional, and physical-property information within a shared representation and uses the available modalities to infer the missing one\cite{ref188}. This framework suggests that nanophotonic geometry, materials, operating conditions, and optical responses could be jointly modeled, allowing a single system to support forward prediction, inverse design, and design-condition completion. 

For Photonic Integrated Circuit (PIC) design, Wu et al. introduced PICBench, which generates simulation-ready PIC netlists from natural-language specifications and evaluates them through automated syntax and functionality tests\cite{ref189}. Simulator errors are returned to the LLM for iterative correction, while recurring failure modes are summarized as reusable restrictions that improve subsequent generations. Sharma et al. developed PhIDO, a modular multi-agent workflow that translates natural-language requests into structured circuit descriptions and GDSII layouts\cite{ref190}. Its benchmark construction also identified and revised prompts whose component or connectivity requirements were ambiguous or incomplete. Together, these studies suggest that nanophotonic agents could maintain reusable memories of solver and workflow failures while separately detecting underspecified design requests, preventing ambiguous inputs from being misclassified as model errors and enabling more reliable improvement across design iterations.

Several materials-science studies provide inspiring workflows that could be transferred to LLM-enabled nanophotonics\cite{ref191,ref192}. Xie et al. introduced structured information inference, in which an LLM maps scientific text directly into a predefined device-level schema that jointly captures normalized attributes, inferred information, and relationships rather than extracting isolated entities\cite{ref193} (Figure 7d). This approach could convert nanophotonics publications and review datasets into machine-actionable records linking geometry, materials, and optical performance, thereby supporting consistent benchmark construction. HoneyComb constructs domain tools by generating functions for representative scientific questions, subjecting them to human verification, decomposing accepted functions into reusable atomic tools, and selecting between curated knowledge and computation for each query\cite{ref194}. This workflow suggests a practical route to mitigate hallucinations in nanophotonic agentic workflows.

\begin{figure}[H]
\centering
\includegraphics[width=\textwidth,height=0.62\textheight,keepaspectratio]{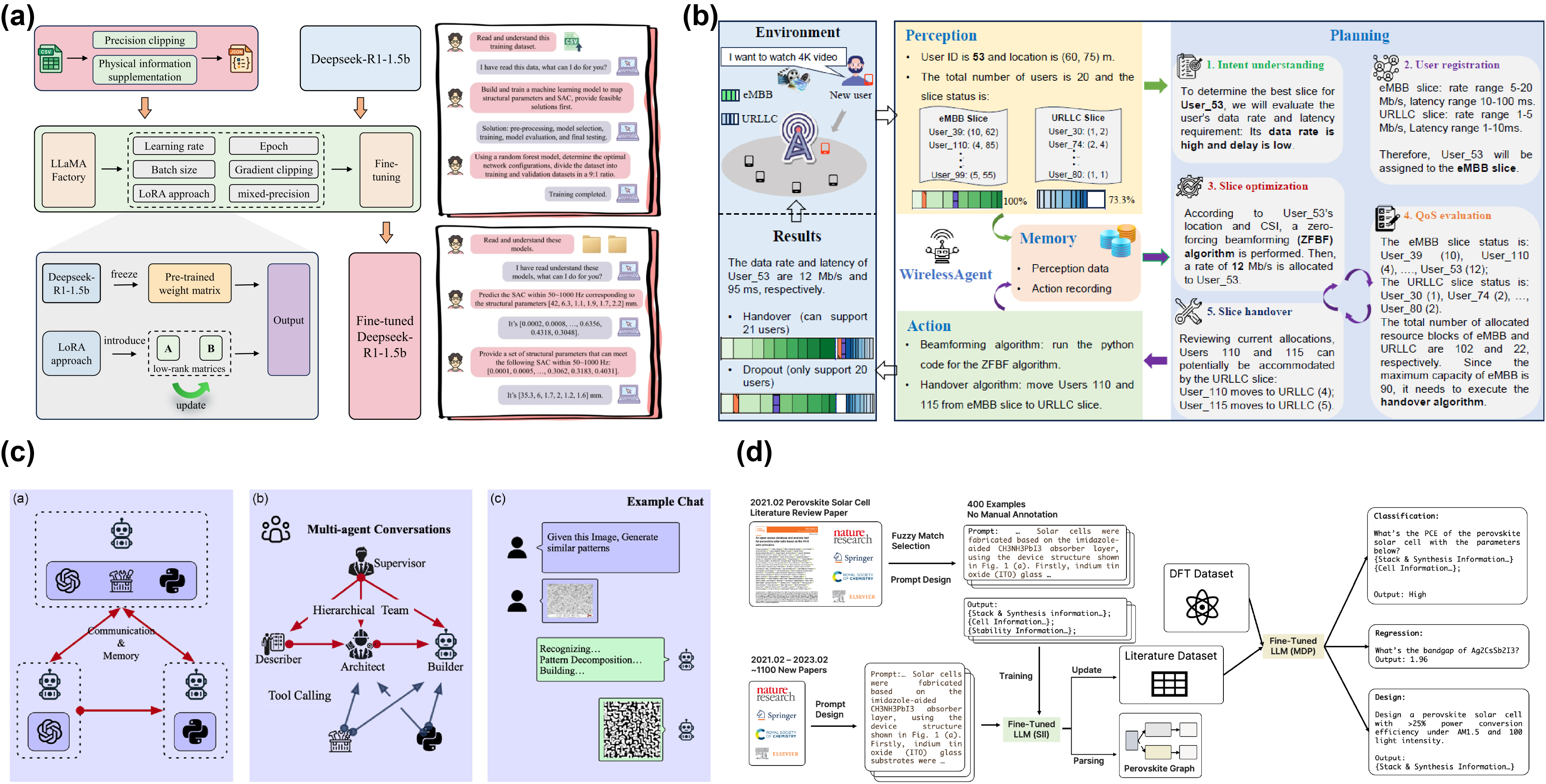}
\caption{Beyond nanophotonics. (a) LLM for acoustic metasurface design. Reproduced from\cite{ref184}, licensed under CC BY 4.0. (b) WirelessAgent. Reproduced with permission from\cite{ref185}. (c) CrossMatAgent. Reproduced from\cite{ref187}, licensed under CC BY 4.0. (d) Structured information inference for materials science. Reproduced from\cite{ref193}, licensed under CC BY-NC-SA 4.0. }
\label{fig:7}
\end{figure}

\section{Perspective}

The integration of LLMs into nanophotonics is advancing rapidly, with developments ranging from surrogate models that predict optical properties to autonomous agents that generate code for designing photonic structures. Future progress will likely depend on hybrid systems in which broadly pre-trained LLMs handle language, retrieval, planning, and code generation, while domain-specific surrogate models and electromagnetic solvers provide quantitative prediction and physical verification. To make this outlook actionable, we propose priorities for benchmark construction, physical-validity testing, solver-in-the-loop evaluation, reproducible reporting, and fabrication-aware experimental data.  

From Text-Based Tokenization to Native Multimodal Physics: Existing LLM surrogates for free-form shape design generally convert 2D geometries into 1D sequences of parameters, shape codes, or flattened pixels. VLMs provide a direct next step for this line of research because they can accept device layouts or topology images through a native visual-input channel, while material properties, wavelengths, polarizations, and boundary conditions are supplied as structured textual or numerical metadata. This removes manual image serialization and task-specific external compression from the surrogate pipeline. Compared with manually flattened pixel sequences, this approach represents geometries through spatially organized visual features and enables an image-native surrogate workflow analogous at the input level to classical CNN-based methods\cite{ref160,ref162,ref195}. Future studies could formulate this approach as an end-to-end multimodal surrogate problem. A practical starting point is to train on geometry-image and electromagnetic-response pairs, and attach a dedicated regression head or continuous numerical decoder for spectra, phases, and complex fields rather than relying on unconstrained text generation. For inverse design, given that general-purpose image generators still struggle with precise spatial relations and compositional constraints\cite{ref196,ref197}, generating fabrication-ready nanophotonic geometries directly from target spectra remains a substantially more demanding and less mature objective. A more practical route to inverse design is to retain established free-form metasurface optimization algorithms while replacing most iterative full-wave evaluations with a VLM-based surrogate, followed by solver-based verification of the final candidates.

Reliability and Validation of LLM-Generated Designs: The "hallucination" and unsupported generation remain major barriers to trust. LLMs might generate Python scripts that are syntactically correct but physically infeasible, such as those that violate the conservation of energy. A promising solution involves neuro-symbolic AI\cite{ref198,ref199}, which integrates the “intuitive” reasoning of LLMs with the “rigorous” logic of symbolic solvers. Under this paradigm, an LLM can propose candidate topologies while external critics reject constraint violations. Such critics may include schema and API validators, fabrication-rule checks, symbolic conservation constraints, Maxwell-residual tests, or independent full-wave solvers. Passing one critic does not prove complete physical validity. Confidence should be based on a documented hierarchy of increasingly stringent tests.

For nanophotonics, validation should be treated as a multidimensional process rather than a single performance metric. We suggest evaluating four complementary aspects: (1) representation and computational validity, including parsability, units, parameter bounds, executable code, and numerical stability; (2) physical and numerical consistency, assessed through problem-appropriate conservation laws, boundary conditions, Maxwell-equation checks where applicable, and independent electromagnetic simulation; (3) generalization and robustness under held-out or out-of-distribution conditions, geometric and material perturbations, and measurement noise; and (4) fabrication feasibility, including minimum feature sizes, aspect ratios, process-compatible materials, and tolerance sensitivity. Future studies can report which forms of validation were performed, together with success rates, retries, human intervention, and verification cost, to improve their robustness and reproducibility.

Reporting Standard for LLM-Enabled Photonics Design: Future studies should report the base model, exact model or checkpoint version, model availability, adaptation method, training or fine-tuning data, prompt templates, tool definitions, output schemas, decoding settings, stopping conditions, random seeds, and permitted repair budget. For hosted models, the access date and available model-version identifier should be recorded. Retrieval-augmented systems should additionally report the retrieval corpus, document versions, embedding or retrieval model, number of retrieved items, reranking procedure, and provenance supplied to the generator. Results should separately report syntactic validity, execution success, geometry-rule compliance, full-wave validation, and target satisfaction. Every percentage should state its denominator, and stochastic systems should be evaluated over multiple independent runs with a distribution or uncertainty interval rather than a single selected example. The community would also benefit from open-source, solver-agnostic agent frameworks, allowing workflows to be reproduced and compared across models, solvers, and computing platforms.

Standardized Experimental Datasets: Simulation-based benchmarks, such as the comparison of tandem networks, VAEs, and GANs by Ma et al. in terms of accuracy, diversity, and robustness, provide useful common testbeds\cite{ref200}. However, nanophotonics still lacks a broadly adopted, standardized real-world experimental dataset. Simulated datasets commonly use nominal geometries and material properties, whereas fabricated devices contain dimensional deviations, thickness variation, surface roughness, material uncertainty, process-dependent defects, calibration uncertainty, and measurement noise. Including these effects is important for evaluating whether a model remains accurate outside an idealized simulation domain. Large, standardized, and well-documented datasets would support the development and evaluation of more transferable models and make comparisons across studies more reliable. 

\section{Conclusion}

The development of nanophotonics has always been driven by the quest for finer control over light, from the early effective medium theory of metamaterials to the complex wavefront shaping of metasurfaces. We are witnessing an ongoing shift in the role of computational tools. Computers are no longer merely tools that solve equations based on human-defined parameters; they are gradually becoming semantic partners capable of understanding design intent, planning experimental logic, and writing and executing code. Although computational efficiency, physical reliability, and the integration of experimental realities remain significant challenges, the next phase of AI-enabled nanophotonics is taking shape. The integration of generative AI with rigorous electromagnetic theory is shortening selected stages of the design cycle and facilitating the discovery of complex optical structures that may be difficult to identify through intuition alone. As these models evolve from passive tools into active collaborators, they are more likely to complement than replace physicists and may allow researchers to devote more effort to scientific judgment and experimental validation. The future of nanophotonics lies not only in designing better structures, but also in building intelligent systems capable of creating such designs.

\section*{Disclosures}

The authors declare that there are no financial interests, commercial affiliations, or other potential conflicts of interest that could have influenced the objectivity of this research or the writing of this paper.

\begin{acknowledgments}
The authors would like to acknowledge the partial support for this work provided by the John L. and Genevieve H. McCain endowed chair professorship. 
\end{acknowledgments}

\bibliographystyle{IEEEtran}
\bibliography{manuscript_v3_2}

@article{ref001,
  author = {Zheludev, Nikolay I. and Kivshar, Yuri S.},
  title = {From metamaterials to metadevices},
  journal = {Nature Materials},
  volume = {11},
  number = {11},
  pages = {917--924},
  month = {November},
  year = {2012},
  doi = {10.1038/nmat3431},
  url = {https://www.nature.com/articles/nmat3431},
  issn = {1476-1122, 1476-4660},
  language = {en},
}

@article{ref002,
  author = {Pendry, J. B.},
  title = {Negative refraction makes a perfect lens},
  journal = {Physical Review Letters},
  volume = {85},
  number = {18},
  pages = {3966--3969},
  month = {October},
  year = {2000},
  doi = {10.1103/PhysRevLett.85.3966},
  url = {https://link.aps.org/doi/10.1103/PhysRevLett.85.3966},
  issn = {0031-9007, 1079-7114},
  language = {en},
}

@article{ref003,
  author = {Shelby, R. A. and Smith, D. R. and Schultz, S.},
  title = {Experimental verification of a negative index of refraction},
  journal = {Science},
  volume = {292},
  number = {5514},
  pages = {77--79},
  month = {April},
  year = {2001},
  doi = {10.1126/science.1058847},
  url = {https://www.science.org/doi/10.1126/science.1058847},
  issn = {0036-8075, 1095-9203},
  language = {en},
}

@article{ref004,
  author = {Valentine, Jason and Zhang, Shuang and Zentgraf, Thomas and Ulin-Avila, Erick and Genov, Dentcho A. and Bartal, Guy and Zhang, Xiang},
  title = {Three-dimensional optical metamaterial with a negative refractive index},
  journal = {Nature},
  volume = {455},
  number = {7211},
  pages = {376--379},
  month = {September},
  year = {2008},
  doi = {10.1038/nature07247},
  url = {https://www.nature.com/articles/nature07247},
  issn = {0028-0836, 1476-4687},
  language = {en},
}

@article{ref005,
  author = {Schurig, D. and Mock, J. J. and Justice, B. J. and Cummer, S. A. and Pendry, J. B. and Starr, A. F. and Smith, D. R.},
  title = {Metamaterial electromagnetic cloak at microwave frequencies},
  journal = {Science},
  volume = {314},
  number = {5801},
  pages = {977--980},
  month = {November},
  year = {2006},
  doi = {10.1126/science.1133628},
  url = {https://www.science.org/doi/10.1126/science.1133628},
  issn = {0036-8075, 1095-9203},
  language = {en},
}

@book{ref006,
  author = {Prasad, Paras N.},
  title = {Nanophotonics},
  publisher = {Wiley},
  address = {Hoboken, NJ},
  year = {2004},
  doi = {10.1002/0471670251},
  isbn = {978-0-471-67024-7 978-0-471-64988-5 978-0-471-67025-4},
  language = {eng},
}

@article{ref007,
  author = {Wang, Hui and Brandl, Daniel W. and Nordlander, Peter and Halas, Naomi J.},
  title = {Plasmonic nanostructures: {Artificial} molecules},
  journal = {Accounts of Chemical Research},
  volume = {40},
  number = {1},
  pages = {53--62},
  month = {January},
  year = {2007},
  doi = {10.1021/ar0401045},
  url = {https://pubs.acs.org/doi/10.1021/ar0401045},
  issn = {0001-4842, 1520-4898},
  language = {en},
}

@article{ref008,
  author = {Kasani, Sujan and Curtin, Kathrine and Wu, Nianqiang},
  title = {A review of {2D} and {3D} plasmonic nanostructure array patterns: {Fabrication}, light management and sensing applications},
  journal = {Nanophotonics},
  volume = {8},
  number = {12},
  pages = {2065--2089},
  month = {December},
  year = {2019},
  doi = {10.1515/nanoph-2019-0158},
  url = {https://www.degruyterbrill.com/document/doi/10.1515/nanoph-2019-0158/html},
  issn = {2192-8614, 2192-8606},
  language = {en},
}

@article{ref009,
  author = {Pan, Meiyan and Fu, Yifei and Zheng, Mengjie and Chen, Hao and Zang, Yujia and Duan, Huigao and Li, Qiang and Qiu, Min and Hu, Yueqiang},
  title = {Dielectric metalens for miniaturized imaging systems: progress and challenges},
  journal = {Light: Science \& Applications},
  volume = {11},
  number = {1},
  pages = {195},
  month = {June},
  year = {2022},
  doi = {10.1038/s41377-022-00885-7},
  url = {https://www.nature.com/articles/s41377-022-00885-7},
  issn = {2047-7538},
  language = {en},
}

@article{ref010,
  author = {Huang, Lingling and Zhang, Shuang and Zentgraf, Thomas},
  title = {Metasurface holography: {From} fundamentals to applications},
  journal = {Nanophotonics},
  volume = {7},
  number = {6},
  pages = {1169--1190},
  month = {June},
  year = {2018},
  doi = {10.1515/nanoph-2017-0118},
  url = {https://www.degruyterbrill.com/document/doi/10.1515/nanoph-2017-0118/html},
  issn = {2192-8614, 2192-8606},
  language = {en},
}

@article{ref011,
  author = {Yu, Nanfang and Genevet, Patrice and Kats, Mikhail A. and Aieta, Francesco and Tetienne, Jean-Philippe and Capasso, Federico and Gaburro, Zeno},
  title = {Light propagation with phase discontinuities: {Generalized} laws of reflection and refraction},
  journal = {Science},
  volume = {334},
  number = {6054},
  pages = {333--337},
  month = {October},
  year = {2011},
  doi = {10.1126/science.1210713},
  url = {https://www.science.org/doi/10.1126/science.1210713},
  issn = {0036-8075, 1095-9203},
  language = {en},
}

@article{ref012,
  author = {Chen, Hou-Tong and Taylor, Antoinette J and Yu, Nanfang},
  title = {A review of metasurfaces: {Physics} and applications},
  journal = {Reports on Progress in Physics},
  volume = {79},
  number = {7},
  pages = {076401},
  month = {July},
  year = {2016},
  doi = {10.1088/0034-4885/79/7/076401},
  url = {https://iopscience.iop.org/article/10.1088/0034-4885/79/7/076401},
  issn = {0034-4885, 1361-6633},
}

@article{ref013,
  author = {Chang, Shengyuan and Guo, Xuexue and Ni, Xingjie},
  title = {Optical metasurfaces: {Progress} and applications},
  journal = {Annual Review of Materials Research},
  volume = {48},
  number = {1},
  pages = {279--302},
  month = {July},
  year = {2018},
  doi = {10.1146/annurev-matsci-070616-124220},
  url = {https://www.annualreviews.org/doi/10.1146/annurev-matsci-070616-124220},
  issn = {1531-7331, 1545-4118},
  language = {en},
}

@article{ref014,
  author = {Ni, Xingjie and Emani, Naresh K. and Kildishev, Alexander V. and Boltasseva, Alexandra and Shalaev, Vladimir M.},
  title = {Broadband light bending with plasmonic nanoantennas},
  journal = {Science},
  volume = {335},
  number = {6067},
  pages = {427--427},
  month = {January},
  year = {2012},
  doi = {10.1126/science.1214686},
  url = {https://www.science.org/doi/10.1126/science.1214686},
  issn = {0036-8075, 1095-9203},
  language = {en},
}

@article{ref015,
  author = {Maguid, Elhanan and Yulevich, Igor and Yannai, Michael and Kleiner, Vladimir and L Brongersma, Mark and Hasman, Erez},
  title = {Multifunctional interleaved geometric-phase dielectric metasurfaces},
  journal = {Light: Science \& Applications},
  volume = {6},
  number = {8},
  pages = {e17027--e17027},
  month = {February},
  year = {2017},
  doi = {10.1038/lsa.2017.27},
  url = {https://www.nature.com/articles/lsa201727},
  issn = {2047-7538},
  language = {en},
}

@article{ref016,
  author = {Maguid, Elhanan and Yulevich, Igor and Veksler, Dekel and Kleiner, Vladimir and Brongersma, Mark L. and Hasman, Erez},
  title = {Photonic spin-controlled multifunctional shared-aperture antenna array},
  journal = {Science},
  volume = {352},
  number = {6290},
  pages = {1202--1206},
  month = {June},
  year = {2016},
  doi = {10.1126/science.aaf3417},
  url = {https://www.science.org/doi/10.1126/science.aaf3417},
  issn = {0036-8075, 1095-9203},
  language = {en},
}

@article{ref017,
  author = {Ni, Xingjie and Kildishev, Alexander V. and Shalaev, Vladimir M.},
  title = {Metasurface holograms for visible light},
  journal = {Nature Communications},
  volume = {4},
  number = {1},
  pages = {2807},
  month = {November},
  year = {2013},
  doi = {10.1038/ncomms3807},
  url = {https://www.nature.com/articles/ncomms3807},
  issn = {2041-1723},
  language = {en},
}

@article{ref018,
  author = {Ye, Weimin and Zeuner, Franziska and Li, Xin and Reineke, Bernhard and He, Shan and Qiu, Cheng-Wei and Liu, Juan and Wang, Yongtian and Zhang, Shuang and Zentgraf, Thomas},
  title = {Spin and wavelength multiplexed nonlinear metasurface holography},
  journal = {Nature Communications},
  volume = {7},
  number = {1},
  pages = {11930},
  month = {June},
  year = {2016},
  doi = {10.1038/ncomms11930},
  url = {https://www.nature.com/articles/ncomms11930},
  issn = {2041-1723},
  language = {en},
}

@article{ref019,
  author = {Li, Guixin and Chen, Shumei and Pholchai, Nitipat and Reineke, Bernhard and Wong, Polis Wing Han and Pun, Edwin Yue Bun and Cheah, Kok Wai and Zentgraf, Thomas and Zhang, Shuang},
  title = {Continuous control of the nonlinearity phase for harmonic generations},
  journal = {Nature Materials},
  volume = {14},
  number = {6},
  pages = {607--612},
  month = {June},
  year = {2015},
  doi = {10.1038/nmat4267},
  url = {https://www.nature.com/articles/nmat4267},
  issn = {1476-1122, 1476-4660},
  language = {en},
}

@article{ref020,
  author = {Chen, Shumei and Zeuner, Franziska and Weismann, Martin and Reineke, Bernhard and Li, Guixin and Valev, Ventsislav Kolev and Cheah, Kok Wai and Panoiu, Nicolae Coriolan and Zentgraf, Thomas and Zhang, Shuang},
  title = {Giant nonlinear optical activity of achiral origin in planar metasurfaces with quadratic and cubic nonlinearities},
  journal = {Advanced Materials},
  volume = {28},
  number = {15},
  pages = {2992--2999},
  month = {April},
  year = {2016},
  doi = {10.1002/adma.201505640},
  url = {https://advanced.onlinelibrary.wiley.com/doi/10.1002/adma.201505640},
  issn = {0935-9648, 1521-4095},
  language = {en},
}

@article{ref021,
  author = {Li, Guixin and Zhang, Shuang and Zentgraf, Thomas},
  title = {Nonlinear photonic metasurfaces},
  journal = {Nature Reviews Materials},
  volume = {2},
  number = {5},
  pages = {17010},
  month = {March},
  year = {2017},
  doi = {10.1038/natrevmats.2017.10},
  url = {https://www.nature.com/articles/natrevmats201710},
  issn = {2058-8437},
  language = {en},
}

@article{ref022,
  author = {Khorasaninejad, Mohammadreza and Chen, Wei Ting and Devlin, Robert C. and Oh, Jaewon and Zhu, Alexander Y. and Capasso, Federico},
  title = {Metalenses at visible wavelengths: {Diffraction}-limited focusing and subwavelength resolution imaging},
  journal = {Science},
  volume = {352},
  number = {6290},
  pages = {1190--1194},
  month = {June},
  year = {2016},
  doi = {10.1126/science.aaf6644},
  url = {https://www.science.org/doi/10.1126/science.aaf6644},
  issn = {0036-8075, 1095-9203},
  language = {en},
}

@article{ref023,
  author = {Chen, Wei Ting and Zhu, Alexander Y. and Sisler, Jared and Bharwani, Zameer and Capasso, Federico},
  title = {A broadband achromatic polarization-insensitive metalens consisting of anisotropic nanostructures},
  journal = {Nature Communications},
  volume = {10},
  number = {1},
  pages = {355},
  month = {January},
  year = {2019},
  doi = {10.1038/s41467-019-08305-y},
  url = {https://www.nature.com/articles/s41467-019-08305-y},
  issn = {2041-1723},
  language = {en},
}

@article{ref024,
  author = {Bao, Yanjun and Yu, Ying and Xu, Haofei and Guo, Chao and Li, Juntao and Sun, Shang and Zhou, Zhang-Kai and Qiu, Cheng-Wei and Wang, Xue-Hua},
  title = {Full-colour nanoprint-hologram synchronous metasurface with arbitrary hue-saturation-brightness control},
  journal = {Light: Science \& Applications},
  volume = {8},
  number = {1},
  pages = {95},
  month = {October},
  year = {2019},
  doi = {10.1038/s41377-019-0206-2},
  url = {https://www.nature.com/articles/s41377-019-0206-2},
  issn = {2047-7538},
  language = {en},
}

@article{ref025,
  author = {Liu, Chao and Zheng, Yi and Wang, Di and Huang, Qian and Li, Xiao‐Wei and Lin, Fan‐Chuan and Zhao, You‐Ran and Zheng, Yi‐Wei and Lu, Xiao‐Ke and Li, Xin‐Ru and Zheng, Xin‐Ru and Xie, Xin and Song, Kun and Li, Zhen‐Fei and Lu, Wei and Tsai, Din Ping and Ji, Ruo‐Nan and Wang, Qiong‐Hua},
  title = {Multi‐wavelength achromatic {3D} meta‐holography with zoom function},
  journal = {Advanced Science},
  volume = {12},
  number = {28},
  pages = {2501881},
  month = {July},
  year = {2025},
  doi = {10.1002/advs.202501881},
  url = {https://advanced.onlinelibrary.wiley.com/doi/10.1002/advs.202501881},
  issn = {2198-3844, 2198-3844},
  language = {en},
}

@article{ref026,
  author = {Arbabi, Amir and Arbabi, Ehsan and Horie, Yu and Kamali, Seyedeh Mahsa and Faraon, Andrei},
  title = {Planar metasurface retroreflector},
  journal = {Nature Photonics},
  volume = {11},
  number = {7},
  pages = {415--420},
  month = {July},
  year = {2017},
  doi = {10.1038/nphoton.2017.96},
  url = {https://www.nature.com/articles/nphoton.2017.96},
  issn = {1749-4885, 1749-4893},
  language = {en},
}

@article{ref027,
  author = {Li, Dancui and Li, Yangyang},
  title = {An aberration-corrected single layer metasurface with large field of view},
  journal = {Optics Communications},
  volume = {530},
  pages = {129195},
  month = {March},
  year = {2023},
  doi = {10.1016/j.optcom.2022.129195},
  url = {https://linkinghub.elsevier.com/retrieve/pii/S0030401822008422},
  issn = {00304018},
  language = {en},
}

@article{ref028,
  author = {Chen, Wei Ting and Zhu, Alexander Y. and Sanjeev, Vyshakh and Khorasaninejad, Mohammadreza and Shi, Zhujun and Lee, Eric and Capasso, Federico},
  title = {A broadband achromatic metalens for focusing and imaging in the visible},
  journal = {Nature Nanotechnology},
  volume = {13},
  number = {3},
  pages = {220--226},
  month = {March},
  year = {2018},
  doi = {10.1038/s41565-017-0034-6},
  url = {https://www.nature.com/articles/s41565-017-0034-6},
  issn = {1748-3387, 1748-3395},
  language = {en},
}

@article{ref029,
  author = {Shrestha, Sajan and Overvig, Adam C. and Lu, Ming and Stein, Aaron and Yu, Nanfang},
  title = {Broadband achromatic dielectric metalenses},
  journal = {Light: Science \& Applications},
  volume = {7},
  number = {1},
  pages = {85},
  month = {November},
  year = {2018},
  doi = {10.1038/s41377-018-0078-x},
  url = {https://www.nature.com/articles/s41377-018-0078-x},
  issn = {2047-7538},
  language = {en},
}

@article{ref030,
  author = {Patoux, Adelin and Agez, Gonzague and Girard, Christian and Paillard, Vincent and Wiecha, Peter R. and Lecestre, Aurélie and Carcenac, Franck and Larrieu, Guilhem and Arbouet, Arnaud},
  title = {Challenges in nanofabrication for efficient optical metasurfaces},
  journal = {Scientific Reports},
  volume = {11},
  number = {1},
  pages = {5620},
  month = {March},
  year = {2021},
  doi = {10.1038/s41598-021-84666-z},
  url = {https://www.nature.com/articles/s41598-021-84666-z},
  issn = {2045-2322},
  language = {en},
}

@article{ref031,
  author = {Tahmi, Yanel and Ansquer, Matthieu and Coulon, Pierre-Marie and Chenot, Sébastien and Brändli, Virginie and Majorel, Clément and Patoux, Adelin and Colom, Rémi and Wattellier, Benoit and Genevet, Patrice and Khadir, Samira},
  title = {Phase characterization of singular metasurfaces},
  journal = {ACS Photonics},
  volume = {12},
  number = {11},
  pages = {6081--6090},
  month = {November},
  year = {2025},
  doi = {10.1021/acsphotonics.5c01539},
  url = {https://pubs.acs.org/doi/10.1021/acsphotonics.5c01539},
  issn = {2330-4022, 2330-4022},
  language = {en},
}

@article{ref032,
  author = {Campbell, Sawyer D. and Sell, David and Jenkins, Ronald P. and Whiting, Eric B. and Fan, Jonathan A. and Werner, Douglas H.},
  title = {Review of numerical optimization techniques for meta-device design},
  journal = {Optical Materials Express},
  volume = {9},
  number = {4},
  pages = {1842},
  month = {April},
  year = {2019},
  doi = {10.1364/OME.9.001842},
  url = {https://opg.optica.org/abstract.cfm?URI=ome-9-4-1842},
  issn = {2159-3930},
  language = {en},
}

@article{ref033,
  author = {Molesky, Sean and Lin, Zin and Piggott, Alexander Y. and Jin, Weiliang and Vucković, Jelena and Rodriguez, Alejandro W.},
  title = {Inverse design in nanophotonics},
  journal = {Nature Photonics},
  volume = {12},
  number = {11},
  pages = {659--670},
  month = {November},
  year = {2018},
  doi = {10.1038/s41566-018-0246-9},
  url = {https://www.nature.com/articles/s41566-018-0246-9},
  issn = {1749-4885, 1749-4893},
  language = {en},
}

@article{ref034,
  author = {Christiansen, Rasmus E. and Sigmund, Ole},
  title = {Inverse design in photonics by topology optimization: {Tutorial}},
  journal = {Journal of the Optical Society of America B},
  volume = {38},
  number = {2},
  pages = {496},
  month = {February},
  year = {2021},
  doi = {10.1364/JOSAB.406048},
  url = {https://opg.optica.org/abstract.cfm?URI=josab-38-2-496},
  issn = {0740-3224, 1520-8540},
  language = {en},
}

@article{ref035,
  author = {Jensen, Jakob S. and Sigmund, Ole},
  title = {Topology optimization of photonic crystal structures: {A} high-bandwidth low-loss {T}-junction waveguide},
  journal = {Journal of the Optical Society of America B},
  volume = {22},
  number = {6},
  pages = {1191},
  month = {June},
  year = {2005},
  doi = {10.1364/JOSAB.22.001191},
  url = {https://opg.optica.org/abstract.cfm?URI=josab-22-6-1191},
  issn = {0740-3224, 1520-8540},
  language = {en},
}

@article{ref036,
  author = {An, Sensong and Zheng, Bowen and Shalaginov, Mikhail Y. and Tang, Hong and Li, Hang and Zhou, Li and Ding, Jun and Agarwal, Anuradha Murthy and Rivero-Baleine, Clara and Kang, Myungkoo and Richardson, Kathleen A. and Gu, Tian and Hu, Juejun and Fowler, Clayton and Zhang, Hualiang},
  title = {Deep learning modeling approach for metasurfaces with high degrees of freedom},
  journal = {Optics Express},
  volume = {28},
  number = {21},
  pages = {31932},
  month = {October},
  year = {2020},
  doi = {10.1364/OE.401960},
  url = {https://opg.optica.org/abstract.cfm?URI=oe-28-21-31932},
  issn = {1094-4087},
  language = {en},
}

@article{ref037,
  author = {Zhang, Huanshu and Kang, Lei and Campbell, Sawyer D. and Zhang, Kaishun and Werner, Douglas H. and Cao, Zhaolong},
  title = {Fixed-attention mechanism for deep-learning-assisted design of high-degree-of-freedom {3D} metamaterials},
  journal = {Optics Express},
  volume = {33},
  number = {9},
  pages = {18928},
  month = {May},
  year = {2025},
  doi = {10.1364/OE.557837},
  url = {https://opg.optica.org/abstract.cfm?URI=oe-33-9-18928},
  issn = {1094-4087},
  language = {en},
}

@article{ref038,
  author = {Yang, Fan and An, Sensong and Shalaginov, Mikhail Y. and Zhang, Hualiang and Rivero-Baleine, Clara and Hu, Juejun and Gu, Tian},
  title = {Design of broadband and wide-field-of-view metalenses},
  journal = {Optics Letters},
  volume = {46},
  number = {22},
  pages = {5735},
  month = {November},
  year = {2021},
  doi = {10.1364/OL.439393},
  url = {https://opg.optica.org/abstract.cfm?URI=ol-46-22-5735},
  issn = {0146-9592, 1539-4794},
  language = {en},
}

@article{ref039,
  author = {Yang, Fan and Shalaginov, Mikhail Y. and Lin, Hung-I and An, Sensong and Agarwal, Anu and Zhang, Hualiang and Rivero-Baleine, Clara and Gu, Tian and Hu, Juejun},
  title = {Wide field-of-view metalens: {A} tutorial},
  journal = {Advanced Photonics},
  volume = {5},
  number = {03},
  month = {May},
  year = {2023},
  doi = {10.1117/1.AP.5.3.033001},
  url = {https://www.spiedigitallibrary.org/journals/advanced-photonics/volume-5/issue-03/033001/Wide-field-of-view-metalens-a-tutorial/10.1117/1.AP.5.3.033001.full},
  issn = {2577-5421},
}

@article{ref040,
  author = {Wang, Yujie and Chen, Qinmiao and Yang, Wenhong and Ji, Ziheng and Jin, Limin and Ma, Xing and Song, Qinghai and Boltasseva, Alexandra and Han, Jiecai and Shalaev, Vladimir M. and Xiao, Shumin},
  title = {High-efficiency broadband achromatic metalens for near-{IR} biological imaging window},
  journal = {Nature Communications},
  volume = {12},
  number = {1},
  pages = {5560},
  month = {September},
  year = {2021},
  doi = {10.1038/s41467-021-25797-9},
  url = {https://www.nature.com/articles/s41467-021-25797-9},
  issn = {2041-1723},
  language = {en},
}

@article{ref041,
  author = {Feng, Fu and Huo, Dewang and Zhang, Ziyang and Lou, Yijie and Wang, Shengyao and Gu, Zhijuan and Liu, Dong-Sheng and Duan, Xinhui and Wang, Daqian and Liu, Xiaowei and Qi, Ji and Yu, Shaoliang and Du, Qingyang and Chen, Guangyong and Lu, Cuicui and Yu, Yu and Ren, Xifeng and Yuan, Xiaocong},
  title = {Symbiotic evolution of photonics and artificial intelligence: {A} comprehensive review},
  journal = {Advanced Photonics},
  volume = {7},
  number = {02},
  month = {April},
  year = {2025},
  doi = {10.1117/1.AP.7.2.024001},
  url = {https://www.spiedigitallibrary.org/journals/advanced-photonics/volume-7/issue-02/024001/Symbiotic-evolution-of-photonics-and-artificial-intelligence--a-comprehensive/10.1117/1.AP.7.2.024001.full},
  issn = {2577-5421},
}

@article{ref042,
  author = {Saifullah, Yasir and Wu, Nanxuan and Wang, Huaping and Zheng, Bin and Qian, Chao and Chen, Hongsheng},
  title = {Deep learning in metasurfaces: from automated design to adaptive metadevices},
  journal = {Advanced Photonics},
  volume = {7},
  number = {03},
  month = {June},
  year = {2025},
  doi = {10.1117/1.AP.7.3.034005},
  url = {https://www.spiedigitallibrary.org/journals/advanced-photonics/volume-7/issue-03/034005/Deep-learning-in-metasurfaces--from-automated-design-to-adaptive/10.1117/1.AP.7.3.034005.full},
  issn = {2577-5421},
}

@article{ref043,
  author = {Ma, Wei and Liu, Zhaocheng and Kudyshev, Zhaxylyk A. and Boltasseva, Alexandra and Cai, Wenshan and Liu, Yongmin},
  title = {Deep learning for the design of photonic structures},
  journal = {Nature Photonics},
  volume = {15},
  number = {2},
  pages = {77--90},
  month = {February},
  year = {2021},
  doi = {10.1038/s41566-020-0685-y},
  url = {https://www.nature.com/articles/s41566-020-0685-y},
  issn = {1749-4885, 1749-4893},
  language = {en},
}

@article{ref044,
  author = {Jiang, Jiaqi and Chen, Mingkun and Fan, Jonathan A.},
  title = {Deep neural networks for the evaluation and design of photonic devices},
  journal = {Nature Reviews Materials},
  volume = {6},
  number = {8},
  pages = {679--700},
  month = {December},
  year = {2020},
  doi = {10.1038/s41578-020-00260-1},
  url = {https://www.nature.com/articles/s41578-020-00260-1},
  issn = {2058-8437},
  language = {en},
}

@article{ref045,
  author = {Zhang, Huanshu and Kang, Lei and Campbell, Sawyer D. and Young, Jacob T. and Werner, Douglas H.},
  title = {Data driven approaches in nanophotonics: a review of {AI}-enabled metadevices},
  journal = {Nanoscale},
  volume = {17},
  number = {41},
  pages = {23788--23803},
  year = {2025},
  doi = {10.1039/D5NR02043C},
  url = {https://xlink.rsc.org/?DOI=D5NR02043C},
  issn = {2040-3364, 2040-3372},
  language = {en},
}

@article{ref046,
  author = {Li, Yongzhong and Wang, Yinpeng and Qi, Shutong and Ren, Qiang and Kang, Lei and Campbell, Sawyer D. and Werner, Pingjuan L. and Werner, Douglas H.},
  title = {Predicting scattering from complex nano-structures via deep learning},
  journal = {IEEE Access},
  volume = {8},
  pages = {139983--139993},
  year = {2020},
  doi = {10.1109/ACCESS.2020.3012132},
  url = {https://ieeexplore.ieee.org/document/9149921/},
  issn = {2169-3536},
}

@article{ref047,
  author = {Kim, Junhyeong and Kim, Jae-Yong and Kim, Jungmin and Hyeong, Yun and Neseli, Berkay and You, Jong-Bum and Shim, Joonsup and Shin, Jonghwa and Park, Hyo-Hoon and Kurt, Hamza},
  title = {Inverse design of nanophotonic devices enabled by optimization algorithms and deep learning: {Recent} achievements and future prospects},
  journal = {Nanophotonics},
  volume = {14},
  number = {2},
  pages = {121--151},
  month = {February},
  year = {2025},
  doi = {10.1515/nanoph-2024-0536},
  url = {https://www.degruyterbrill.com/document/doi/10.1515/nanoph-2024-0536/html},
  issn = {2192-8614},
  language = {en},
}

@article{ref048,
  author = {Fu, Yunlai and Zhou, Xuxi and Yu, Yiwan and Chen, Jiawang and Wang, Shuming and Zhu, Shining and Wang, Zhenlin},
  title = {Unleashing the potential: {AI} empowered advanced metasurface research},
  journal = {Nanophotonics},
  volume = {13},
  number = {8},
  pages = {1239--1278},
  month = {April},
  year = {2024},
  doi = {10.1515/nanoph-2023-0759},
  url = {https://www.degruyter.com/document/doi/10.1515/nanoph-2023-0759/html},
  issn = {2192-8614},
  language = {en},
}

@article{ref049,
  author = {Kang, Minsung and Choi, Seokju and Fu, Kaixi and Liu, Xiaoyuan and Wei, Zhun and Jin, Lei and Wang, Hao and J. F. Martin, Olivier and K. W. Yang, Joel and So, Sunae and Badloe, Trevon},
  title = {{AI}-assisted metaphotonics},
  journal = {Opto-Electronic Advances},
  volume = {9},
  number = {4},
  pages = {250263},
  year = {2026},
  doi = {10.29026/oea.2026.250263},
  url = {https://www.oejournal.org/oea/article/doi/10.29026/oea.2026.250263},
  issn = {2096-4579, 2097-3993},
}

@article{ref050,
  author = {Ma, Qian and Feng, Zi Rui and Hou, Junming and Liu, Che and Ning, Yu Ming and Li, Rui Si and Wu, Qian Wen and Huang, Yi Su and Zhang, Yi and Gu, Ze and You, Jian Wei and Cui, Tie Jun},
  title = {Artificial intelligence with metasurfaces: {From} intelligent design to intelligent computing},
  journal = {PhotoniX},
  volume = {7},
  number = {1},
  pages = {23},
  month = {April},
  year = {2026},
  doi = {10.1186/s43074-026-00239-1},
  url = {https://link.springer.com/10.1186/s43074-026-00239-1},
  issn = {2662-1991},
  language = {en},
}

@article{ref051,
  author = {Khaireh-Walieh, Abdourahman and Langevin, Denis and Bennet, Pauline and Teytaud, Olivier and Moreau, Antoine and Wiecha, Peter R.},
  title = {A newcomer’s guide to deep learning for inverse design in nano-photonics},
  journal = {Nanophotonics},
  volume = {12},
  number = {24},
  pages = {4387--4414},
  month = {December},
  year = {2023},
  doi = {10.1515/nanoph-2023-0527},
  url = {https://www.degruyter.com/document/doi/10.1515/nanoph-2023-0527/html},
  issn = {2192-8614},
  language = {en},
}

@article{ref052,
  author = {Wiecha, Peter R. and Arbouet, Arnaud and Girard, Christian and Muskens, Otto L.},
  title = {Deep learning in nano-photonics: {Inverse} design and beyond},
  journal = {Photonics Research},
  volume = {9},
  number = {5},
  pages = {B182},
  month = {May},
  year = {2021},
  doi = {10.1364/PRJ.415960},
  url = {https://opg.optica.org/abstract.cfm?URI=prj-9-5-B182},
  issn = {2327-9125},
  language = {en},
}

@misc{ref053,
  author = {Vaswani, Ashish and Shazeer, Noam and Parmar, Niki and Uszkoreit, Jakob and Jones, Llion and Gomez, Aidan N. and Kaiser, Lukasz and Polosukhin, Illia},
  title = {Attention is all you need},
  publisher = {arXiv},
  year = {2017},
  doi = {10.48550/ARXIV.1706.03762},
  url = {https://arxiv.org/abs/1706.03762},
  eprint = {1706.03762},
  archiveprefix = {arXiv},
  howpublished = {arXiv preprint arXiv:1706.03762},
}

@incollection{ref054,
  author = {Vapnik, V. N. and Chervonenkis, A. Ya.},
  editor = {Vovk, Vladimir and Papadopoulos, Harris and Gammerman, Alexander},
  title = {On the uniform convergence of relative frequencies of events to their probabilities},
  booktitle = {Measures of {Complexity}},
  pages = {11--30},
  publisher = {Springer International Publishing},
  address = {Cham},
  year = {2015},
  doi = {10.1007/978-3-319-21852-6_3},
  url = {https://link.springer.com/10.1007/978-3-319-21852-6_3},
  isbn = {978-3-319-21851-9 978-3-319-21852-6},
  language = {en},
}

@book{ref055,
  author = {Vapnik, Vladimir Naumovich},
  title = {The {Nature} of {Statistical} {Learning} {Theory}},
  series = {Statistics for engineering and information science},
  publisher = {Springer},
  address = {New York, NY},
  edition = {Second Edition},
  year = {2000},
  doi = {10.1007/978-1-4757-3264-1},
  isbn = {978-1-4757-3264-1},
  language = {eng},
}

@book{ref056,
  author = {Morozov, V. A.},
  title = {Methods for {Solving} {Incorrectly} {Posed} {Problems}},
  publisher = {Springer New York},
  address = {New York, NY},
  year = {1984},
  doi = {10.1007/978-1-4612-5280-1},
  isbn = {978-1-4612-5280-1},
  language = {eng},
}

@article{ref057,
  author = {Cortes, Corinna and Vapnik, Vladimir},
  title = {Support-vector networks},
  journal = {Machine Learning},
  volume = {20},
  number = {3},
  pages = {273--297},
  month = {September},
  year = {1995},
  doi = {10.1007/BF00994018},
  url = {http://link.springer.com/10.1007/BF00994018},
  issn = {0885-6125, 1573-0565},
  language = {en},
}

@article{ref058,
  author = {Hearst, M.A. and Dumais, S.T. and Osuna, E. and Platt, J. and Scholkopf, B.},
  title = {Support vector machines},
  journal = {IEEE Intelligent Systems and their Applications},
  volume = {13},
  number = {4},
  pages = {18--28},
  month = {July},
  year = {1998},
  doi = {10.1109/5254.708428},
  url = {http://ieeexplore.ieee.org/document/708428/},
  issn = {1094-7167},
}

@article{ref059,
  author = {Li, Hongwei and Chen, Hailiang and Li, Yuxin and Chen, Qiang and Fan, Xiaoya and Li, Shuguang and Ma, Mingjian},
  title = {Prediction of the optical properties in photonic crystal fiber using support vector machine based on radial basis functions},
  journal = {Optik},
  volume = {275},
  pages = {170603},
  month = {March},
  year = {2023},
  doi = {10.1016/j.ijleo.2023.170603},
  url = {https://linkinghub.elsevier.com/retrieve/pii/S0030402623000992},
  issn = {00304026},
  language = {en},
}

@article{ref060,
  author = {Bansal, Malti and Goyal, Apoorva and Choudhary, Apoorva},
  title = {A comparative analysis of {K}-nearest neighbor, genetic, support vector machine, decision tree, and long short term memory algorithms in machine learning},
  journal = {Decision Analytics Journal},
  volume = {3},
  pages = {100071},
  month = {June},
  year = {2022},
  doi = {10.1016/j.dajour.2022.100071},
  url = {https://linkinghub.elsevier.com/retrieve/pii/S2772662222000261},
  issn = {27726622},
  language = {en},
}

@book{ref061,
  author = {Chen, Bingqi and Chen, Siyao},
  title = {Machine {Vision} {Technology}},
  series = {Advanced and {Intelligent} {Manufacturing} in {China}},
  publisher = {Springer Nature Singapore},
  address = {Singapore},
  edition = {1st ed. 2026},
  year = {2026},
  doi = {10.1007/978-981-95-1184-6},
  isbn = {9789819511846},
  language = {eng},
}

@article{ref062,
  author = {LeCun, Yann and Bengio, Yoshua and Hinton, Geoffrey},
  title = {Deep learning},
  journal = {Nature},
  volume = {521},
  number = {7553},
  pages = {436--444},
  month = {May},
  year = {2015},
  doi = {10.1038/nature14539},
  url = {https://www.nature.com/articles/nature14539},
  issn = {0028-0836, 1476-4687},
  language = {en},
}

@article{ref063,
  author = {Lippmann, Richard P.},
  title = {An introduction to computing with neural nets},
  journal = {ACM SIGARCH Computer Architecture News},
  volume = {16},
  number = {1},
  pages = {7--25},
  month = {March},
  year = {1988},
  doi = {10.1145/44571.44572},
  url = {https://dl.acm.org/doi/10.1145/44571.44572},
  issn = {0163-5964},
  language = {en},
}

@article{ref064,
  author = {Gu, Jiuxiang and Wang, Zhenhua and Kuen, Jason and Ma, Lianyang and Shahroudy, Amir and Shuai, Bing and Liu, Ting and Wang, Xingxing and Wang, Gang and Cai, Jianfei and Chen, Tsuhan},
  title = {Recent advances in convolutional neural networks},
  journal = {Pattern Recognition},
  volume = {77},
  pages = {354--377},
  month = {May},
  year = {2018},
  doi = {10.1016/j.patcog.2017.10.013},
  url = {https://linkinghub.elsevier.com/retrieve/pii/S0031320317304120},
  issn = {00313203},
  language = {en},
}

@article{ref065,
  author = {Lecun, Y. and Bottou, L. and Bengio, Y. and Haffner, P.},
  title = {Gradient-based learning applied to document recognition},
  journal = {Proceedings of the IEEE},
  volume = {86},
  number = {11},
  pages = {2278--2324},
  month = {November},
  year = {1998},
  doi = {10.1109/5.726791},
  url = {http://ieeexplore.ieee.org/document/726791/},
  issn = {00189219},
}

@inproceedings{ref066,
  author = {Deng, Jia and Dong, Wei and Socher, Richard and Li, Li-Jia and {Kai Li} and {Li Fei-Fei}},
  title = {{ImageNet}: {A} large-scale hierarchical image database},
  booktitle = {2009 {IEEE} {Conference} on {Computer} {Vision} and {Pattern} {Recognition}},
  pages = {248--255},
  publisher = {IEEE},
  address = {Miami, FL},
  month = {June},
  year = {2009},
  doi = {10.1109/CVPR.2009.5206848},
  url = {https://ieeexplore.ieee.org/document/5206848/},
  isbn = {978-1-4244-3992-8},
}

@article{ref067,
  author = {Elman, Jeffrey L.},
  title = {Finding structure in time},
  journal = {Cognitive Science},
  volume = {14},
  number = {2},
  pages = {179--211},
  month = {March},
  year = {1990},
  doi = {10.1207/s15516709cog1402_1},
  url = {https://onlinelibrary.wiley.com/doi/10.1207/s15516709cog1402_1},
  issn = {0364-0213, 1551-6709},
  language = {en},
}

@article{ref068,
  author = {Hochreiter, Sepp and Schmidhuber, Jürgen},
  title = {Long short-term memory},
  journal = {Neural Computation},
  volume = {9},
  number = {8},
  pages = {1735--1780},
  month = {November},
  year = {1997},
  doi = {10.1162/neco.1997.9.8.1735},
  url = {https://direct.mit.edu/neco/article/9/8/1735-1780/6109},
  issn = {0899-7667, 1530-888X},
  language = {en},
}

@inproceedings{ref069,
  author = {Cho, Kyunghyun and Van Merrienboer, Bart and Gulcehre, Caglar and Bahdanau, Dzmitry and Bougares, Fethi and Schwenk, Holger and Bengio, Yoshua},
  title = {Learning phrase representations using {RNN} encoder-decoder for statistical machine translation},
  booktitle = {Proceedings of the 2014 {Conference} on {Empirical} {Methods} in {Natural} {Language} {Processing} ({EMNLP})},
  pages = {1724--1734},
  publisher = {Association for Computational Linguistics},
  address = {Doha, Qatar},
  year = {2014},
  doi = {10.3115/v1/D14-1179},
  url = {http://aclweb.org/anthology/D14-1179},
  language = {en},
}

@article{ref070,
  author = {Malkiel, Itzik and Mrejen, Michael and Nagler, Achiya and Arieli, Uri and Wolf, Lior and Suchowski, Haim},
  title = {Plasmonic nanostructure design and characterization via {Deep} {Learning}},
  journal = {Light: Science \& Applications},
  volume = {7},
  number = {1},
  pages = {60},
  month = {September},
  year = {2018},
  doi = {10.1038/s41377-018-0060-7},
  url = {https://www.nature.com/articles/s41377-018-0060-7},
  issn = {2047-7538},
  language = {en},
}

@article{ref071,
  author = {Deng, Wenqiang and Xu, Zhengji and Wang, Jinhao and Lv, Jinwen},
  title = {Long short-term memory neural network for directly inverse design of nanofin metasurface},
  journal = {Optics Letters},
  volume = {47},
  number = {13},
  pages = {3239},
  month = {July},
  year = {2022},
  doi = {10.1364/OL.458453},
  url = {https://opg.optica.org/abstract.cfm?URI=ol-47-13-3239},
  issn = {0146-9592, 1539-4794},
  language = {en},
}

@article{ref072,
  author = {An, Sensong and Zheng, Bowen and Tang, Hong and Shalaginov, Mikhail Y. and Zhou, Li and Li, Hang and Kang, Myungkoo and Richardson, Kathleen A. and Gu, Tian and Hu, Juejun and Fowler, Clayton and Zhang, Hualiang},
  title = {Multifunctional metasurface design with a generative adversarial network},
  journal = {Advanced Optical Materials},
  volume = {9},
  number = {5},
  pages = {2001433},
  month = {March},
  year = {2021},
  doi = {10.1002/adom.202001433},
  url = {https://advanced.onlinelibrary.wiley.com/doi/10.1002/adom.202001433},
  issn = {2195-1071, 2195-1071},
  language = {en},
}

@article{ref073,
  author = {Zhang, Zezhou and Yang, Chuanchuan and Qin, Yifeng and Feng, Hao and Feng, Jiqiang and Li, Hongbin},
  title = {Diffusion probabilistic model based accurate and high-degree-of-freedom metasurface inverse design},
  journal = {Nanophotonics},
  volume = {12},
  number = {20},
  pages = {3871--3881},
  month = {October},
  year = {2023},
  doi = {10.1515/nanoph-2023-0292},
  url = {https://www.degruyter.com/document/doi/10.1515/nanoph-2023-0292/html},
  issn = {2192-8614},
  language = {en},
}

@article{ref074,
  author = {Peurifoy, John and Shen, Yichen and Jing, Li and Yang, Yi and Cano-Renteria, Fidel and DeLacy, Brendan G. and Joannopoulos, John D. and Tegmark, Max and Soljačić, Marin},
  title = {Nanophotonic particle simulation and inverse design using artificial neural networks},
  journal = {Science Advances},
  volume = {4},
  number = {6},
  pages = {eaar4206},
  month = {June},
  year = {2018},
  doi = {10.1126/sciadv.aar4206},
  url = {https://www.science.org/doi/10.1126/sciadv.aar4206},
  issn = {2375-2548},
  language = {en},
}

@article{ref075,
  author = {Liu, Dianjing and Tan, Yixuan and Khoram, Erfan and Yu, Zongfu},
  title = {Training deep neural networks for the inverse design of nanophotonic structures},
  journal = {ACS Photonics},
  volume = {5},
  number = {4},
  pages = {1365--1369},
  month = {April},
  year = {2018},
  doi = {10.1021/acsphotonics.7b01377},
  url = {https://pubs.acs.org/doi/10.1021/acsphotonics.7b01377},
  issn = {2330-4022, 2330-4022},
  language = {en},
}

@inproceedings{ref076,
  author = {He, Kaiming and Zhang, Xiangyu and Ren, Shaoqing and Sun, Jian},
  title = {Deep residual learning for image recognition},
  booktitle = {2016 {IEEE} {Conference} on {Computer} {Vision} and {Pattern} {Recognition} ({CVPR})},
  pages = {770--778},
  publisher = {IEEE},
  address = {Las Vegas, NV, USA},
  month = {June},
  year = {2016},
  doi = {10.1109/CVPR.2016.90},
  url = {http://ieeexplore.ieee.org/document/7780459/},
  isbn = {978-1-4673-8851-1},
}

@incollection{ref077,
  author = {Ronneberger, Olaf and Fischer, Philipp and Brox, Thomas},
  editor = {Navab, Nassir and Hornegger, Joachim and Wells, William M. and Frangi, Alejandro F.},
  title = {U-{Net}: {Convolutional} networks for biomedical image segmentation},
  booktitle = {Medical {Image} {Computing} and {Computer}-{Assisted} {Intervention} – {MICCAI} 2015},
  series = {Lecture Notes in Computer Science},
  volume = {9351},
  pages = {234--241},
  publisher = {Springer International Publishing},
  address = {Cham},
  year = {2015},
  doi = {10.1007/978-3-319-24574-4_28},
  url = {http://link.springer.com/10.1007/978-3-319-24574-4_28},
  isbn = {978-3-319-24573-7 978-3-319-24574-4},
  language = {en},
}

@misc{ref078,
  author = {Goodfellow, Ian J. and Pouget-Abadie, Jean and Mirza, Mehdi and Xu, Bing and Warde-Farley, David and Ozair, Sherjil and Courville, Aaron and Bengio, Yoshua},
  title = {Generative adversarial networks},
  publisher = {arXiv},
  year = {2014},
  doi = {10.48550/ARXIV.1406.2661},
  url = {https://arxiv.org/abs/1406.2661},
  eprint = {1406.2661},
  archiveprefix = {arXiv},
  howpublished = {arXiv preprint arXiv:1406.2661},
}

@misc{ref079,
  author = {Ho, Jonathan and Jain, Ajay and Abbeel, Pieter},
  title = {Denoising diffusion probabilistic models},
  publisher = {arXiv},
  year = {2020},
  doi = {10.48550/ARXIV.2006.11239},
  url = {https://arxiv.org/abs/2006.11239},
  eprint = {2006.11239},
  archiveprefix = {arXiv},
  howpublished = {arXiv preprint arXiv:2006.11239},
}

@article{ref080,
  author = {Liu, Zhaocheng and Zhu, Dayu and Rodrigues, Sean P. and Lee, Kyu-Tae and Cai, Wenshan},
  title = {Generative model for the inverse design of metasurfaces},
  journal = {Nano Letters},
  volume = {18},
  number = {10},
  pages = {6570--6576},
  month = {October},
  year = {2018},
  doi = {10.1021/acs.nanolett.8b03171},
  url = {https://pubs.acs.org/doi/10.1021/acs.nanolett.8b03171},
  issn = {1530-6984, 1530-6992},
  language = {en},
}

@article{ref081,
  author = {Raissi, M. and Perdikaris, P. and Karniadakis, G.E.},
  title = {Physics-informed neural networks: {A} deep learning framework for solving forward and inverse problems involving nonlinear partial differential equations},
  journal = {Journal of Computational Physics},
  volume = {378},
  pages = {686--707},
  month = {February},
  year = {2019},
  doi = {10.1016/j.jcp.2018.10.045},
  url = {https://linkinghub.elsevier.com/retrieve/pii/S0021999118307125},
  issn = {00219991},
  language = {en},
}

@article{ref082,
  author = {Lim, Joowon and Psaltis, Demetri},
  title = {{MaxwellNet}: {Physics}-driven deep neural network training based on {Maxwell}’s equations},
  journal = {APL Photonics},
  volume = {7},
  number = {1},
  pages = {011301},
  month = {January},
  year = {2022},
  doi = {10.1063/5.0071616},
  url = {https://pubs.aip.org/app/article/7/1/011301/2835095/MaxwellNet-Physics-driven-deep-neural-network},
  issn = {2378-0967},
  language = {en},
}

@misc{ref083,
  author = {Li, Zongyi and Kovachki, Nikola and Azizzadenesheli, Kamyar and Liu, Burigede and Bhattacharya, Kaushik and Stuart, Andrew and Anandkumar, Anima},
  title = {Fourier neural operator for parametric partial differential equations},
  publisher = {arXiv},
  month = {May},
  year = {2021},
  doi = {10.48550/arXiv.2010.08895},
  url = {http://arxiv.org/abs/2010.08895},
  eprint = {2010.08895},
  archiveprefix = {arXiv},
  primaryclass = {cs.LG},
  howpublished = {arXiv preprint arXiv:2010.08895},
}

@article{ref084,
  author = {Lu, Lu and Jin, Pengzhan and Pang, Guofei and Zhang, Zhongqiang and Karniadakis, George Em},
  title = {Learning nonlinear operators via {DeepONet} based on the universal approximation theorem of operators},
  journal = {Nature Machine Intelligence},
  volume = {3},
  number = {3},
  pages = {218--229},
  month = {March},
  year = {2021},
  doi = {10.1038/s42256-021-00302-5},
  url = {https://www.nature.com/articles/s42256-021-00302-5},
  issn = {2522-5839},
  language = {en},
}

@article{ref085,
  author = {Augenstein, Yannick and Repän, Taavi and Rockstuhl, Carsten},
  title = {Neural operator-based surrogate solver for free-form electromagnetic inverse design},
  journal = {ACS Photonics},
  volume = {10},
  number = {5},
  pages = {1547--1557},
  month = {May},
  year = {2023},
  doi = {10.1021/acsphotonics.3c00156},
  url = {https://pubs.acs.org/doi/10.1021/acsphotonics.3c00156},
  issn = {2330-4022, 2330-4022},
  language = {en},
}

@book{ref086,
  editor = {Campbell, Sawyer D. and Werner, Douglas H.},
  title = {Advances in {Electromagnetics} {Empowered} by {Artificial} {Intelligence} and {Deep} {Learning}},
  publisher = {Wiley-IEEE Press},
  address = {Hoboken, New Jersey},
  year = {2023},
  doi = {10.1002/9781119853923},
  isbn = {978-1-119-85391-6 978-1-119-85390-9},
}

@article{ref087,
  author = {Campbell, Sawyer D. and Jenkins, Ronald P. and O'Connor, Philip J. and Werner, Douglas},
  title = {The explosion of artificial intelligence in antennas and propagation: {How} deep learning is advancing our state of the art},
  journal = {IEEE Antennas and Propagation Magazine},
  volume = {63},
  number = {3},
  pages = {16--27},
  month = {June},
  year = {2021},
  doi = {10.1109/MAP.2020.3021433},
  url = {https://ieeexplore.ieee.org/document/9211490/},
  issn = {1045-9243, 1558-4143},
}

@article{ref088,
  author = {Joshi, Aravind K.},
  title = {Natural language processing},
  journal = {Science},
  volume = {253},
  number = {5025},
  pages = {1242--1249},
  month = {September},
  year = {1991},
  doi = {10.1126/science.253.5025.1242},
  url = {https://www.science.org/doi/10.1126/science.253.5025.1242},
  issn = {0036-8075, 1095-9203},
  language = {en},
}

@article{ref089,
  author = {Qiu, XiPeng and Sun, TianXiang and Xu, YiGe and Shao, YunFan and Dai, Ning and Huang, XuanJing},
  title = {Pre-trained models for natural language processing: {A} survey},
  journal = {Science China Technological Sciences},
  volume = {63},
  number = {10},
  pages = {1872--1897},
  month = {October},
  year = {2020},
  doi = {10.1007/s11431-020-1647-3},
  url = {https://link.springer.com/10.1007/s11431-020-1647-3},
  issn = {1674-7321, 1869-1900},
  language = {en},
}

@article{ref090,
  author = {Bengio, Y. and Simard, P. and Frasconi, P.},
  title = {Learning long-term dependencies with gradient descent is difficult},
  journal = {IEEE Transactions on Neural Networks},
  volume = {5},
  number = {2},
  pages = {157--166},
  month = {March},
  year = {1994},
  doi = {10.1109/72.279181},
  url = {https://ieeexplore.ieee.org/document/279181/},
  issn = {1045-9227, 1941-0093},
}

@inproceedings{ref091,
  author = {Devlin, Jacob and Chang, Ming-Wei and Lee, Kenton and Toutanova, Kristina},
  editor = {Burstein, Jill and Doran, Christy and Solorio, Thamar},
  title = {{BERT}: {Pre}-training of deep bidirectional transformers for language understanding},
  booktitle = {Proceedings of the 2019 {Conference} of the {North} {American} {Chapter} of the {Association} for {Computational} {Linguistics}: {Human} {Language} {Technologies}, {Volume} 1 ({Long} and {Short} {Papers})},
  pages = {4171--4186},
  publisher = {Association for Computational Linguistics},
  address = {Minneapolis, Minnesota},
  month = {June},
  year = {2019},
  doi = {10.18653/v1/N19-1423},
  url = {https://aclanthology.org/N19-1423/},
}

@misc{ref092,
  author = {Zhao, Wayne Xin and Zhou, Kun and Li, Junyi and Tang, Tianyi and Wang, Xiaolei and Hou, Yupeng and Min, Yingqian and Zhang, Beichen and Zhang, Junjie and Dong, Zican and Du, Yifan and Yang, Chen and Chen, Yushuo and Chen, Zhipeng and Jiang, Jinhao and Ren, Ruiyang and Li, Yifan and Tang, Xinyu and Liu, Zikang and Liu, Peiyu and Nie, Jian-Yun and Wen, Ji-Rong},
  title = {A survey of large language models},
  publisher = {arXiv},
  year = {2023},
  doi = {10.48550/ARXIV.2303.18223},
  url = {https://arxiv.org/abs/2303.18223},
  eprint = {2303.18223},
  archiveprefix = {arXiv},
  howpublished = {arXiv preprint arXiv:2303.18223},
}

@misc{ref093,
  author = {Kaplan, Jared and McCandlish, Sam and Henighan, Tom and Brown, Tom B. and Chess, Benjamin and Child, Rewon and Gray, Scott and Radford, Alec and Wu, Jeffrey and Amodei, Dario},
  title = {Scaling laws for neural language models},
  publisher = {arXiv},
  month = {January},
  year = {2020},
  doi = {10.48550/arXiv.2001.08361},
  url = {http://arxiv.org/abs/2001.08361},
  eprint = {2001.08361},
  archiveprefix = {arXiv},
  primaryclass = {cs},
  howpublished = {arXiv preprint arXiv:2001.08361},
}

@misc{ref094,
  author = {Hoffmann, Jordan and Borgeaud, Sebastian and Mensch, Arthur and Buchatskaya, Elena and Cai, Trevor and Rutherford, Eliza and Casas, Diego de Las and Hendricks, Lisa Anne and Welbl, Johannes and Clark, Aidan and Hennigan, Tom and Noland, Eric and Millican, Katie and Driessche, George van den and Damoc, Bogdan and Guy, Aurelia and Osindero, Simon and Simonyan, Karen and Elsen, Erich and Rae, Jack W. and Vinyals, Oriol and Sifre, Laurent},
  title = {Training compute-optimal large language models},
  publisher = {arXiv},
  month = {March},
  year = {2022},
  doi = {10.48550/arXiv.2203.15556},
  url = {http://arxiv.org/abs/2203.15556},
  eprint = {2203.15556},
  archiveprefix = {arXiv},
  primaryclass = {cs},
  howpublished = {arXiv preprint arXiv:2203.15556},
}

@misc{ref095,
  author = {Bommasani, Rishi and Hudson, Drew A. and Adeli, Ehsan and Altman, Russ and Arora, Simran and von Arx, Sydney and Bernstein, Michael S. and Bohg, Jeannette and Bosselut, Antoine and Brunskill, Emma and Brynjolfsson, Erik and Buch, Shyamal and Card, Dallas and Castellon, Rodrigo and Chatterji, Niladri and Chen, Annie and Creel, Kathleen and Davis, Jared Quincy and Demszky, Dora and Donahue, Chris and Doumbouya, Moussa and Durmus, Esin and Ermon, Stefano and Etchemendy, John and Ethayarajh, Kawin and Fei-Fei, Li and Finn, Chelsea and Gale, Trevor and Gillespie, Lauren and Goel, Karan and Goodman, Noah and Grossman, Shelby and Guha, Neel and Hashimoto, Tatsunori and Henderson, Peter and Hewitt, John and Ho, Daniel E. and Hong, Jenny and Hsu, Kyle and Huang, Jing and Icard, Thomas and Jain, Saahil and Jurafsky, Dan and Kalluri, Pratyusha and Karamcheti, Siddharth and Keeling, Geoff and Khani, Fereshte and Khattab, Omar and Koh, Pang Wei and Krass, Mark and Krishna, Ranjay and Kuditipudi, Rohith and Kumar, Ananya and Ladhak, Faisal and Lee, Mina and Lee, Tony and Leskovec, Jure and Levent, Isabelle and Li, Xiang Lisa and Li, Xuechen and Ma, Tengyu and Malik, Ali and Manning, Christopher D. and Mirchandani, Suvir and Mitchell, Eric and Munyikwa, Zanele and Nair, Suraj and Narayan, Avanika and Narayanan, Deepak and Newman, Ben and Nie, Allen and Niebles, Juan Carlos and Nilforoshan, Hamed and Nyarko, Julian and Ogut, Giray and Orr, Laurel and Papadimitriou, Isabel and Park, Joon Sung and Piech, Chris and Portelance, Eva and Potts, Christopher and Raghunathan, Aditi and Reich, Rob and Ren, Hongyu and Rong, Frieda and Roohani, Yusuf and Ruiz, Camilo and Ryan, Jack and Ré, Christopher and Sadigh, Dorsa and Sagawa, Shiori and Santhanam, Keshav and Shih, Andy and Srinivasan, Krishnan and Tamkin, Alex and Taori, Rohan and Thomas, Armin W. and Tramèr, Florian and Wang, Rose E. and Wang, William and Wu, Bohan and Wu, Jiajun and Wu, Yuhuai and Xie, Sang Michael and Yasunaga, Michihiro and You, Jiaxuan and Zaharia, Matei and Zhang, Michael and Zhang, Tianyi and Zhang, Xikun and Zhang, Yuhui and Zheng, Lucia and Zhou, Kaitlyn and Liang, Percy},
  title = {On the opportunities and risks of foundation models},
  publisher = {arXiv},
  year = {2021},
  doi = {10.48550/ARXIV.2108.07258},
  url = {https://arxiv.org/abs/2108.07258},
  eprint = {2108.07258},
  archiveprefix = {arXiv},
  howpublished = {arXiv preprint arXiv:2108.07258},
}

@article{ref096,
  author = {Zhuang, Fuzhen and Qi, Zhiyuan and Duan, Keyu and Xi, Dongbo and Zhu, Yongchun and Zhu, Hengshu and Xiong, Hui and He, Qing},
  title = {A comprehensive survey on transfer learning},
  journal = {Proceedings of the IEEE},
  volume = {109},
  number = {1},
  pages = {43--76},
  month = {January},
  year = {2021},
  doi = {10.1109/JPROC.2020.3004555},
  url = {https://ieeexplore.ieee.org/document/9134370/},
  issn = {0018-9219, 1558-2256},
}

@misc{ref097,
  author = {Howard, Jeremy and Ruder, Sebastian},
  title = {Universal language model fine-tuning for text classification},
  publisher = {arXiv},
  month = {May},
  year = {2018},
  doi = {10.48550/arXiv.1801.06146},
  url = {http://arxiv.org/abs/1801.06146},
  eprint = {1801.06146},
  archiveprefix = {arXiv},
  primaryclass = {cs},
  howpublished = {arXiv preprint arXiv:1801.06146},
}

@misc{ref098,
  author = {Wei, Jason and Tay, Yi and Bommasani, Rishi and Raffel, Colin and Zoph, Barret and Borgeaud, Sebastian and Yogatama, Dani and Bosma, Maarten and Zhou, Denny and Metzler, Donald and Chi, Ed H. and Hashimoto, Tatsunori and Vinyals, Oriol and Liang, Percy and Dean, Jeff and Fedus, William},
  title = {Emergent abilities of large language models},
  publisher = {arXiv},
  year = {2022},
  doi = {10.48550/ARXIV.2206.07682},
  url = {https://arxiv.org/abs/2206.07682},
  eprint = {2206.07682},
  archiveprefix = {arXiv},
  howpublished = {arXiv preprint arXiv:2206.07682},
}

@misc{ref099,
  author = {Shoeybi, Mohammad and Patwary, Mostofa and Puri, Raul and LeGresley, Patrick and Casper, Jared and Catanzaro, Bryan},
  title = {Megatron-{LM}: {Training} multi-billion parameter language models using model parallelism},
  publisher = {arXiv},
  month = {March},
  year = {2020},
  doi = {10.48550/arXiv.1909.08053},
  url = {http://arxiv.org/abs/1909.08053},
  eprint = {1909.08053},
  archiveprefix = {arXiv},
  primaryclass = {cs},
  howpublished = {arXiv preprint arXiv:1909.08053},
}

@inproceedings{ref100,
  author = {Rajbhandari, Samyam and Rasley, Jeff and Ruwase, Olatunji and He, Yuxiong},
  title = {{ZeRO}: {Memory} optimizations toward training trillion parameter models},
  booktitle = {{SC20}: {International} {Conference} for {High} {Performance} {Computing}, {Networking}, {Storage} and {Analysis}},
  pages = {1--16},
  publisher = {IEEE},
  address = {Atlanta, GA, USA},
  month = {November},
  year = {2020},
  doi = {10.1109/SC41405.2020.00024},
  url = {https://ieeexplore.ieee.org/document/9355301/},
  isbn = {978-1-7281-9998-6},
}

@inproceedings{ref101,
  author = {Bender, Emily M. and Gebru, Timnit and McMillan-Major, Angelina and Shmitchell, Shmargaret},
  title = {On the dangers of stochastic parrots: {Can} language models be too big? \texttwemoji{parrot}},
  booktitle = {Proceedings of the 2021 {ACM} {Conference} on {Fairness}, {Accountability}, and {Transparency}},
  pages = {610--623},
  publisher = {ACM},
  address = {Virtual Event Canada},
  month = {March},
  year = {2021},
  doi = {10.1145/3442188.3445922},
  url = {https://dl.acm.org/doi/10.1145/3442188.3445922},
  isbn = {978-1-4503-8309-7},
  language = {en},
}

@inproceedings{ref102,
  author = {Ouyang, Long and Wu, Jeffrey and Jiang, Xu and Almeida, Diogo and Wainwright, Carroll and Mishkin, Pamela and Zhang, Chong and Agarwal, Sandhini and Slama, Katarina and Ray, Alex and Schulman, John and Hilton, Jacob and Kelton, Fraser and Miller, Luke and Simens, Maddie and Askell, Amanda and Welinder, Peter and Christiano, Paul F and Leike, Jan and Lowe, Ryan},
  editor = {Koyejo, S. and Mohamed, S. and Agarwal, A. and Belgrave, D. and Cho, K. and Oh, A.},
  title = {Training language models to follow instructions with human feedback},
  booktitle = {Advances in {Neural} {Information} {Processing} {Systems}},
  volume = {35},
  pages = {27730--27744},
  publisher = {Curran Associates, Inc.},
  year = {2022},
  doi = {10.52202/068431-2011},
  url = {https://proceedings.neurips.cc/paper_files/paper/2022/file/b1efde53be364a73914f58805a001731-Paper-Conference.pdf},
}

@article{ref103,
  author = {Huang, Lei and Yu, Weijiang and Ma, Weitao and Zhong, Weihong and Feng, Zhangyin and Wang, Haotian and Chen, Qianglong and Peng, Weihua and Feng, Xiaocheng and Qin, Bing and Liu, Ting},
  title = {A survey on hallucination in large language models: {Principles}, taxonomy, challenges, and open questions},
  journal = {ACM Transactions on Information Systems},
  volume = {43},
  number = {2},
  pages = {1--55},
  month = {March},
  year = {2025},
  doi = {10.1145/3703155},
  url = {https://dl.acm.org/doi/10.1145/3703155},
  issn = {1046-8188, 1558-2868},
  language = {en},
}

@misc{ref104,
  author = {Lialin, Vladislav and Deshpande, Vijeta and Yao, Xiaowei and Rumshisky, Anna},
  title = {Scaling down to scale up: {A} guide to parameter-efficient fine-tuning},
  publisher = {arXiv},
  year = {2023},
  doi = {10.48550/ARXIV.2303.15647},
  url = {https://arxiv.org/abs/2303.15647},
  eprint = {2303.15647},
  archiveprefix = {arXiv},
  howpublished = {arXiv preprint arXiv:2303.15647},
}

@misc{ref105,
  author = {Han, Zeyu and Gao, Chao and Liu, Jinyang and Zhang, Jeff and Zhang, Sai Qian},
  title = {Parameter-efficient fine-tuning for large models: {A} comprehensive survey},
  publisher = {arXiv},
  year = {2024},
  doi = {10.48550/ARXIV.2403.14608},
  url = {https://arxiv.org/abs/2403.14608},
  eprint = {2403.14608},
  archiveprefix = {arXiv},
  howpublished = {arXiv preprint arXiv:2403.14608},
}

@article{ref106,
  author = {Xu, Lingling and Xie, Haoran and Qin, S. Joe and Tao, Xiaohui and Wang, Fu Lee},
  title = {Parameter-efficient fine-tuning methods for pretrained language models: {A} critical review and assessment},
  journal = {IEEE Transactions on Pattern Analysis and Machine Intelligence},
  volume = {48},
  number = {6},
  pages = {6107--6126},
  year = {2026},
  doi = {10.1109/TPAMI.2026.3657354},
  url = {https://ieeexplore.ieee.org/document/11364256/},
  issn = {0162-8828, 2160-9292, 1939-3539},
}

@inproceedings{ref107,
  author = {Hu, Edward J. and Shen, Yelong and Wallis, Phillip and Allen-Zhu, Zeyuan and Li, Yuanzhi and Wang, Shean and Wang, Lu and Chen, Weizhu},
  title = {{LoRA}: {Low}-rank adaptation of large language models},
  booktitle = {{ICLR} 2022},
  month = {April},
  year = {2022},
  url = {https://www.microsoft.com/en-us/research/publication/lora-low-rank-adaptation-of-large-language-models/},
}

@misc{ref108,
  author = {Yang, Menglin and Chen, Jialin and Tao, Jinkai and Zhang, Yifei and Liu, Jiahong and Zhang, Jiasheng and Ma, Qiyao and Verma, Harshit and Zhang, Regina and Zhou, Min and King, Irwin and Ying, Rex},
  title = {Low-rank adaptation for foundation models: {A} comprehensive review},
  publisher = {arXiv},
  year = {2025},
  doi = {10.48550/ARXIV.2501.00365},
  url = {https://arxiv.org/abs/2501.00365},
  eprint = {2501.00365},
  archiveprefix = {arXiv},
  howpublished = {arXiv preprint arXiv:2501.00365},
}

@article{ref109,
  author = {Liu, Pengfei and Yuan, Weizhe and Fu, Jinlan and Jiang, Zhengbao and Hayashi, Hiroaki and Neubig, Graham},
  title = {Pre-train, prompt, and predict: {A} systematic survey of prompting methods in natural language processing},
  journal = {ACM Computing Surveys},
  volume = {55},
  number = {9},
  pages = {1--35},
  month = {September},
  year = {2023},
  doi = {10.1145/3560815},
  url = {https://dl.acm.org/doi/10.1145/3560815},
  issn = {0360-0300, 1557-7341},
  language = {en},
}

@misc{ref110,
  author = {Dong, Qingxiu and Li, Lei and Dai, Damai and Zheng, Ce and Ma, Jingyuan and Li, Rui and Xia, Heming and Xu, Jingjing and Wu, Zhiyong and Liu, Tianyu and Chang, Baobao and Sun, Xu and Sui, Zhifang},
  title = {A survey on in-context learning},
  publisher = {arXiv},
  year = {2023},
  doi = {10.48550/ARXIV.2301.00234},
  url = {https://arxiv.org/abs/2301.00234},
  eprint = {2301.00234},
  archiveprefix = {arXiv},
  howpublished = {arXiv preprint arXiv:2301.00234},
}

@inproceedings{ref111,
  author = {Brown, Tom and Mann, Benjamin and Ryder, Nick and Subbiah, Melanie and Kaplan, Jared D and Dhariwal, Prafulla and Neelakantan, Arvind and Shyam, Pranav and Sastry, Girish and Askell, Amanda and Agarwal, Sandhini and Herbert-Voss, Ariel and Krueger, Gretchen and Henighan, Tom and Child, Rewon and Ramesh, Aditya and Ziegler, Daniel and Wu, Jeffrey and Winter, Clemens and Hesse, Chris and Chen, Mark and Sigler, Eric and Litwin, Mateusz and Gray, Scott and Chess, Benjamin and Clark, Jack and Berner, Christopher and McCandlish, Sam and Radford, Alec and Sutskever, Ilya and Amodei, Dario},
  editor = {Larochelle, H. and Ranzato, M. and Hadsell, R. and Balcan, M. F. and Lin, H.},
  title = {Language models are few-shot learners},
  booktitle = {Advances in {Neural} {Information} {Processing} {Systems}},
  volume = {33},
  pages = {1877--1901},
  publisher = {Curran Associates, Inc.},
  year = {2020},
  url = {https://proceedings.neurips.cc/paper_files/paper/2020/file/1457c0d6bfcb4967418bfb8ac142f64a-Paper.pdf},
}

@misc{ref112,
  author = {Gao, Yunfan and Xiong, Yun and Gao, Xinyu and Jia, Kangxiang and Pan, Jinliu and Bi, Yuxi and Dai, Yi and Sun, Jiawei and Wang, Meng and Wang, Haofen},
  title = {Retrieval-augmented generation for large language models: {A} survey},
  publisher = {arXiv},
  year = {2023},
  doi = {10.48550/ARXIV.2312.10997},
  url = {https://arxiv.org/abs/2312.10997},
  eprint = {2312.10997},
  archiveprefix = {arXiv},
  howpublished = {arXiv preprint arXiv:2312.10997},
}

@misc{ref113,
  author = {Lewis, Patrick and Perez, Ethan and Piktus, Aleksandra and Petroni, Fabio and Karpukhin, Vladimir and Goyal, Naman and Küttler, Heinrich and Lewis, Mike and Yih, Wen-tau and Rocktäschel, Tim and Riedel, Sebastian and Kiela, Douwe},
  title = {Retrieval-augmented generation for knowledge-intensive {NLP} tasks},
  publisher = {arXiv},
  year = {2020},
  doi = {10.48550/ARXIV.2005.11401},
  url = {https://arxiv.org/abs/2005.11401},
  eprint = {2005.11401},
  archiveprefix = {arXiv},
  howpublished = {arXiv preprint arXiv:2005.11401},
}

@misc{ref114,
  author = {Wang, Lei and Ma, Chen and Feng, Xueyang and Zhang, Zeyu and Yang, Hao and Zhang, Jingsen and Chen, Zhiyuan and Tang, Jiakai and Chen, Xu and Lin, Yankai and Zhao, Wayne Xin and Wei, Zhewei and Wen, Ji-Rong},
  title = {A survey on large language model based autonomous agents},
  publisher = {arXiv},
  year = {2023},
  doi = {10.48550/ARXIV.2308.11432},
  url = {https://arxiv.org/abs/2308.11432},
  eprint = {2308.11432},
  archiveprefix = {arXiv},
  howpublished = {arXiv preprint arXiv:2308.11432},
}

@misc{ref115,
  author = {Hou, Xinyi and Zhao, Yanjie and Wang, Shenao and Wang, Haoyu},
  title = {Model context protocol ({MCP}): {Landscape}, security threats, and future research directions},
  publisher = {arXiv},
  year = {2025},
  doi = {10.48550/ARXIV.2503.23278},
  url = {https://arxiv.org/abs/2503.23278},
  eprint = {2503.23278},
  archiveprefix = {arXiv},
  howpublished = {arXiv preprint arXiv:2503.23278},
}

@misc{ref116,
  author = {Yao, Shunyu and Zhao, Jeffrey and Yu, Dian and Du, Nan and Shafran, Izhak and Narasimhan, Karthik and Cao, Yuan},
  title = {{ReAct}: {Synergizing} reasoning and acting in language models},
  publisher = {arXiv},
  year = {2022},
  doi = {10.48550/ARXIV.2210.03629},
  url = {https://arxiv.org/abs/2210.03629},
  eprint = {2210.03629},
  archiveprefix = {arXiv},
  howpublished = {arXiv preprint arXiv:2210.03629},
}

@misc{ref117,
  author = {Schick, Timo and Dwivedi-Yu, Jane and Dessì, Roberto and Raileanu, Roberta and Lomeli, Maria and Zettlemoyer, Luke and Cancedda, Nicola and Scialom, Thomas},
  title = {Toolformer: {Language} models can teach themselves to use tools},
  publisher = {arXiv},
  year = {2023},
  doi = {10.48550/ARXIV.2302.04761},
  url = {https://arxiv.org/abs/2302.04761},
  eprint = {2302.04761},
  archiveprefix = {arXiv},
  howpublished = {arXiv preprint arXiv:2302.04761},
}

@misc{ref118,
  author = {Shinn, Noah and Cassano, Federico and Berman, Edward and Gopinath, Ashwin and Narasimhan, Karthik and Yao, Shunyu},
  title = {Reflexion: {Language} agents with verbal reinforcement learning},
  publisher = {arXiv},
  year = {2023},
  doi = {10.48550/ARXIV.2303.11366},
  url = {https://arxiv.org/abs/2303.11366},
  eprint = {2303.11366},
  archiveprefix = {arXiv},
  howpublished = {arXiv preprint arXiv:2303.11366},
}

@misc{ref119,
  author = {Madaan, Aman and Tandon, Niket and Gupta, Prakhar and Hallinan, Skyler and Gao, Luyu and Wiegreffe, Sarah and Alon, Uri and Dziri, Nouha and Prabhumoye, Shrimai and Yang, Yiming and Gupta, Shashank and Majumder, Bodhisattwa Prasad and Hermann, Katherine and Welleck, Sean and Yazdanbakhsh, Amir and Clark, Peter},
  title = {Self-{Refine}: {Iterative} refinement with self-feedback},
  publisher = {arXiv},
  year = {2023},
  doi = {10.48550/ARXIV.2303.17651},
  url = {https://arxiv.org/abs/2303.17651},
  eprint = {2303.17651},
  archiveprefix = {arXiv},
  howpublished = {arXiv preprint arXiv:2303.17651},
}

@misc{ref120,
  author = {Yao, Shunyu and Yu, Dian and Zhao, Jeffrey and Shafran, Izhak and Griffiths, Thomas L. and Cao, Yuan and Narasimhan, Karthik},
  title = {Tree of {Thoughts}: {Deliberate} problem solving with large language models},
  publisher = {arXiv},
  year = {2023},
  doi = {10.48550/ARXIV.2305.10601},
  url = {https://arxiv.org/abs/2305.10601},
  eprint = {2305.10601},
  archiveprefix = {arXiv},
  howpublished = {arXiv preprint arXiv:2305.10601},
}

@misc{ref121,
  author = {Wu, Qingyun and Bansal, Gagan and Zhang, Jieyu and Wu, Yiran and Li, Beibin and Zhu, Erkang and Jiang, Li and Zhang, Xiaoyun and Zhang, Shaokun and Liu, Jiale and Awadallah, Ahmed Hassan and White, Ryen W and Burger, Doug and Wang, Chi},
  title = {{AutoGen}: {Enabling} next-gen {LLM} applications via multi-agent conversation},
  publisher = {arXiv},
  year = {2023},
  doi = {10.48550/ARXIV.2308.08155},
  url = {https://arxiv.org/abs/2308.08155},
  eprint = {2308.08155},
  archiveprefix = {arXiv},
  howpublished = {arXiv preprint arXiv:2308.08155},
}

@misc{ref122,
  author = {Chen, Mark and Tworek, Jerry and Jun, Heewoo and Yuan, Qiming and Pinto, Henrique Ponde de Oliveira and Kaplan, Jared and Edwards, Harri and Burda, Yuri and Joseph, Nicholas and Brockman, Greg and Ray, Alex and Puri, Raul and Krueger, Gretchen and Petrov, Michael and Khlaaf, Heidy and Sastry, Girish and Mishkin, Pamela and Chan, Brooke and Gray, Scott and Ryder, Nick and Pavlov, Mikhail and Power, Alethea and Kaiser, Lukasz and Bavarian, Mohammad and Winter, Clemens and Tillet, Philippe and Such, Felipe Petroski and Cummings, Dave and Plappert, Matthias and Chantzis, Fotios and Barnes, Elizabeth and Herbert-Voss, Ariel and Guss, William Hebgen and Nichol, Alex and Paino, Alex and Tezak, Nikolas and Tang, Jie and Babuschkin, Igor and Balaji, Suchir and Jain, Shantanu and Saunders, William and Hesse, Christopher and Carr, Andrew N. and Leike, Jan and Achiam, Josh and Misra, Vedant and Morikawa, Evan and Radford, Alec and Knight, Matthew and Brundage, Miles and Murati, Mira and Mayer, Katie and Welinder, Peter and McGrew, Bob and Amodei, Dario and McCandlish, Sam and Sutskever, Ilya and Zaremba, Wojciech},
  title = {Evaluating large language models trained on code},
  publisher = {arXiv},
  year = {2021},
  doi = {10.48550/ARXIV.2107.03374},
  url = {https://arxiv.org/abs/2107.03374},
  eprint = {2107.03374},
  archiveprefix = {arXiv},
  howpublished = {arXiv preprint arXiv:2107.03374},
}

@inproceedings{ref123,
  author = {Houlsby, Neil and Giurgiu, Andrei and Jastrzebski, Stanislaw and Morrone, Bruna and De Laroussilhe, Quentin and Gesmundo, Andrea and Attariyan, Mona and Gelly, Sylvain},
  editor = {Chaudhuri, Kamalika and Salakhutdinov, Ruslan},
  title = {Parameter-efficient transfer learning for {NLP}},
  booktitle = {Proceedings of the 36th {International} {Conference} on {Machine} {Learning}},
  series = {Proceedings of {Machine} {Learning} {Research}},
  volume = {97},
  pages = {2790--2799},
  publisher = {PMLR},
  month = {June},
  year = {2019},
  url = {https://proceedings.mlr.press/v97/houlsby19a.html},
}

@inproceedings{ref124,
  author = {Li, Xiang Lisa and Liang, Percy},
  title = {Prefix-tuning: {Optimizing} continuous prompts for generation},
  booktitle = {Proceedings of the 59th {Annual} {Meeting} of the {Association} for {Computational} {Linguistics} and the 11th {International} {Joint} {Conference} on {Natural} {Language} {Processing} ({Volume} 1: {Long} {Papers})},
  pages = {4582--4597},
  publisher = {Association for Computational Linguistics},
  address = {Online},
  year = {2021},
  doi = {10.18653/v1/2021.acl-long.353},
  url = {https://aclanthology.org/2021.acl-long.353},
  language = {en},
}

@inproceedings{ref125,
  author = {Zhou, Yuxiang and Li, Jiazheng and Xiang, Yanzheng and Yan, Hanqi and Gui, Lin and He, Yulan},
  title = {The mystery of in-context learning: {A} comprehensive survey on interpretation and analysis},
  booktitle = {Proceedings of the 2024 {Conference} on {Empirical} {Methods} in {Natural} {Language} {Processing}},
  pages = {14365--14378},
  publisher = {Association for Computational Linguistics},
  address = {Miami, Florida, USA},
  year = {2024},
  doi = {10.18653/v1/2024.emnlp-main.795},
  url = {https://aclanthology.org/2024.emnlp-main.795},
  language = {en},
}

@article{ref126,
  author = {Zhao, Penghao and Zhang, Hailin and Yu, Qinhan and Wang, Zhengren and Geng, Yunteng and Fu, Fangcheng and Yang, Ling and Zhang, Wentao and Jiang, Jie and Cui, Bin},
  title = {Retrieval-augmented generation for {AI}-generated content: {A} survey},
  journal = {Data Science and Engineering},
  volume = {11},
  number = {1},
  pages = {1--29},
  month = {January},
  year = {2026},
  doi = {10.1007/s41019-025-00335-5},
  url = {https://link.springer.com/10.1007/s41019-025-00335-5},
  issn = {2364-1185, 2364-1541},
  language = {en},
}

@misc{ref127,
  author = {Minaee, Shervin and Mikolov, Tomas and Nikzad, Narjes and Chenaghlu, Meysam and Socher, Richard and Amatriain, Xavier and Gao, Jianfeng},
  title = {Large language models: {A} survey},
  publisher = {arXiv},
  month = {March},
  year = {2025},
  doi = {10.48550/arXiv.2402.06196},
  url = {http://arxiv.org/abs/2402.06196},
  eprint = {2402.06196},
  archiveprefix = {arXiv},
  primaryclass = {cs},
  howpublished = {arXiv preprint arXiv:2402.06196},
}

@misc{ref128,
  author = {Liu, Ze and Lin, Yutong and Cao, Yue and Hu, Han and Wei, Yixuan and Zhang, Zheng and Lin, Stephen and Guo, Baining},
  title = {Swin transformer: {Hierarchical} vision transformer using shifted windows},
  publisher = {arXiv},
  month = {August},
  year = {2021},
  doi = {10.48550/arXiv.2103.14030},
  url = {http://arxiv.org/abs/2103.14030},
  eprint = {2103.14030},
  archiveprefix = {arXiv},
  primaryclass = {cs},
  howpublished = {arXiv preprint arXiv:2103.14030},
}

@article{ref129,
  author = {Chen, Wei and Gao, Yuan and Li, Yuyang and Yan, Yiming and Ou, Jun‐Yu and Ma, Wenzhuang and Zhu, Jinfeng},
  title = {Broadband solar metamaterial absorbers empowered by transformer‐based deep learning},
  journal = {Advanced Science},
  volume = {10},
  number = {13},
  pages = {2206718},
  month = {May},
  year = {2023},
  doi = {10.1002/advs.202206718},
  url = {https://advanced.onlinelibrary.wiley.com/doi/10.1002/advs.202206718},
  issn = {2198-3844, 2198-3844},
  language = {en},
}

@article{ref130,
  author = {Niu, Chen and Phaneuf, Mario and Qiu, Tianke and Mojabi, Puyan},
  title = {A deep learning-based approach to design metasurfaces from desired far-field specifications},
  journal = {IEEE Open Journal of Antennas and Propagation},
  volume = {4},
  pages = {641--653},
  year = {2023},
  doi = {10.1109/OJAP.2023.3292108},
  url = {https://ieeexplore.ieee.org/document/10172261/},
  issn = {2637-6431},
}

@article{ref131,
  author = {Cai, Yijun and Huang, Yangpeng and Feng, Naixing and Huang, Zhixiang},
  title = {Improved transformer-based target matching of terahertz broadband reflective metamaterials with monolayer graphene},
  journal = {IEEE Transactions on Microwave Theory and Techniques},
  volume = {71},
  number = {8},
  pages = {3284--3293},
  month = {August},
  year = {2023},
  doi = {10.1109/TMTT.2023.3249357},
  url = {https://ieeexplore.ieee.org/document/10061702/},
  issn = {0018-9480, 1557-9670},
}

@article{ref132,
  author = {Chen, Wei and Li, Yuyang and Liu, Yineng and Gao, Yuan and Yan, Yiming and Dong, Zhaogang and Zhu, Jinfeng},
  title = {All‐dielectric {SERS} metasurface with strong coupling quasi‐{BIC} energized by transformer‐based deep learning},
  journal = {Advanced Optical Materials},
  volume = {12},
  number = {4},
  pages = {2301697},
  month = {February},
  year = {2024},
  doi = {10.1002/adom.202301697},
  url = {https://advanced.onlinelibrary.wiley.com/doi/10.1002/adom.202301697},
  issn = {2195-1071, 2195-1071},
  language = {en},
}

@article{ref133,
  author = {Lin, Chia-Hsiang and Huang, Shih-Hsiu and Lin, Ting-Hsuan and Wu, Pin Chieh},
  title = {Metasurface-empowered snapshot hyperspectral imaging with convex/deep ({CODE}) small-data learning theory},
  journal = {Nature Communications},
  volume = {14},
  number = {1},
  pages = {6979},
  month = {November},
  year = {2023},
  doi = {10.1038/s41467-023-42381-5},
  url = {https://www.nature.com/articles/s41467-023-42381-5},
  issn = {2041-1723},
  language = {en},
}

@article{ref134,
  author = {Gao, Yuan and Chen, Wei and Li, Fajun and Zhuang, Mingyong and Yan, Yiming and Wang, Jun and Wang, Xiang and Dong, Zhaogang and Ma, Wei and Zhu, Jinfeng},
  title = {Meta‐attention deep learning for smart development of metasurface sensors},
  journal = {Advanced Science},
  volume = {11},
  number = {42},
  pages = {2405750},
  month = {November},
  year = {2024},
  doi = {10.1002/advs.202405750},
  url = {https://onlinelibrary.wiley.com/doi/10.1002/advs.202405750},
  issn = {2198-3844, 2198-3844},
  language = {en},
}

@article{ref135,
  author = {Zhang, Jie and Qian, Chao and You, Guangfeng and Wang, Tao and Saifullah, Yasir and Abdi‐Ghaleh, Reza and Chen, Hongsheng},
  title = {Harnessing the missing spectral correlation for metasurface inverse design},
  journal = {Advanced Science},
  volume = {11},
  number = {33},
  pages = {2308807},
  month = {September},
  year = {2024},
  doi = {10.1002/advs.202308807},
  url = {https://advanced.onlinelibrary.wiley.com/doi/10.1002/advs.202308807},
  issn = {2198-3844, 2198-3844},
  language = {en},
}

@misc{ref136,
  author = {Sun, Zhi and Li, Tianyue and Kuang, Shiqi and Yun, Xue and He, Minru and Fu, Boyan and Fu, Yunlai and Zhao, Tianyu and Wang, Shaowei and Liang, Yansheng and Wang, Shuming and Lei, Ming},
  title = {On-demand quick metasurface design with neighborhood attention transformer},
  publisher = {arXiv},
  month = {December},
  year = {2024},
  doi = {10.48550/arXiv.2412.08405},
  url = {http://arxiv.org/abs/2412.08405},
  eprint = {2412.08405},
  archiveprefix = {arXiv},
  primaryclass = {physics},
  howpublished = {arXiv preprint arXiv:2412.08405},
}

@article{ref137,
  author = {Ma, Jindong and Ma, Zihang and Li, Muyang and Li, Yujie and Tan, Benying and Ding, Shuxue},
  title = {{TRMD}: {A} transformer-based reverse design model for quad-band metasurface absorbers},
  journal = {Physica Scripta},
  volume = {100},
  number = {1},
  pages = {016003},
  month = {January},
  year = {2025},
  doi = {10.1088/1402-4896/ad9558},
  url = {https://iopscience.iop.org/article/10.1088/1402-4896/ad9558},
  issn = {0031-8949, 1402-4896},
}

@article{ref138,
  author = {Yin, Shan and Zhong, Haotian and Huang, Wei and Zhang, Wentao},
  title = {Deep learning enabled design of terahertz high-{Q} metamaterials},
  journal = {Optics \& Laser Technology},
  volume = {181},
  pages = {111684},
  month = {February},
  year = {2025},
  doi = {10.1016/j.optlastec.2024.111684},
  url = {https://linkinghub.elsevier.com/retrieve/pii/S0030399224011423},
  issn = {00303992},
  language = {en},
}

@article{ref139,
  author = {Belonovskii, Aleksei and Girshova, Elizaveta and Lähderanta, Erkki and Kaliteevski, Mikhail},
  title = {Predicting {VCSEL} emission properties using transformer neural networks},
  journal = {Laser \& Photonics Reviews},
  volume = {19},
  number = {12},
  pages = {2401636},
  month = {June},
  year = {2025},
  doi = {10.1002/lpor.202401636},
  url = {https://onlinelibrary.wiley.com/doi/10.1002/lpor.202401636},
  issn = {1863-8880, 1863-8899},
  language = {en},
}

@article{ref140,
  author = {Yan, Jiahao and Yi, Jilong and Ma, Churong and Bao, Yanjun and Chen, Qin and Li, Baojun},
  title = {Metasurface vision transformer: {A} generic {AI} model for metasurface inverse design},
  journal = {Nanophotonics},
  volume = {15},
  number = {1},
  pages = {e70001},
  year = {2026},
  doi = {10.1002/nap2.70001},
  url = {https://onlinelibrary.wiley.com/doi/abs/10.1002/nap2.70001},
}

@misc{ref141,
  author = {Chen, Wei and Gao, Yuan and Yan, Yiming and Shen, Jiaqing and Lin, Yongxiang and Zhuang, Mingyong and Dong, Zhaogang and Zhu, Jinfeng},
  title = {Reality-infused deep learning for angle-resolved quasi-optical {Fourier} surfaces},
  publisher = {arXiv},
  month = {May},
  year = {2025},
  doi = {10.48550/arXiv.2505.05011},
  url = {http://arxiv.org/abs/2505.05011},
  eprint = {2505.05011},
  archiveprefix = {arXiv},
  primaryclass = {physics},
  howpublished = {arXiv preprint arXiv:2505.05011},
}

@article{ref142,
  author = {Ma, Jindong and Ma, Zihang and Li, Muyang and Li, Yujie and Tan, Benying and Ding, Shuxue},
  title = {Inverse design of multi-band absorbers driven by peak features: {TIMD}-based local feature enhancement},
  journal = {Physica Scripta},
  volume = {100},
  number = {8},
  pages = {086013},
  month = {August},
  year = {2025},
  doi = {10.1088/1402-4896/adfbae},
  url = {https://iopscience.iop.org/article/10.1088/1402-4896/adfbae},
  issn = {0031-8949, 1402-4896},
}

@misc{ref143,
  author = {Li, Hao and Bogdanov, Andrey},
  title = {{MetaDiT}: {Enabling} fine-grained constraints in high-degree-of freedom metasurface design},
  publisher = {arXiv},
  month = {August},
  year = {2025},
  doi = {10.48550/arXiv.2508.05076},
  url = {http://arxiv.org/abs/2508.05076},
  eprint = {2508.05076},
  archiveprefix = {arXiv},
  primaryclass = {physics},
  howpublished = {arXiv preprint arXiv:2508.05076},
}

@article{ref144,
  author = {Liao, Jiahui and Bian, Xucong and Cheng, Xiang’ai and Li, Quanjiang and Jiang, Yuting and Lou, Shaozhen and Wang, Haoqian and Hua, Zixiao and Li, Teng and Zhang, Jiangbin and Xu, Zhongjie and Hu, Yueqiang and Xing, Zhongyang},
  title = {{GLSaT}: {A} spectral-aware transformer-based network enabling highly efficient and precise inverse design in metasurface optical filters},
  journal = {Advanced Photonics Nexus},
  volume = {4},
  number = {05},
  month = {October},
  year = {2025},
  doi = {10.1117/1.APN.4.5.056014},
  url = {https://www.spiedigitallibrary.org/journals/advanced-photonics-nexus/volume-4/issue-05/056014/GLSaT--a-spectral-aware-transformer-based-network-enabling-highly/10.1117/1.APN.4.5.056014.full},
  issn = {2791-1519},
}

@article{ref145,
  author = {Bian, Xucong and Cheng, Xiang’ai and Liao, Jiahui and Hua, Zixiao and Xu, Zhongjie and Zhang, Jiangbin and Xing, Zhongyang},
  title = {A transformer-based approach to facilitate inverse design of achromatic metasurfaces},
  journal = {Photonics},
  volume = {12},
  number = {9},
  pages = {913},
  month = {September},
  year = {2025},
  doi = {10.3390/photonics12090913},
  url = {https://www.mdpi.com/2304-6732/12/9/913},
  issn = {2304-6732},
  language = {en},
}

@article{ref146,
  author = {Xin, Qi and Huang, Hai and Li, Chenyu and Shi, Kewei and Zhang, Zhaoyu},
  title = {{POST}: {Photonic} swin transformer for automated and efficient prediction of {PCSEL}},
  journal = {Nanophotonics},
  volume = {14},
  number = {22},
  pages = {3599--3610},
  month = {November},
  year = {2025},
  doi = {10.1515/nanoph-2025-0317},
  url = {https://www.degruyterbrill.com/document/doi/10.1515/nanoph-2025-0317/html},
  issn = {2192-8614},
  language = {en},
}

@article{ref147,
  author = {Chu, Bo and Li, Yan and Zhong, Facheng and Chen, Zhuo},
  title = {On-demand design of strong unidirectional scattering dominated by electric resonances via transformer-based network},
  journal = {Optics Communications},
  volume = {616},
  pages = {133321},
  month = {October},
  year = {2026},
  doi = {10.1016/j.optcom.2026.133321},
  url = {https://linkinghub.elsevier.com/retrieve/pii/S0030401826004608},
  issn = {00304018},
  language = {en},
}

@article{ref148,
  author = {Zeng, Zhenjia and Wang, Lei and Wu, Yiran and Hu, Zhipeng and Evans, Julian and Zhu, Xinhua and Ye, Gaoao and He, Sailing},
  title = {Utilizing mixed training and multi-head attention to address data shift in {AI}-based electromagnetic solvers for nano-structured metamaterials},
  journal = {Nanomaterials},
  volume = {13},
  number = {20},
  pages = {2778},
  month = {October},
  year = {2023},
  doi = {10.3390/nano13202778},
  url = {https://www.mdpi.com/2079-4991/13/20/2778},
  issn = {2079-4991},
  language = {en},
}

@article{ref149,
  author = {Huang, Yangpeng and Feng, Naixing and Cai, Yijun},
  title = {Artificial intelligence-generated terahertz multi-resonant metasurfaces via improved transformer and {CGAN} neural networks},
  journal = {Journal of Lightwave Technology},
  volume = {42},
  number = {5},
  pages = {1518--1525},
  month = {March},
  year = {2024},
  doi = {10.1109/JLT.2023.3325156},
  url = {https://ieeexplore.ieee.org/document/10286417/},
  issn = {0733-8724, 1558-2213},
}

@article{ref150,
  author = {Yuan, Xiaogen and Wei, Zhongchao and Ma, Qiongxiong and Ding, Wen and Guo, Jianping},
  title = {Multitask learning deep neural networks enable embedded design of active metamaterials},
  journal = {ACS Applied Materials \& Interfaces},
  volume = {16},
  number = {20},
  pages = {26500--26511},
  month = {May},
  year = {2024},
  doi = {10.1021/acsami.4c01730},
  url = {https://pubs.acs.org/doi/10.1021/acsami.4c01730},
  issn = {1944-8244, 1944-8252},
  language = {en},
}

@article{ref151,
  author = {Mao, Bingxuan and Chang, Honghao and Xing, Xiaohua and Zhang, Qiankun and Zou, Die and Liu, Yin and Yao, Jianquan and Bi, Haixia and Wu, Liang},
  title = {Designing ultra-broadband terahertz polarization converters based on the transformer model},
  journal = {Optics Communications},
  volume = {559},
  pages = {130434},
  month = {May},
  year = {2024},
  doi = {10.1016/j.optcom.2024.130434},
  url = {https://linkinghub.elsevier.com/retrieve/pii/S0030401824001718},
  issn = {00304018},
  language = {en},
}

@article{ref152,
  author = {Ma, Taigao and Wang, Haozhu and Guo, L. Jay},
  title = {{OptoGPT}: {A} foundation model for inverse design in optical multilayer thin film structures},
  journal = {Opto-Electronic Advances},
  volume = {7},
  number = {7},
  pages = {240062},
  year = {2024},
  doi = {10.29026/oea.2024.240062},
  url = {http://www.oejournal.org//article/doi/10.29026/oea.2024.240062},
  issn = {2096-4579},
}

@misc{ref153,
  author = {Dang, David and Love, Stuart and Salib, Meena and Dang, Quynh and Rothfarb, Samuel and Alnatour, Mysk and Salij, Andrew and Chen, Hou-Tong and Lee, Ho Wai and Kort-Kamp, Wilton J. M.},
  title = {Meta-{GPT}: {Decoding} the metasurface genome with generative artificial intelligence},
  publisher = {arXiv},
  year = {2025},
  doi = {10.48550/ARXIV.2512.12888},
  url = {https://arxiv.org/abs/2512.12888},
  eprint = {2512.12888},
  archiveprefix = {arXiv},
  howpublished = {arXiv preprint arXiv:2512.12888},
}

@article{ref154,
  author = {Wu, Qi and Fan, Wen-Hui and Qin, Chong and Jiang, Xiao-Qiang},
  title = {Inverse design of reconfigurable terahertz metasurface with wide phase modulation range via a high–precision {CNN}–transformer framework},
  journal = {Journal of Physics D: Applied Physics},
  volume = {58},
  number = {47},
  pages = {475103},
  month = {November},
  year = {2025},
  doi = {10.1088/1361-6463/ae1e89},
  url = {https://iopscience.iop.org/article/10.1088/1361-6463/ae1e89},
  issn = {0022-3727, 1361-6463},
}

@article{ref155,
  author = {Shan, Hanyu and Jiang, Tao},
  title = {{IGPSO}-{ViTNet}: {A} vision transformer network-assisted {IGPSO} algorithm for dynamic optimization design of absorption and phase cancellation metasurface},
  journal = {IEEE Transactions on Microwave Theory and Techniques},
  volume = {74},
  number = {2},
  pages = {1211--1224},
  year = {2026},
  doi = {10.1109/TMTT.2025.3623612},
  url = {https://ieeexplore.ieee.org/document/11224335/},
  issn = {0018-9480, 1557-9670},
}

@article{ref156,
  author = {Marzban, Reza and Abiri, Hamed and Pestourie, Raphaël and Adibi, Ali},
  title = {{HiLAB}: {A} hybrid inverse‐design framework},
  journal = {Small Methods},
  volume = {9},
  number = {11},
  pages = {e00975},
  month = {November},
  year = {2025},
  doi = {10.1002/smtd.202500975},
  url = {https://onlinelibrary.wiley.com/doi/10.1002/smtd.202500975},
  issn = {2366-9608, 2366-9608},
  language = {en},
}

@misc{ref157,
  author = {Beltagy, Iz and Peters, Matthew E. and Cohan, Arman},
  title = {Longformer: {The} long-document transformer},
  publisher = {arXiv},
  year = {2020},
  doi = {10.48550/ARXIV.2004.05150},
  url = {https://arxiv.org/abs/2004.05150},
  eprint = {2004.05150},
  archiveprefix = {arXiv},
  howpublished = {arXiv preprint arXiv:2004.05150},
}

@misc{ref158,
  author = {Choromanski, Krzysztof and Likhosherstov, Valerii and Dohan, David and Song, Xingyou and Gane, Andreea and Sarlos, Tamas and Hawkins, Peter and Davis, Jared and Mohiuddin, Afroz and Kaiser, Lukasz and Belanger, David and Colwell, Lucy and Weller, Adrian},
  title = {Rethinking attention with performers},
  publisher = {arXiv},
  year = {2020},
  doi = {10.48550/ARXIV.2009.14794},
  url = {https://arxiv.org/abs/2009.14794},
  eprint = {2009.14794},
  archiveprefix = {arXiv},
  howpublished = {arXiv preprint arXiv:2009.14794},
}

@misc{ref159,
  author = {Dao, Tri and Fu, Daniel Y. and Ermon, Stefano and Rudra, Atri and Ré, Christopher},
  title = {{FlashAttention}: {Fast} and memory-efficient exact attention with {IO}-awareness},
  publisher = {arXiv},
  year = {2022},
  doi = {10.48550/ARXIV.2205.14135},
  url = {https://arxiv.org/abs/2205.14135},
  eprint = {2205.14135},
  archiveprefix = {arXiv},
  howpublished = {arXiv preprint arXiv:2205.14135},
}

@misc{ref160,
  author = {Dosovitskiy, Alexey and Beyer, Lucas and Kolesnikov, Alexander and Weissenborn, Dirk and Zhai, Xiaohua and Unterthiner, Thomas and Dehghani, Mostafa and Minderer, Matthias and Heigold, Georg and Gelly, Sylvain and Uszkoreit, Jakob and Houlsby, Neil},
  title = {An image is worth 16x16 words: transformers for image recognition at scale},
  publisher = {arXiv},
  month = {June},
  year = {2021},
  doi = {10.48550/arXiv.2010.11929},
  url = {http://arxiv.org/abs/2010.11929},
  eprint = {2010.11929},
  archiveprefix = {arXiv},
  primaryclass = {cs},
  howpublished = {arXiv preprint arXiv:2010.11929},
}

@inproceedings{ref161,
  author = {Jaegle, Andrew and Gimeno, Felix and Brock, Andy and Vinyals, Oriol and Zisserman, Andrew and Carreira, Joao},
  editor = {Meila, Marina and Zhang, Tong},
  title = {Perceiver: {General} perception with iterative attention},
  booktitle = {Proceedings of the 38th {International} {Conference} on {Machine} {Learning}},
  series = {Proceedings of {Machine} {Learning} {Research}},
  volume = {139},
  pages = {4651--4664},
  publisher = {PMLR},
  month = {July},
  year = {2021},
  url = {https://proceedings.mlr.press/v139/jaegle21a.html},
}

@misc{ref162,
  author = {Alayrac, Jean-Baptiste and Donahue, Jeff and Luc, Pauline and Miech, Antoine and Barr, Iain and Hasson, Yana and Lenc, Karel and Mensch, Arthur and Millican, Katie and Reynolds, Malcolm and Ring, Roman and Rutherford, Eliza and Cabi, Serkan and Han, Tengda and Gong, Zhitao and Samangooei, Sina and Monteiro, Marianne and Menick, Jacob and Borgeaud, Sebastian and Brock, Andrew and Nematzadeh, Aida and Sharifzadeh, Sahand and Binkowski, Mikolaj and Barreira, Ricardo and Vinyals, Oriol and Zisserman, Andrew and Simonyan, Karen},
  title = {Flamingo: {A} visual language model for few-shot learning},
  publisher = {arXiv},
  year = {2022},
  doi = {10.48550/ARXIV.2204.14198},
  url = {https://arxiv.org/abs/2204.14198},
  eprint = {2204.14198},
  archiveprefix = {arXiv},
  howpublished = {arXiv preprint arXiv:2204.14198},
}

@article{ref163,
  author = {Zhang, Huanshu and Kang, Lei and Campbell, Sawyer D. and Werner, Douglas H.},
  title = {Chat to chip: {Large} language model based design of arbitrarily shaped metasurfaces},
  journal = {Nanophotonics},
  volume = {14},
  number = {22},
  pages = {3625--3633},
  month = {November},
  year = {2025},
  doi = {10.1515/nanoph-2025-0343},
  url = {https://www.degruyterbrill.com/document/doi/10.1515/nanoph-2025-0343/html},
  issn = {2192-8614},
  language = {en},
}

@article{ref164,
  author = {Kim, Myungjoon and Park, Hyeonjin and Shin, Jonghwa},
  title = {Nanophotonic device design based on large language models: {Multilayer} and metasurface examples},
  journal = {Nanophotonics},
  volume = {14},
  number = {8},
  pages = {1273--1282},
  month = {April},
  year = {2025},
  doi = {10.1515/nanoph-2024-0674},
  url = {https://www.degruyterbrill.com/document/doi/10.1515/nanoph-2024-0674/html},
  issn = {2192-8614},
  language = {en},
}

@article{ref165,
  author = {Lu, Darui and Deng, Yang and Malof, Jordan M. and Padilla, Willie J.},
  title = {Learning electromagnetic metamaterial physics with {ChatGPT}},
  journal = {IEEE Access},
  volume = {13},
  pages = {51513--51526},
  year = {2025},
  doi = {10.1109/ACCESS.2025.3552418},
  url = {https://ieeexplore.ieee.org/document/10930934/},
  issn = {2169-3536},
}

@article{ref166,
  author = {Lupoiu, Robert and Shao, Yixuan and Dai, Tianxiang and Mao, Chenkai and Edée, Kofi and Fan, Jonathan A.},
  title = {A multi-agentic framework for real-time, autonomous freeform metasurface design},
  journal = {Science Advances},
  volume = {11},
  number = {44},
  pages = {eadx8006},
  month = {October},
  year = {2025},
  doi = {10.1126/sciadv.adx8006},
  url = {https://www.science.org/doi/10.1126/sciadv.adx8006},
  issn = {2375-2548},
  language = {en},
}

@article{ref167,
  author = {Iwanaga, Masanobu and Watanabe, Keisuke},
  title = {Generative-{AI}-assisted knowledge-based augmented exploration in nanophotonics},
  journal = {Science and Technology of Advanced Materials: Methods},
  volume = {5},
  number = {1},
  pages = {2596940},
  month = {December},
  year = {2025},
  doi = {10.1080/27660400.2025.2596940},
  url = {https://www.tandfonline.com/doi/full/10.1080/27660400.2025.2596940},
  issn = {2766-0400},
  language = {en},
}

@article{ref168,
  author = {Jenkins, Ronald P. and Campbell, Sawyer D. and Werner, Douglas H.},
  title = {Establishing exhaustive metasurface robustness against fabrication uncertainties through deep learning},
  journal = {Nanophotonics},
  volume = {10},
  number = {18},
  pages = {4497--4509},
  month = {December},
  year = {2021},
  doi = {10.1515/nanoph-2021-0428},
  url = {https://www.degruyter.com/document/doi/10.1515/nanoph-2021-0428/html},
  issn = {2192-8614},
  language = {en},
}

@article{ref169,
  author = {Li, Jiahao and Wang, Chenxu and Zhou, Qi and Tang, Junyang and Luo, Ming and Wu, Lin and Tao, Jin and Huang, Tianye and Li, Hanbing and Li, Xiang and Qiu, Ying},
  title = {An ultra-compact efficient silicon power beam splitter based on large language model by inverse design},
  journal = {Optics Communications},
  volume = {601},
  pages = {132733},
  month = {March},
  year = {2026},
  doi = {10.1016/j.optcom.2025.132733},
  url = {https://linkinghub.elsevier.com/retrieve/pii/S0030401825012611},
  issn = {00304018},
  language = {en},
}

@misc{ref170,
  author = {Yang, Shujie and Zhao, Xuzhe and Zhang, Yuqi and Tang, Yansong and Dong, Kaichen},
  title = {{CoSP}: {Reconfigurable} multi-state metamaterial inverse design via contrastive pretrained large language model},
  publisher = {arXiv},
  month = {November},
  year = {2025},
  doi = {10.48550/arXiv.2511.16135},
  url = {http://arxiv.org/abs/2511.16135},
  eprint = {2511.16135},
  archiveprefix = {arXiv},
  primaryclass = {physics},
  howpublished = {arXiv preprint arXiv:2511.16135},
}

@misc{ref171,
  author = {Wu, Bei and Xiong, Bo and Luo, Haiyao and Li, Yaqi and Zhang, Li and Chen, Qiaolu and Chen, Hongsheng and Yang, Yihao},
  title = {Agentic metasurface design with self-correcting language-model systems},
  publisher = {arXiv},
  month = {May},
  year = {2026},
  doi = {10.48550/arXiv.2605.22647},
  url = {http://arxiv.org/abs/2605.22647},
  eprint = {2605.22647},
  archiveprefix = {arXiv},
  primaryclass = {physics.optics},
  howpublished = {arXiv preprint arXiv:2605.22647},
}

@article{ref172,
  author = {Lu, Darui and Malof, Jordan M. and Padilla, Willie J.},
  title = {An agentic framework for autonomous metamaterial modeling and inverse design},
  journal = {ACS Photonics},
  volume = {12},
  number = {11},
  pages = {6071--6080},
  month = {November},
  year = {2025},
  doi = {10.1021/acsphotonics.5c01514},
  url = {https://pubs.acs.org/doi/10.1021/acsphotonics.5c01514},
  issn = {2330-4022, 2330-4022},
  language = {en},
}

@misc{ref173,
  author = {Li, Renjie and Zhang, Ceyao and Mao, Sixuan and Zhou, Xiyuan and Yin, Feng and Theodoridis, Sergios and Zhang, Zhaoyu},
  title = {{LLM4Laser}: {Large} language models automate the design of lasers},
  publisher = {arXiv},
  year = {2021},
  doi = {10.48550/ARXIV.2104.12145},
  url = {https://arxiv.org/abs/2104.12145},
  eprint = {2104.12145},
  archiveprefix = {arXiv},
  howpublished = {arXiv preprint arXiv:2104.12145},
}

@inproceedings{ref174,
  author = {Zhao, Yizhe and Chen, Lili and Li, Xinyi and Dong, Yue-Jiang and Ren, Ju and Zhang, Yaoxue},
  title = {{MetaGen}: {LLM}-driven generative framework for intelligent metasurface element},
  booktitle = {Proceedings of the 23rd {Annual} {International} {Conference} on {Mobile} {Systems}, {Applications} and {Services}},
  pages = {333--346},
  publisher = {ACM},
  address = {Hilton Anaheim Anaheim CA USA},
  month = {June},
  year = {2025},
  doi = {10.1145/3711875.3729129},
  url = {https://dl.acm.org/doi/10.1145/3711875.3729129},
  isbn = {979-8-4007-1453-5},
  language = {en},
}

@misc{ref175,
  author = {Lai, Zheyuan and Pu, Yingming},
  title = {{PriM}: {Principle}-inspired material discovery through multi-agent collaboration},
  publisher = {arXiv},
  month = {April},
  year = {2025},
  doi = {10.48550/arXiv.2504.08810},
  url = {http://arxiv.org/abs/2504.08810},
  eprint = {2504.08810},
  archiveprefix = {arXiv},
  primaryclass = {cs},
  howpublished = {arXiv preprint arXiv:2504.08810},
}

@article{ref176,
  author = {Huang, Yi and Zheng, Bowen and Dong, Yunxi and Tang, Hong and Zhao, Huan and Shawon, Rakibul Hasan and An, Sensong and Zhang, Hualiang},
  title = {{MCP}-enabled {LLM} for meta-optics inverse design: leveraging differentiable solver without {LLM} expertise},
  journal = {Nanophotonics},
  volume = {14},
  number = {27},
  pages = {5589--5602},
  month = {December},
  year = {2025},
  doi = {10.1515/nanoph-2025-0507},
  url = {https://www.degruyterbrill.com/document/doi/10.1515/nanoph-2025-0507/html},
  issn = {2192-8614},
  language = {en},
}

@article{ref177,
  author = {Huang, Yi and Zheng, Bowen and Dong, Yunxi and Tang, Hong and Zhao, Huan and Shawon, S. M. Rakibul Hasan and Zhang, Hualiang},
  title = {A self‐evolving agentic framework for metasurface inverse design},
  journal = {Laser \& Photonics Reviews},
  pages = {e71739},
  month = {August},
  year = {2026},
  doi = {10.1002/lpor.71739},
  url = {https://onlinelibrary.wiley.com/doi/10.1002/lpor.71739},
  issn = {1863-8880, 1863-8899},
  language = {en},
}

@article{ref178,
  author = {Hu, Shengguo and Li, Mingyi and Xu, Jiawen and Zhang, Hongrui and Zhang, Shanghang and Cui, Tie Jun and Del Hougne, Philipp and Li, Lianlin},
  title = {Electromagnetic metamaterial agent},
  journal = {Light: Science \& Applications},
  volume = {14},
  number = {1},
  pages = {12},
  month = {January},
  year = {2025},
  doi = {10.1038/s41377-024-01678-w},
  url = {https://www.nature.com/articles/s41377-024-01678-w},
  issn = {2047-7538},
  language = {en},
}

@inproceedings{ref179,
  author = {Yin, Haoran and Kononova, Anna V. and Bäck, Thomas and Van Stein, Niki},
  title = {Optimizing photonic structures with large language model driven algorithm discovery},
  booktitle = {Proceedings of the {Genetic} and {Evolutionary} {Computation} {Conference} {Companion}},
  pages = {2354--2362},
  publisher = {ACM},
  address = {NH Malaga Hotel Malaga Spain},
  month = {July},
  year = {2025},
  doi = {10.1145/3712255.3734288},
  url = {https://dl.acm.org/doi/10.1145/3712255.3734288},
  isbn = {979-8-4007-1464-1},
  language = {en},
}

@misc{ref180,
  author = {Maman, Roy and Ohana, David and Engelberg, Jacob and Levy, Uriel},
  title = {Prompt-to-prescription: {Towards} generative design of diffraction-limited refractive optics},
  publisher = {arXiv},
  year = {2026},
  doi = {10.48550/ARXIV.2602.13893},
  url = {https://arxiv.org/abs/2602.13893},
  eprint = {2602.13893},
  archiveprefix = {arXiv},
  howpublished = {arXiv preprint arXiv:2602.13893},
}

@article{ref181,
  author = {Papa, Lorenzo and Russo, Paolo and Amerini, Irene and Zhou, Luping},
  title = {A survey on efficient vision transformers: {Algorithms}, techniques, and performance benchmarking},
  journal = {IEEE Transactions on Pattern Analysis and Machine Intelligence},
  volume = {46},
  number = {12},
  pages = {7682--7700},
  month = {December},
  year = {2024},
  doi = {10.1109/TPAMI.2024.3392941},
  url = {https://ieeexplore.ieee.org/document/10508091/},
  issn = {0162-8828, 2160-9292, 1939-3539},
}

@article{ref182,
  author = {Rosy, N. Arockia and Balasubadra, K. and Deepa, K.},
  title = {Are vision transformers replacing convolutional neural networks in scene interpretation?: {A} review},
  journal = {Discover Applied Sciences},
  volume = {7},
  number = {9},
  pages = {932},
  month = {August},
  year = {2025},
  doi = {10.1007/s42452-025-07574-1},
  url = {https://link.springer.com/10.1007/s42452-025-07574-1},
  issn = {3004-9261},
  language = {en},
}

@article{ref183,
  author = {Han, Kai and Wang, Yunhe and Chen, Hanting and Chen, Xinghao and Guo, Jianyuan and Liu, Zhenhua and Tang, Yehui and Xiao, An and Xu, Chunjing and Xu, Yixing and Yang, Zhaohui and Zhang, Yiman and Tao, Dacheng},
  title = {A survey on vision transformer},
  journal = {IEEE Transactions on Pattern Analysis and Machine Intelligence},
  volume = {45},
  number = {1},
  pages = {87--110},
  month = {January},
  year = {2023},
  doi = {10.1109/TPAMI.2022.3152247},
  url = {https://ieeexplore.ieee.org/document/9716741/},
  issn = {0162-8828, 2160-9292, 1939-3539},
}

@article{ref184,
  author = {Jiang, Yongfeng and Cao, Siyang and Meng, Han and Zhou, Runze and Ren, Jianwei and Feng, Xiangchao and Shen, Cheng and Lu, Tianjian},
  title = {A data-driven design for sound absorption of acoustic metamaterials based on large language models},
  journal = {Scientific Reports},
  volume = {16},
  number = {1},
  pages = {517},
  month = {December},
  year = {2025},
  doi = {10.1038/s41598-025-29930-2},
  url = {https://www.nature.com/articles/s41598-025-29930-2},
  issn = {2045-2322},
  language = {en},
}

Huanshu Zhang is a Ph.D. candidate in Electrical Engineering at The Pennsylvania State University. He received his BS degree in Electrical Engineering from Sun Yat-sen University in 2023. His current research interests focus on AI (especially LLM) driven design of metasurfaces.

Kegeng Tang is currently a PhD student in Computer Science at the University of Tennessee at Chattanooga. He received his bachelor’s degree from Sun Yat-sen University. His research focuses on large language models and vision–language models, with an emphasis on reliable reasoning and efficient adaptation, and their applications in AI for science.

Lei Kang received the Ph.D. degree in Material Science and Engineering from Tsinghua University, China, in 2010. He is currently an Associate Research Professor in the Department of Electrical Engineering at The Pennsylvania State University. Dr. Kang’s research interests include metadevices at both RF and optical frequencies, nanophotonics, and nonlinear optics. He has published over 90 journal articles and has authored 1 book and 4 book chapters (h-index = 41, total citations \textasciitilde{} 6,400). He is a member of the IEEE and the OSA.

Sawyer D. Campbell received the B.S. degree in physics from Illinois Wesleyan University, Bloomington, IL, USA, in 2008, and the M.S. and Ph.D. degrees in optical sciences from the University of Arizona, Tucson, AZ, USA, in 2010 and 2013, respectively. In 2014, he joined the Computational Electromagnetics and Antennas Research Laboratory, Department of Electrical Engineering, The Pennsylvania State University. Since 2025 he is an Associate Professor in the Department of Electrical Engineering and Director of Workforce Development in the Center of Excellence in Directed Energy (CEDE) at Penn State. He is a senior member of the IEEE, OPTICA, SPIE, and UNSC-URSI Commission B. He is the past Chair and vice-Chair/Treasurer of the IEEE Central Pennsylvania Section. Dr. Campbell has extensive experience in the computational modeling and inverse-design of metamaterial- and transformation optics-based devices at RF, infrared, and optical frequencies. Additionally, he is an expert in the application of multi-objective and surrogate-assisted optimization techniques to challenging problems in both the RF and optical regimes. He has published over 220 technical papers and proceedings articles, 2 books, and is the author/coauthor of 5 book chapters. He serves as Associate Editor for IEEE Access and IEEE Transactions on Antennas and Propagation. His current research interests include metasurfaces, gradient-index lenses, high power microwave antennas, optimization, nanophotonics, and applications of deep learning to RF and optical inverse-design problems.

Zihao Wang (Member, IEEE) received the Diplôme d’ingénieur degree in computer science from École Supérieure d’Ingénieurs en Électrotechnique et Électronique, Paris, France, in 2016, and the Ph.D. degree from Inria, Valbonne, France. He was with Inria, Valbonne, France, and served as a Research Engineer with Philips Healthcare, Suzhou, China. He is currently an Assistant Professor with the Department of Computer Science at the University of Tennessee, Chattanooga, USA, where he leads the Laplace Lab. His research interests include reliable and data-efficient artificial intelligence for signal processing and biomedical engineering, with an emphasis on physics-grounded generative learning, computational anatomy priors, and robust medical image reconstruction and analysis under domain shift and acquisition constraints.

Douglas H. Werner holds the John L. and Genevieve H. McCain Chair Professorship with the Department of Electrical Engineering, The Pennsylvania State University, University Park, PA, USA. He is the Director of the Computational Electromagnetics and Antennas Research Laboratory (CEARL) and the Center of Excellence in Directed Energy (CEDE). He holds 21 patents, has published over 1175 technical papers and proceedings articles, 8 books, and 31 book chapters (h-index = 91, total citations > 40,000). Prof. Werner received the IEEE Antennas and Propagation Society Edward E. Altshuler Prize Paper Award and the Harold A. Wheeler Applications Prize Paper Award in 2011 and 2014, respectively. In 2018, he received the DoD Ordnance Technology Consortium (DOTC) Outstanding Technical Achievement Award. He also received the 2015 ACES Technical Achievement Award, the 2019 ACES Computational Electromagnetics Award, the IEEE Antennas and Propagation Society 2006 R. W. P. King Award, the 2019 Chen-To Tai Distinguished Educator Award, the 2023 John Kraus Antenna Award, and the 2024 Harrington-Mittra Award in Computational Electromagnetics. He is currently the editor for the IEEE Press Series on Electromagnetic Wave Theory \& Applications. He is also a senior editor of the newly launched IEEE Journal of Selected Topics in Electromagnetics, Antennas, and Propagation (JSTEAP). He is a fellow of nine professional societies including IEEE, IET, NAI, AAAS, ACES, SPIE, OSA/Optica, AAIA, and the PIER Electromagnetics Academy. He is also a Senior Member of the URSI and a Life Member of the Association of Old Crows (AOC).

\end{spacing}
\clearpage

\includepdf[
pages=1,
fitpaper=true
]{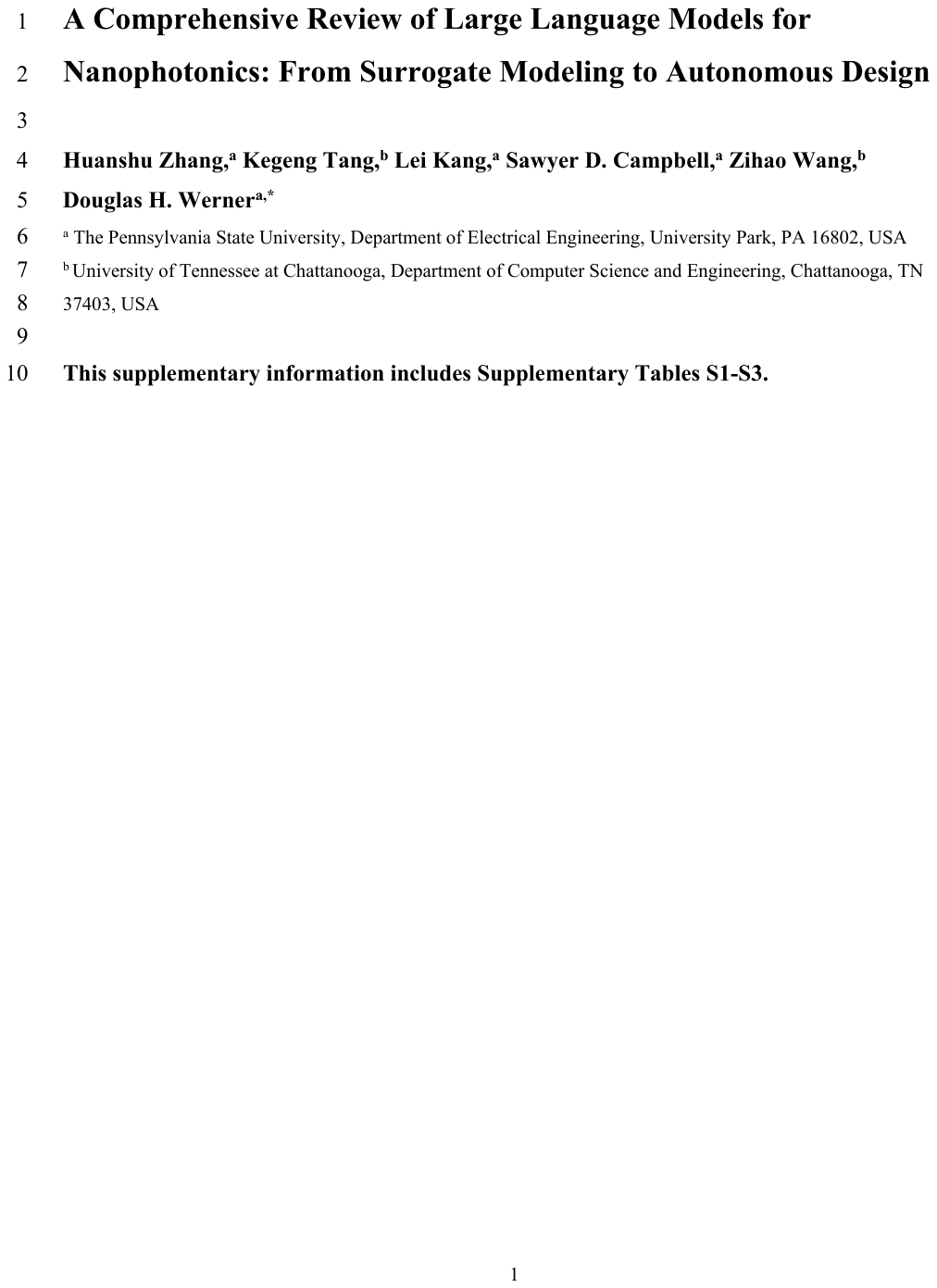}

\includepdf[
pages=2-25,
fitpaper=true
]{supplementary_vSub.pdf}
\end{document}